\documentclass[usenatbib]{eas}
\usepackage{graphicx}
\usepackage{mesjournaux}
\usepackage{natbib}
\newcommand{\be}{\begin{equation}}
\newcommand{\ee}{\end{equation}}

\begin{document}

%%-----------------------------
%%      the top matter
%%-----------------------------
\title{Braking down an accreting protostar: disc-locking, disc winds, stellar winds, X-winds and Magnetospheric Ejecta} 
\author{Jonathan Ferreira}\address{UJF-Grenoble 1/CNRS-INSU, Institut de Plan\'etologie et d'Astrophysique de Grenoble (IPAG) UMR 5274, Grenoble, F-38041, France}
\begin{abstract}
Classical T~Tauri stars are low mass young forming stars that are surrounded by a circumstellar accretion disc from which they gain mass. Despite this accretion and their own contraction that should both lead to their spin up, these stars seem to conserve instead an almost constant rotational period as long as the disc is maintained. Several scenarios have been proposed in the literature in order to explain this puzzling  "disc-locking" situation: either deposition in the disc of the stellar angular momentum by the stellar magnetosphere or its ejection through winds, providing thereby an explanation of jets from Young Stellar Objects.

In this lecture, these various mechanisms will be critically detailed, from the physics of the star-disc interaction to the launching of self-confined jets (disc winds, stellar winds, X-winds, conical winds). It will be shown that no simple model can account alone for the whole bulk of observational data and that "disc locking" requires a combination of some of them. 

\end{abstract}
\maketitle
%%-----------------------------

%%%%%%%%%%%%%%%%%%%%%%%%%%%%%%%%%%%%%%%%%%%%%%%%%%%%%%%%%%%%%
\section{Introduction}
%%%%%%%%%%%%%%%%%%%%%%%%%%%%%%%%%%%%%%%%%%%%%%%%%%%%%%%%%%%%%

For more than ten years, the circumstellar accretion disc in a classical T~Tauri star (hereafter cTTS) was envisioned to extend down to the stellar surface, where an equatorial boundary layer would mark the star-disc transition (see eg. \citealt{rege95} and references therein). This boundary layer would be responsible for the UV excess seen in Young Stellar Objects (hereafter YSO). However, it would also always provide a stellar spin up. Indeed, the stellar radius $R_*$ is always smaller than the co-rotation radius $r_{co}= (\sqrt{GM_*/\Omega_*^2})^{1/3}$, defined as the radius where the stellar angular velocity $\Omega_*$ is equal to the disc angular Keplerian velocity. This was highly problematic, especially when new observations came up showing that accreting TTS were slowly rotating, at roughly 10\% of their break-up speed despite accretion (see J. Bouvier's contribution, this book). 

One way out of this puzzle was to assume a totally different star-disc interaction. T Tauri stars would possess strong dipolar magnetic fields, able to truncate the disc at a radius $r_{t} > R_*$ and maintain their magnetosphere connected to the outer accretion disc. This magnetized star-disc interaction would be able to transfer the stellar angular momentum back to the disc without stopping accretion. In this new scenario, the UV excess would be explained by the accretion shock onto the (hard) stellar surface by the infalling disc material, enforced to follow the stellar magnetospheric field lines. These ideas, originally developed in the context of neutron stars (eg. \citealt{prin72, ghos77, elsn77}), were first proposed by  \citet{came90} and \citet{koni91} in the context of YSOs.

But some YSOs have also powerful jets and explaining their launching, power and kinematic properties is quite challenging (see \citealt{ray07} and references therein). In the eighties, after in particular the disc wind model of \citet{blan82}, it has been realized that (invisible) large scale magnetic fields could indeed be the main agent for driving and collimating these outflows \citep{pudr86}.    One alternative and tempting idea was to relate the two puzzling phenomena, namely the stellar angular momentum problem and the energy source of these winds. The "X-celerator" of \citet{shu88} was one such suggestion, although it appeared that it could not work in cTTS because of their too low rotation. Thus, for quite a long time (about thirty years) the jet formation problem and the stellar spin down problem remained very closely related.  It seems that we are about to better understand now these two phenomena and how they are coupled. 

This lecture is organized as follows.  Section 2 addresses the \citet{ghos78} paradigm of a dipolar magnetosphere interacting with a circumstellar accretion disc.  After almost two decades of theoretical investigations and especially thanks to the use of numerical experiments, one can fairly safely conclude that, contrary to the early expectations, such a magnetic configuration never leads to a stellar spin equilibrium and that the star gets spun up instead. Since the star cannot deposit its angular momentum in the disc, it must be expelled away. Section 3 then briefly recalls some properties of YSO jets. It will be argued that steady-state MHD models can be used as a first attempt to grasp jet properties. The general equations of steady-state, axisymmetric jet models are then derived and a list of such YSO jet models is given. Section 4 is devoted to the magnetized disc wind model \citep{ferr97}, Section 5 to a new class of Accretion Powered Stellar Wind models \citep{matt05b}, Section 6 to the X-wind model \citep{shu94a}. It will be argued that none of these models can efficiently brake down a cTTS. Time-dependent numerical simulations where unsteady Magnetospheric Ejections have been reported \citep{zann13} will be described in Section 7. It will be shown that such MEs may well be the answer to the stellar spin down problem. We finally conclude by proposing a paradigm for the magnetic environment  of cTTS and their related ejection properties.

%%%%%%%%%%%%%%%%%%%%%%%%%%%%%%%%%%%%%%%%%%%%%%%%%%%%%%%%%%%%%
\section{The disc-locking scenario}
%%%%%%%%%%%%%%%%%%%%%%%%%%%%%%%%%%%%%%%%%%%%%%%%%%%%%%%%%%%%%
\label{sec:GL}

\subsection{The Ghosh \& Lamb paradigm} 
%%%%%%%%%%%%%%%%%%%%%%%

%%%%%%%%%%
\begin{figure}[t]
\centering   \includegraphics[width=.8\textwidth]{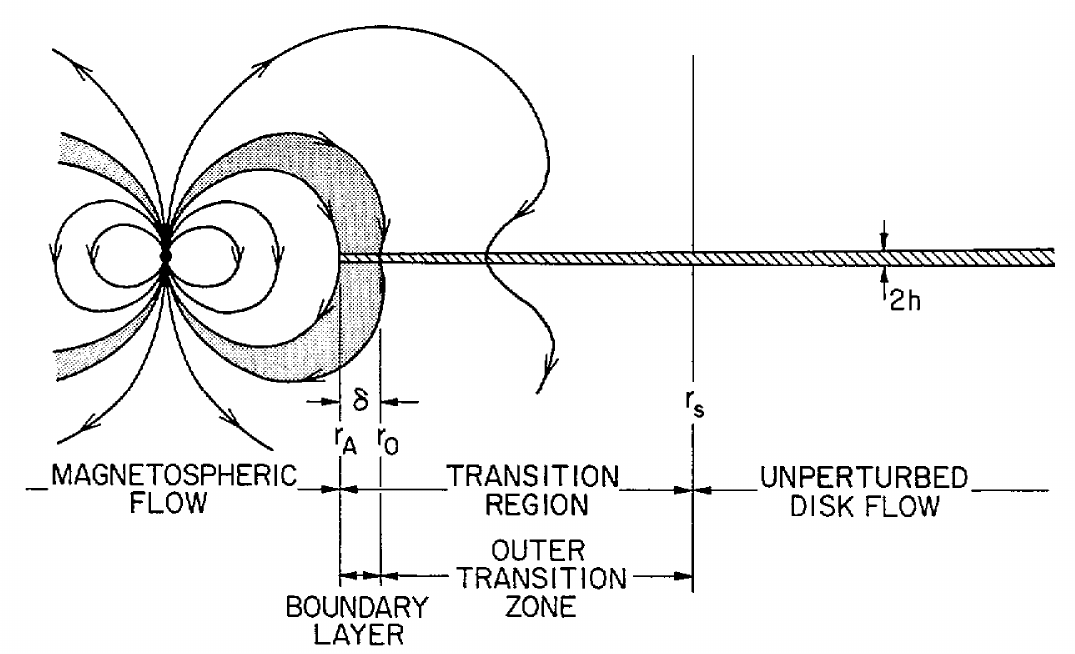}
  \caption{The \citet{ghos78} paradigm of a magnetic star-disc interaction.}
   \label{fig_1}
\end{figure}
%%%%%%%%

Since the early work of \citet{koni91}, there have been many analytical studies of the interaction between a stellar dipole field and its accretion disc in the context of YSOs (e.g. \citealt{came93, came95, love95, armi96, liJ96a, liJ96b, kold02a} to cite only a few). They all share the same assumptions (see Fig~\ref{fig_1}):\\
i- The model is axisymmetric, with a stellar magnetic dipole aligned with the rotation axis of the star and the disc;\\
ii- The disc is truncated at a radius $r_t$, whose value is estimated or a free parameter;\\ 
iii- Accretion onto the star occurs along a funnel flow originating mostly from $r_t$;\\
iv- The magnetosphere below $r_{in}= r_t$ is an unperturbed dipole field devoid of mass;\\
v- The magnetosphere beyond $r_t$ remains connected to the disc up to an outer radius $r_{out} \gg r_{t}$, whose value is estimated or a free parameter;\\
vi- The turbulent accretion disc remains mostly Keplerian (ie. thin) and viscosity transports radially outwardly the angular momentum;\\
vii- There is no mass loss from this system.\\ 

In order to compute the stellar spin evolution, one has to solve the angular momentum conservation equation
\be
\frac{\partial \rho \Omega r^2}{\partial t} \, +\,  div \, \left ( \rho \Omega r^2 \vec u_p - \frac{r B_\phi}{\mu_o} \vec B_p \right ) = 0
\label{eq:starevol}
\ee
and integrate it over the stellar volume. The first term writes $ \int_V dV \frac{\partial \rho \Omega r^2}{\partial t} = \frac{d I_* \Omega_*}{dt}$ where the stellar moment of inertia is $I_*= k^2 M_* R_*^2$ and $k$ is a constant of order unity depending on the stellar structure. The integration of the second term involves magnetic and mass flux contributions (torques) at the stellar surface, that vary according to the magnetic field lines considered: \\
- zone 1: The inner unperturbed dipole field below $r_t$ provides (by definition) a zero net torque.\\
- zone 2: At higher latitudes, closed magnetic field lines carry the funnel flow, from the disc truncation radius $r_t$ to the base of the funnel flow $r_{bf}$ (labelled respectively $r_A$ and $r_0$ in Fig~\ref{fig_1}). This part is responsible for the so called accretion torque $\Gamma_{acc}$.\\
- zone 3: Higher, closed magnetospheric field lines intersect the accretion disc from $r_{bf}$ to $r_{out}$. This part is responsible for the magnetic torque $\Gamma_{mag}$.\\
- zone 4: Beyond, at even higher latitudes, open field lines with an outflowing stellar wind are expected. But within the Ghosh \& Lamb picture, this wind contribution to the torque $\Gamma_{w}$ is neglected. We will come back to this contribution in Section 5.       

Given these assumptions, the spin evolution of a star follows from
\be
\frac{d I_* \Omega_*}{dt} = \Gamma_{acc} +  \Gamma_{mag}
\ee
but computing exactly the torques acting on the star is not an easy task. For instance, to compute the contribution due to the inflowing mass requires the knowledge of the torques that acted upon it all the way, from the disc to the stellar surface. And to compute the magnetic contribution, that can be written as $\Gamma_{mag}= I \Phi/2\pi$, where $I$ is the poloidal electric current flowing at the surface of the star and $\Phi$ the poloidal magnetic flux related to zone 3, requires to know both that flux and the toroidal magnetic field at the disc surface. Unless a global solution is actually computed, one is left with estimates that depend on several assumptions.   

Let us define $S_*$ and $S_d$ as, respectively, the surfaces on the star and on the disc corresponding to the same flux tube defining the accretion funnel. Given the assumptions listed above, one can estimate the accretion torque (zone 2) as
\be
\Gamma_{acc}= \left [ \dot M_a \Omega r^2\right ]_{S_d}^{S_*}  = \dot M_a \Omega_K r_t^2 \left ( 1 - \delta \sin^2_o (R_*/r_t)^{1/2} \right )
\simeq \dot M_a \sqrt{GM_* r_t} 
\ee
where $\delta = \Omega_*/\Omega_K(R_*) \sim 0.1$ for a TTS. In this estimate, the magnetic contribution has been neglected wrt to the matter term. However, it must have played a role by allowing mass to accrete. Precisely, matter is assumed to have been spun from a Keplerian rotation velocity at $r_t$ down to the stellar angular velocity $\Omega_*$ at $R_*$ (hence $B_\phi=0$ at  $R_*$, the magnetic shear being distributed along the funnel). But this can only be done if the star rotates slower than the disc material, namely if $r_t < r_{co}$. This is an absolute requirement. Even if some disc material could be loaded onto closed stellar field lines, gravity alone would not lead to an accretion funnel flow. This uplifted material would fall in only if its angular momentum is removed. This is done by a magnetic braking torque along the accretion funnel itself.  

\begin{figure}[t]
\centering   \includegraphics[width=.3\textwidth]{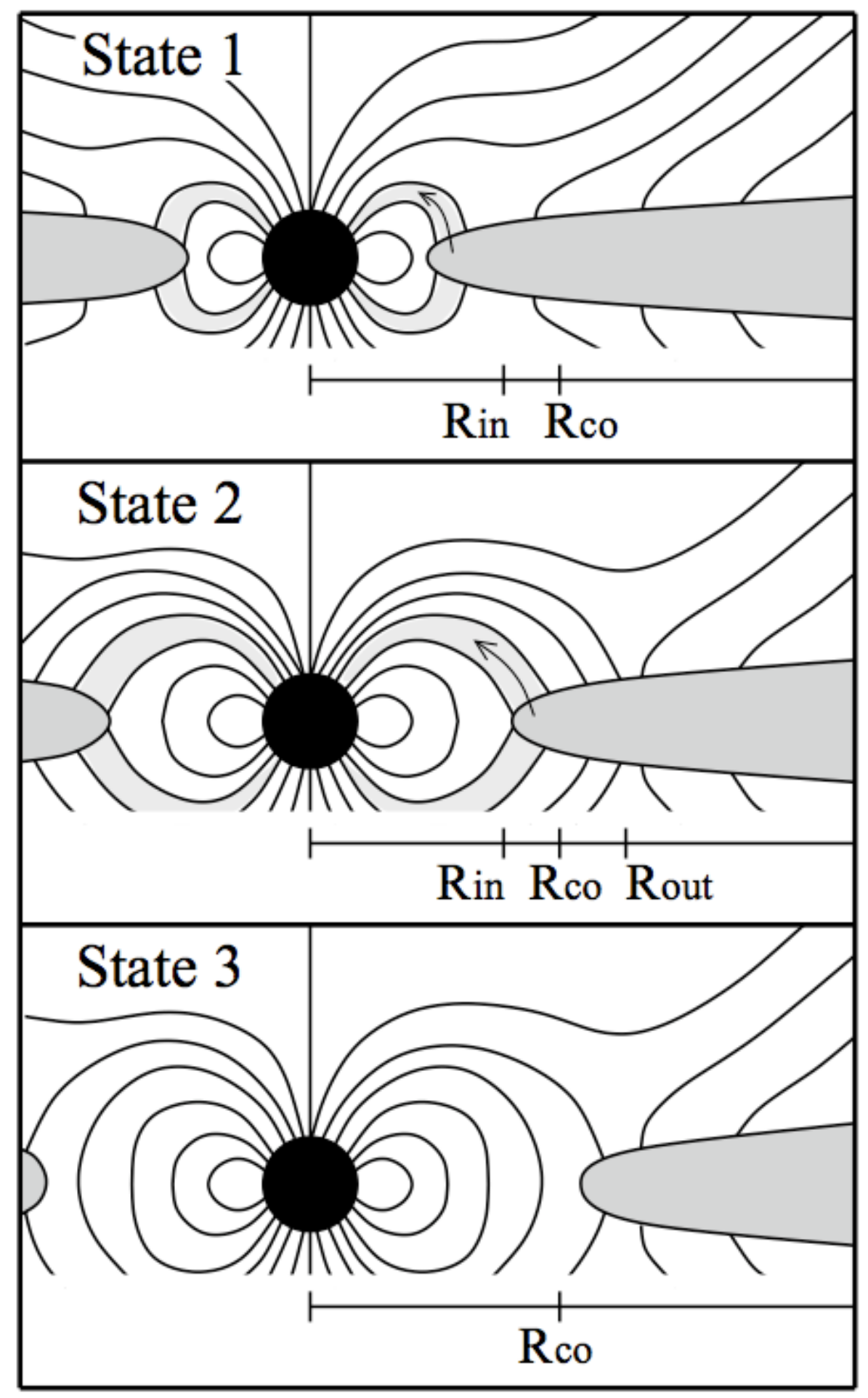}
  \caption{The 3 configurations of a star-disc interaction for an aligned dipole. States 1 and 2 allow accretion ($r_t < r_{co}$) and differ only by the extent of the magnetosphere that remains connected with the disc. Only State 2, with $r_{out} > r_{co}$, could allow a net braking torque on the star and represents therefore the GL scenario. State 3 corresponds to the propeller case where no accretion can in principle occur in steady-state. From \citet{matt05a}.}
   \label{fig_2}
\end{figure}

Such a magnetic torque is conveyed by the stellar magnetosphere, which is co-rotating with the star. Thus, the magnetic braking which is necessary for accretion can {\em only occur at radii where the disc material rotates faster than the star, namely below $r_{co}$}. Figure~(\ref{fig_2}) shows three possible states for an aligned dipole. In two of them $r_t < r_{co}$ and the formation of accretion funnel flows is theoretically possible (accretor case). No accretion funnel flows can be formed if   $r_t > r_{co}$ and all the incoming mass is ultimately expelled away by the fastly rotating star (propeller case, \citealt{love99b, roma03a, usty06}). Hereafter, we consider only the accretor case as no TTS has been observed (or identified) yet in a propeller configuration.  
%%%%%%%%%
%%%%%%%

If no other torque balances the torque $\Gamma_{acc}$ due to the accreting material, a cTTs would spin up on a time scale  
\begin{eqnarray}
\tau &\sim &\frac{I_* \Delta \Omega_*}{\Gamma_{acc}} \sim k^2 \frac{M_*}{\dot M_a} \left (\frac{R_*}{r_t}\right )^{1/2} \Delta \delta \nonumber \\
& \sim & 10^6 yr \left ( \frac{\Delta \delta}{0.1} \right)  \left ( \frac{M_*}{0.5} \right)  \left (\frac{R_*}{r_t} \right )^{1/2} \left ( \frac{\dot M_a}{10^{-8} M_\odot/yr} \right)^{-1}
\end{eqnarray}
smaller than the disc lifetime. If one takes into account the stellar contraction, this is even worse and all TTs should thus be close to their break-up speed ($\delta =1$), which is in contradiction with observations. Thus, an extra braking torque must be present.

Within the GL paradigm, this extra torque is due to the zone 3, mostly thanks to the magnetic contribution as no mass is assumed to fill in this extended magnetosphere. It can be written
\be
 \Gamma_{mag} = 2 \times \int_{r_{t}}^{r_{out}} dr 2\pi r^2 \frac{B_\phi^+  B_z}{\mu_o}
 \ee
where the toroidal field $B_\phi^+$ is evaluated at the disc surface. Looking pole-on, stellar field lines threading the disc below the co-rotation radius would provide a pattern of leading spirals whereas field lines beyond $r_{co}$ provide trailing spirals. For a negative $B_z$ field at the disc surface, this translates into $B_\phi^+ >0$ below $r_{co}$ and $B_\phi^+ <0$ beyond, thereby a positive (spin-up) torque onto the star below  $r_{co}$ and negative (spin-down) torque beyond. All the game here is to compute the overall magnetic torque. Note that, correspondingly, at radii where the star is spun-down (beyond  $r_{co}$), the disc is being azimuthaly accelerated and that accretion can proceed only if the turbulent viscosity is able to radially expel away this angular momentum excess. In most (if not all) accretion models published so far, this extra torque onto the disc has not being taken into account in the disc structure. 

Estimating the toroidal field $B_\phi^+$ is usually done the following way. Above the disc surface, ideal MHD prevails and $\bf E + \bf u \times \bf B = 0$ holds, leading to a radial electric field $E^+_r = - u^+_\phi B^+_z = - \Omega_* r B^+_z$. In the underlying resistive MHD disc (in the plasma frame) one has instead $E_r = \eta_m J_r = - \nu_m \frac{\partial B_\phi}{\partial z} \sim - \nu_m \frac{B_\phi^+}{h}$, where $\nu_m= \eta_m/\mu_o$ is a magnetic diffusivity of turbulent origin. The continuity of the electric field leads to a magnetic shear as seen in the fixed frame \citep{ghos79b, came93, armi96,matt05a}
\begin{eqnarray}
q &=& - \frac{B_\phi^+}{B_z} \simeq \frac{(\Omega - \Omega_*) r h}{\nu_m} 
= \frac{\Omega_K rh}{\nu_m} \left ( 1 - \frac{\Omega_*}{\Omega_K} \right ) \nonumber  \\
&=& \beta^{-1} \left ( 1 - \left(  \frac{r}{r_{co}}\right )^{3/2} \right )
\label{eq:q(r)}
\end{eqnarray}
Inserting this expression in the magnetic torque provides
\be
\Gamma_{mag} = 2 \times \int_{r_{t}}^{r_{out}} dr 2\pi r^2 \frac{B_\phi^+  B_z}{\mu_o} = 4 \pi \beta^{-1} \int_{r_{t}}^{r_{out}} dr r^2 \frac{B_z^2}{\mu_o} \left ( 1 - \left(  \frac{r}{r_{co}}\right )^{3/2} \right )
\ee
where $B_z$ is assumed to vary as $r^{-3}$ (dipole). Within this simplified approach, the strength of the magnetic torque depends on (1)  the truncation radius  $r_t$,  (2) the magnetic diffusion parameter $\beta$ and (3) the extent $r_{out}$ of the interaction zone.

\subsubsection{The disc truncation radius}
%%%%%%%%%%%%%%%%%%%%%%%%%

There have been several attempts to provide an accurate estimate of the truncation radius of a thin disc by an aligned dipole. There are two "historical" criteria. The first one defines $r_t$ as the radius where the local magnetic torque due to the stellar field equals the viscous torque \citep{ghos79a, koni91}
\be 
F_\phi= - J_r B_z \simeq - \frac{B_\phi^+ B_z}{\mu_o h} \sim \frac{1}{r^2} \frac{\partial}{\partial r} r^2 \tau_{r\phi}\sim \frac{\tau_{r\phi}}{r} \sim - \alpha_v \frac{P}{r}  
\ee
Defining the local disc magnetization as $\mu = \frac{B_z^2/\mu_o}{P}= V_A^2/C_s^2$ the ratio of the Alfv\'en speed to the sound speed at the disc midplane, allows to translate the above criterion into the following condition
\be
\mu  \sim \left | \frac{B_z}{B_\phi^+} \right | \alpha_v \frac{h}{r} \ll 1
\ee
However, this formula does not catch the physics of the disc truncation but provides instead an upper limit on $r_t$ (since the magnetic field strength steeply increases towards the star). A second criterion has been put forward by \citet{roma02, kold02a}. It states that the disc is truncated whenever the magnetic field strength dominates the dynamics namely when $\frac{B_z^2}{\mu_o} \sim \rho v^2 + P $. This criterion, although indeed verified in numerical simulations, does not provide a predictive formula unless assuming $v \simeq \Omega_Kr$. In that case, this criterion says that the disc gets truncated when the radial magnetic tension becomes comparable to the gravity $ F_r = J_\phi B_z \sim - B_z\frac{\partial B_z}{\partial r} \sim \frac{B_z^2}{\mu_o r} \sim \rho \Omega_K^2 r$, which translates into a disc magnetization 
\be
\mu  \sim \frac{r^2}{h^2} \gg 1 
\ee
Although predictive, this criterion provides this time only a lower limit on $r_t$. A more accurate criterion has been provided by \citet{bess08} and results from several considerations. First, disc accretion must be dominated by the stellar torque (requiring to be below $r_{rco}$). As a result the accretion speed, as measured by the sonic Mach number $m_s = u_r /C_s$ increases from its canonical value $m_s \sim \alpha_v h/r \ll 1$ in the outer disc to a much larger value $m_s = 2 q \mu $ in the magnetically dominated zone. This fast radial flow must then be deflected in the vertical direction in order to give rise to two funnel flows directed towards the stellar magnetic poles. This is done in two ways. In the radial direction, the flow eventually encounters a magnetic wall so that it cannot move further inwards, namely  $\rho u_r^2 \sim \frac{B_z^2}{\mu_o}$ or  $m_s^2 \sim \mu$.  The disc thermal pressure builds up around $r_t$ until it becomes able to lift the disc material in the vertical direction, despite the initial compression due to the magnetic field pressure and gravity. Meeting these two constraints is optimized only when the field is close to equipartition, namely 
\be
\mu \sim 1
\ee
Note that this condition is the same as for driving magnetized super-slow magnetosonic flows from a thin Keplerian accretion disc \citep{ferr95}. This is no surprise as such a critical point is also present in the physics of accretion columns.

%%%%%%%%%
\begin{figure}[t]
\[ \begin{array}{cc}
 \includegraphics[width=.5\textwidth]{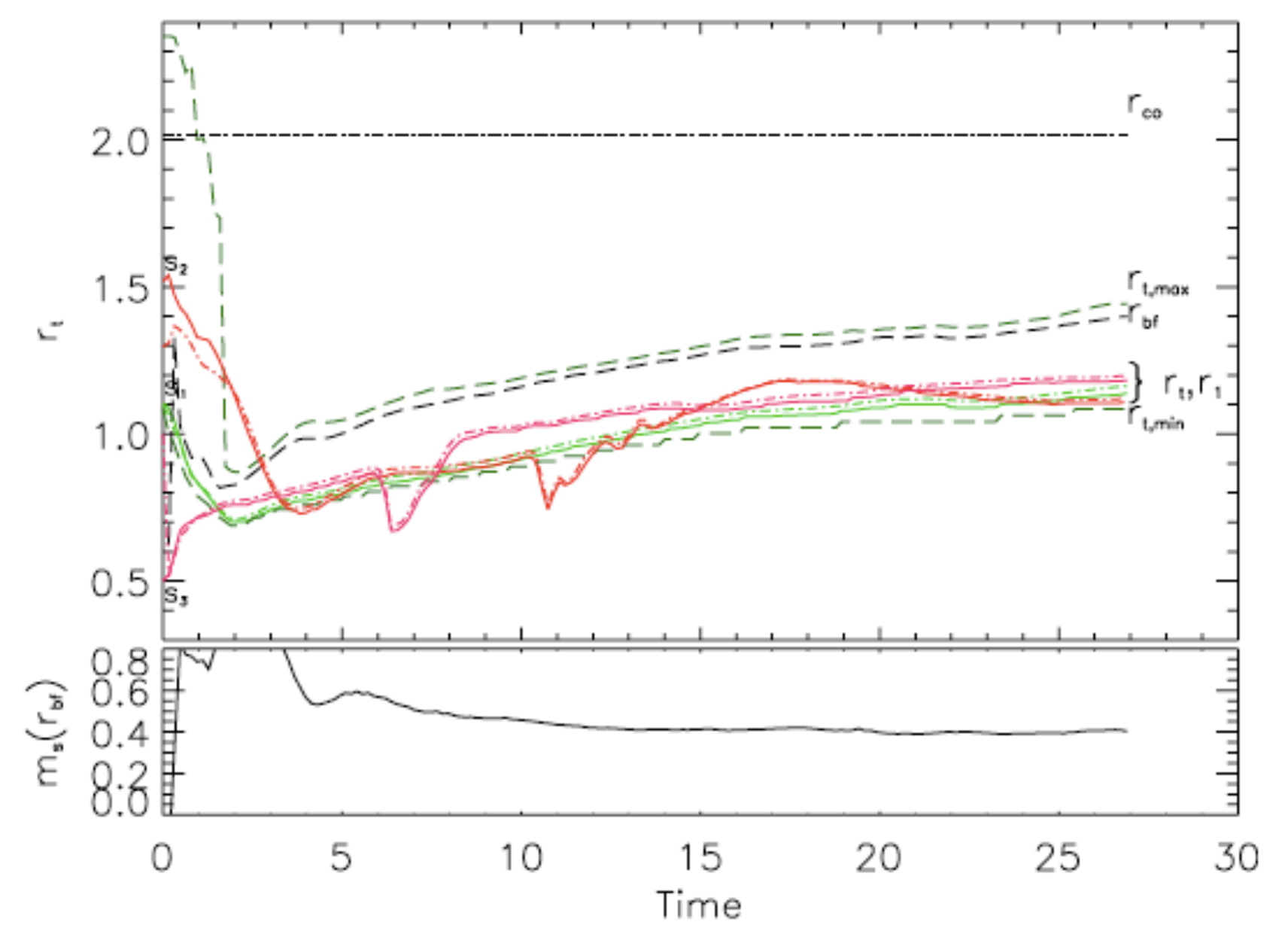}
  &    
 \includegraphics[width=.5\textwidth]{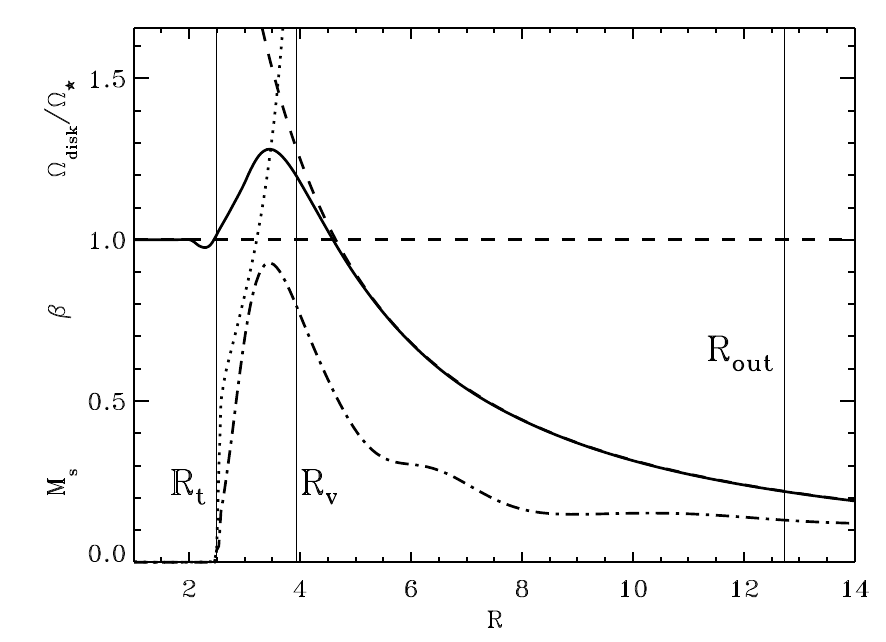}
\end{array} \]
  \caption{2D numerical simulations of star-disc interaction aimed at probing the disc truncation radius. {\bf Left:} Time evolution of $r_{t,max}$ (criterion 1), $r_{t,min}$ (criterion 2), $r_t$ (as defined by $\mu \sim 1$), the outermost radius of the funnel flow $r_{bf}$ and $m_s$ measured at $r_{bf}$ \citep{bess08}.  {\bf Right:} Radial profiles of various quantities measured at the disc midplane: angular velocity of the accreting material (normalized to $\Omega_*$, solid line), sonic Mach number $m_s$ (dash-dotted line), plasma beta $\beta= 2/\mu$ (dotted line). The three vertical lines mark the position of the truncation radius $r_t$, $R_V= r_{t,max}$ and $R_{out}$\citep{zann09}.}
  \label{fig_3}
\end{figure}
%%%%%%%

Figure~(\ref{fig_3}) shows that this physical picture is indeed confirmed by numerical experiments of 2D MHD simulations of a dipole field interacting with a viscous resistive accretion disc. Actually, the various analytical estimates do not differ that much. This is because of the steep radial decrease of the dipole field (thereby $\mu$). In fact, it turns out that the interaction with the disc leads to a compression of the stellar field so that its radial dependency is closer to $r^{-4.5}$ rather than $r^{-3}$ (see Fig~\ref{fig_8}, \citealt{zann09}). Nevertheless, using a dipole dependency provides the following scaling
\be 
    r_t \simeq m_s^{2/7} r_A  \sim r_A
    \label{eq:rt}
\ee
where 
\begin{eqnarray}
\frac{r_A}{R_*} &= &\left ( \frac{4 \pi^2}{\mu_o^2}  \right )^{1/7}  \left (  \frac{B_*^4 R_*^5}{GM_* \dot M_a^2} \right )^{1/7} \nonumber \\
&\simeq & 6 \left( \frac{B_*}{1 kG}  \right )^{4/7} \left( \frac{\dot M_a}{10^{-8} M_\odot/yr}  \right )^{-2/7} \left(\frac{M_*}{0.8 M_\odot}  \right)^{-1/7} \left(\frac{ R_*}{2 R_\odot} \right)^{5/7} 
\end{eqnarray}
is the spherical Alfv\'en radius found in the case of a free-falling plasma onto a dipole field \citep{elsn77}. This is remarkable as the physics involved are totally different (the magic of similarity properties of fluid dynamics\footnote{In this last case, one uses $\dot M = 4 \pi r_A^2 \rho_A u_r$, $u_r = V_A = \frac{B_A}{\sqrt{\mu_o \rho_A}} $ and $B_A = B_* (r_A/R_*)^{-3}$}).  Many works actually used this scaling, namely $r_t= \alpha r_A$, with $\alpha$ ranging from 0.5 \citep{ghos79a, koni91} to unity \citep{ostr95}. Putting numbers, one finds truncation radii that seem consistent with current observations, with $r_t$ typically 0.2 to 0.5 times $r_{co}$. But this is actually hard to say.  Indeed, the actual truncation radius $r_t$ is smaller than the disc inner edge inferred from spectral energy distributions, as the latter depends on the presence of dust while the former is usually in a gas region where dust has been sublimated.

\subsubsection{The magnetic diffusion}
%%%%%%%%%%%%%%%%%%%%%%%%%

Most GL models differ in the way $\beta= \Omega_K rh/\nu_m$ is estimated. In a standard viscous accretion disc the turbulent viscosity is parametrized using $\nu_v = \alpha_v C_sh =\alpha_v \Omega_K h^2$, with $\alpha_v < 1$  \citep{shak73}. Defining a turbulent (effective) magnetic Prandtl number $ {\cal P}_m = \nu_v/\nu_m$, one can write $\beta= \alpha_v  \frac{h}{r} {\cal P}_m^{-1}$. One would therefore expect $\beta << 1$ in a thin (Keplerian) disc with an effective magnetic Prandtl number of order unity. Such a small diffusivity translates of course into a large magnetic shear $q$. Note  for instance that, following other arguments, \citet{armi96} used $\beta=1$.   

%%%%%%%%%
\begin{figure}[t]
\centering   \includegraphics[width=.8\textwidth]{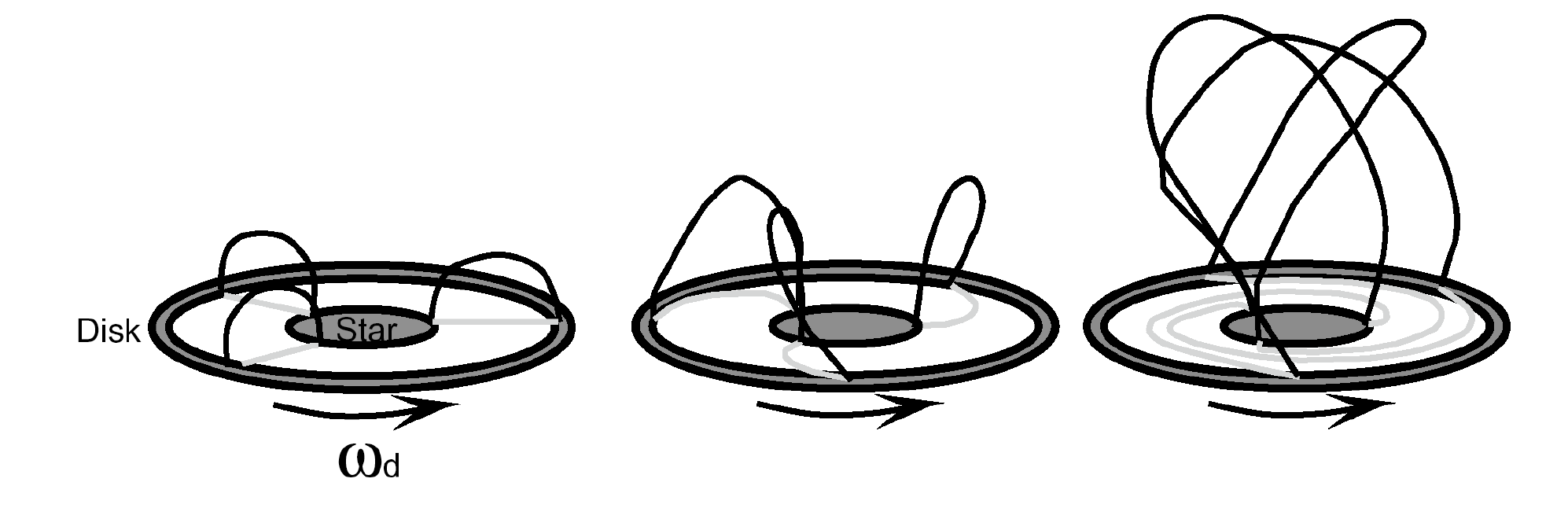}
  \caption{Cartoon of the inflation of magnetospheric field lines due to the differential rotation between the star and the disc. In order to minimize its energy, the magnetic configuration tends to occupy all space and distribute its shear along the field lines. Adapted from \citet{love95}.}
   \label{fig_4}
\end{figure}
%%%%%%%%%

\subsubsection{The maximum magnetic shear}
%%%%%%%%%%%%%%%%%%%%%%%%%

The main limitation comes actually from the maximum magnetic shear $q$ that is allowed in a quasi-steady state. As first pointed out by 
 \citet{aly84, aly91}, a force-free magnetic flux tube which undergoes an increase in its helicity tends to inflate (increase its volume, see Fig~\ref{fig_4}). Now, this is exactly what differential rotation does for a stellar field line threading the disc. This results in an ever growing winding up of the magnetic configuration that expands simultaneously in the poloidal direction. This process is faster as one moves away from the co-rotation radius and is therefore controlled by the innermost radius with the shortest time scales and higher energies. The outcome, as usual in MHD, is determined by microphysics. 
%%%%%%%%%
\begin{figure}[t]
\centering   \includegraphics[width=.8\textwidth]{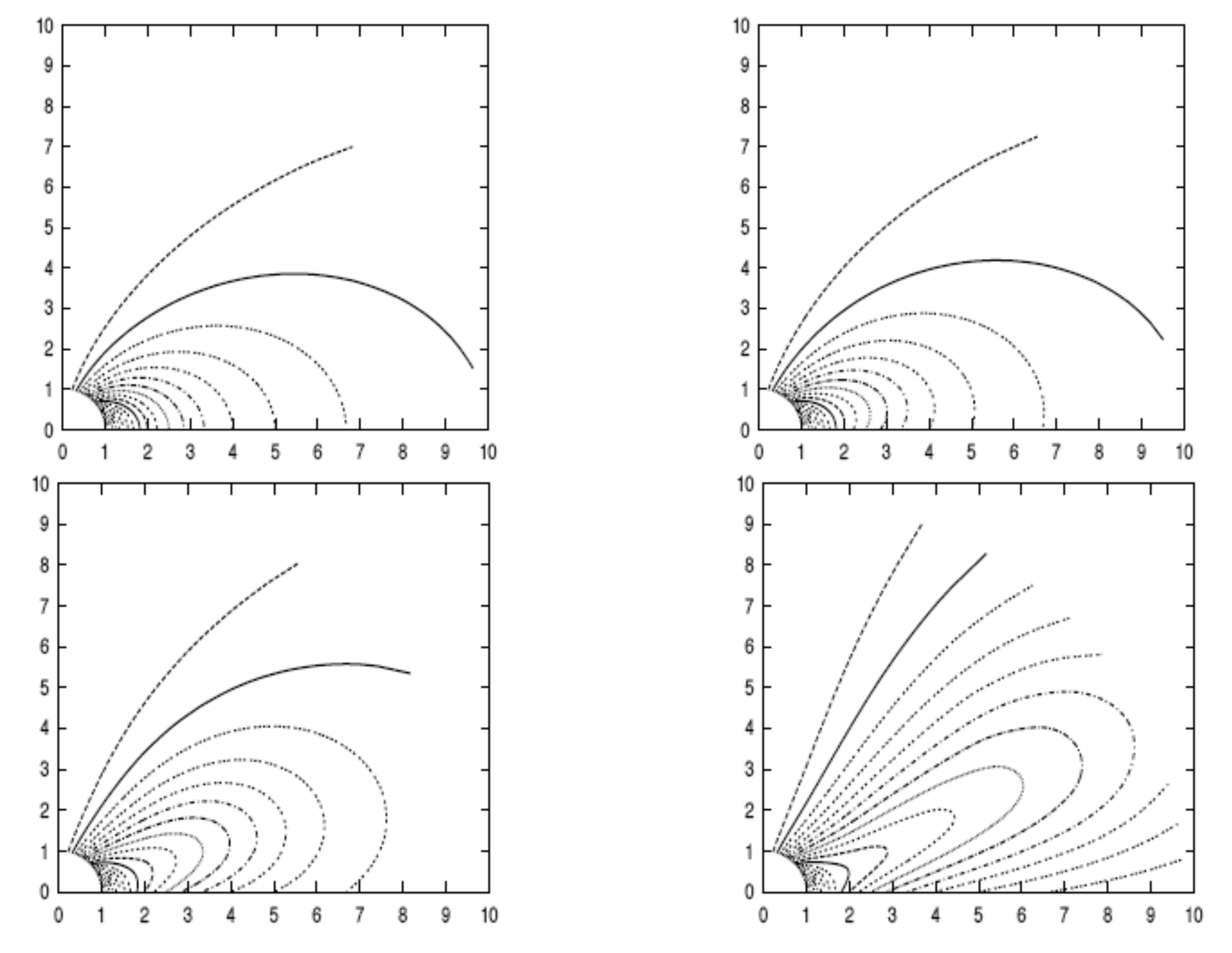}
  \caption{Sequences of magnetic field contours for the Keplerian disk model, mimicking a time evolution (from top left to bottom right). In these steady-state equilibria, the field line footpoints are not allowed to move \citep{uzde02a, uzde02b}. Allowing field diffusion across the disc (as done in eg. \citealt{bard96}) would not help much as the time scales involved for field inflation and reconnection are much smaller than field diffusion.}
   \label{fig_5}
\end{figure}
%%%%%%%%%
 
If there is no outer pressure forbidding the field lines to occupy all space (as in the case of a magnetic tower, \citealt{lynd94, love96, ciar09}), the magnetic surfaces will tend to spread radially with an opening angle of roughly $60^o$ (see Fig~\ref{fig_5}). As this process goes on, the oppositely directed field lines will undergo an attraction leading to reconnection and thereby an opening of the field lines \citep{love95, uzde02a, uzde02b}. The outcome depends on microphysics, but seems nevertheless unavoidable: on very short time scales, field lines not loaded with accreting material (hence force-free) will inflate and enforce the outer field lines to do the same. This will lead to reconnection, breaking the causal connection (the torque) between the star and its disc beyond some radius. The maximum twist that a force-free field seems to absorb without leading to reconnection seems to be a toroidal field comparable to the poloidal one, namely $q_{max}$ around unity.

According to this picture, the maximum radius $r_{out}$ is therefore directly related to this maximum magnetic shear $q_{max}$. Indeed, inverting Eq.(\ref{eq:q(r)}), one gets $ r_{out} = (1 + \beta q_{max})^{2/3} r_{co}$. Note that there is also an innermost radius  $r_{in} = (1 - \beta q_{max})^{2/3} r_{co}$. But a consistent picture requires $r_t > r_{in}$, so that the closed magnetosphere extends from $r_t$ to $r_{out}$.

\subsubsection{Can the GL scenario provide a disc-locking?}
%%%%%%%%%%%%%%%%%%%%%%%%%%%%%%%

%%%%%%%%%
\begin{figure}[t]
\[ \begin{array}{cc}
 \includegraphics[width=.5\textwidth]{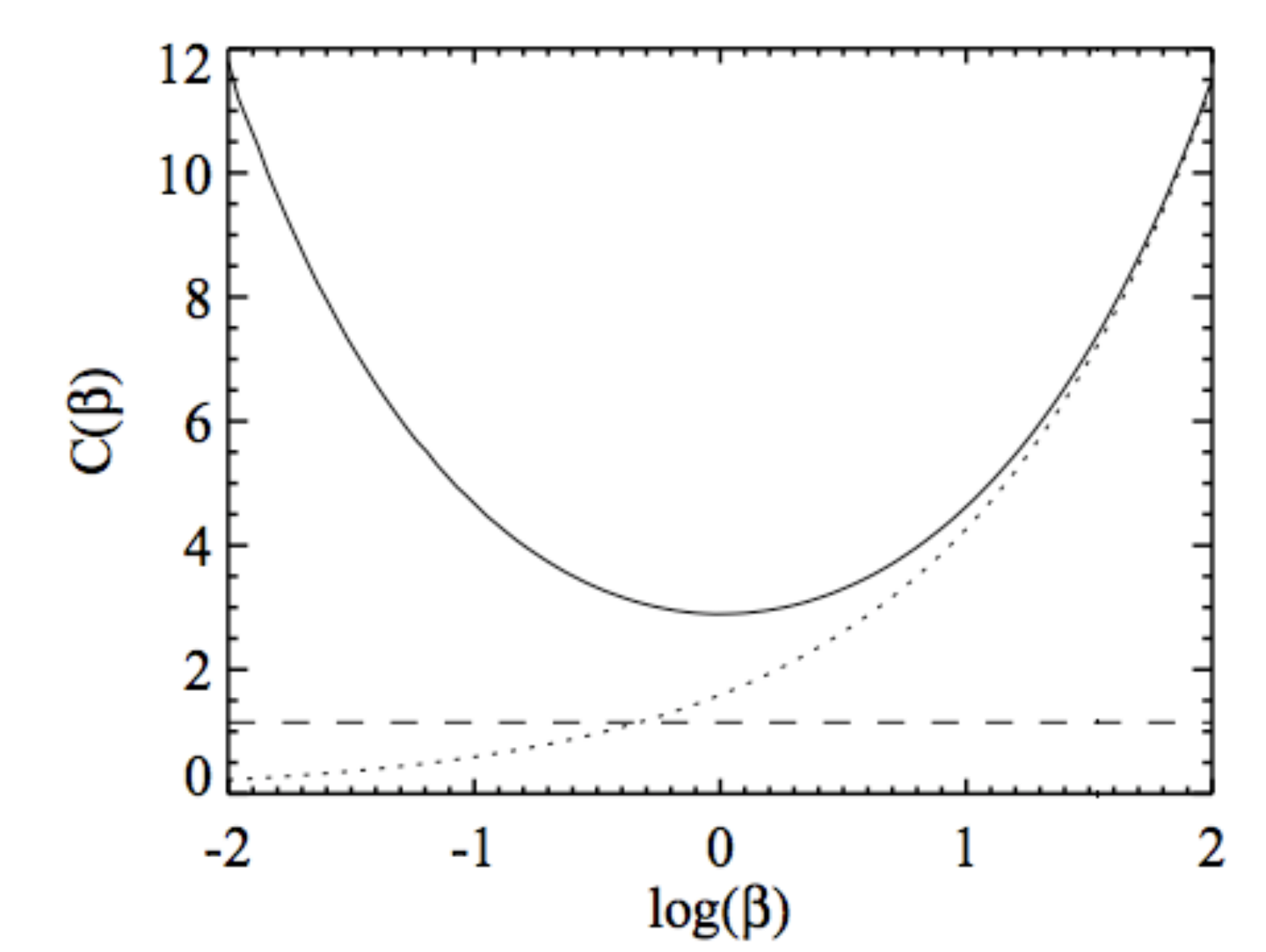}
  &    
 \includegraphics[width=.5\textwidth]{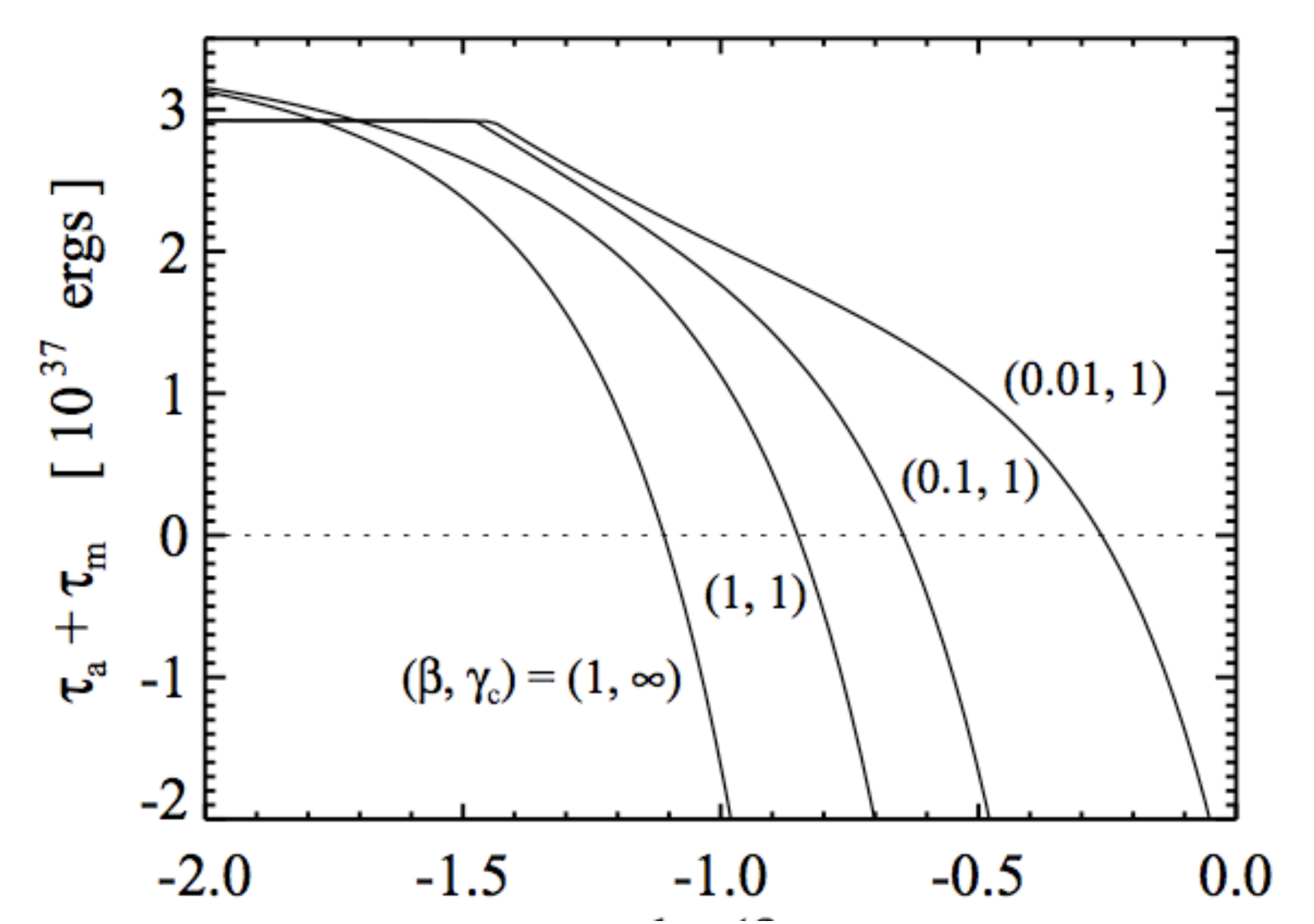}
\end{array} \]
  \caption{{\bf Left:} Constant $C$ in Eq.~(\ref{eq:C}) as function of the magnetic diffusion parameter $\beta$, computed for $q_{max}=1$. The dotted line is for $q_{max} \rightarrow \infty$, while the dashed line is the value $C=1.15$ used by \citet{koni91}. {\bf Right:} Total torque acting on the star as function of the log of $\delta= \Omega_*/\Omega_K(R_*)$. Here $\gamma_c \equiv q_{max}$, see text. From \citet{matt05a}.}
\label{fig_6}
\end{figure}
%%%%%%%

Disc-locking has been actually a minimal requirement of a spin equilibrium solution $\dot \Omega_*=0$ {\em neglecting the stellar contraction}. In the GL framework this translates into $\Gamma_{mag} = - \Gamma_{acc}$ providing $\delta_{eq} = C (GM_*)^{3/14}\,  R_*^{3/2}\,  \dot M_a^{3/7} \, \mu_*^{-6/7} $ or equivalently 
\be
\Omega_{eq} = C (GM_*)^{5/7}\,   \dot M_a^{3/7} \, \mu_*^{-6/7} 
\label{eq:C}
\ee
where $\mu_*  = B_* R_*^3$ is the stellar magnetic moment and $C$ a constant depending on the model parameters ($q_{max}, \beta, r_t$). This last expression is remarkable as it is independent of the stellar radius and, to match observations, requires quite strong fields ($>$ kG).  Figure~(\ref{fig_6}a) shows the value of $C$ as a function of the magnetic diffusion parameter $\beta$ computed by \citet{matt05a}. For small $\beta$, the magnetic shear tends to dramatically increase as one goes away from the co-rotation radius range. This leads to an opening of the magnetic structure, thereby lowering $\Gamma_{mag}$ and leading to a higher equilibrium spin. At large $\beta$ the same trend appears because although the magnetosphere is this time extended over a large domain, the actual value of $q$ is small, lowering the magnetic torque. Thus, the torque is the largest ($\Omega_{eq}$ smallest) only for $\beta$ of order unity. And, indeed, all models so far that succeeded in providing a disc locking within the GL framework used $\beta$ (hence $C$) of order unity. As shown previously, this is troublesome in a standard turbulent accretion disc, where $\beta << 1$ would be better expected.
  
\citet{matt05a} computed the global torque $\Gamma_{mag} + \Gamma_{acc}$ acting on the star as a function of the stellar rotation (Fig~\ref{fig_6}b). It can be seen that a spin equilibrium with $\delta =0.1$ (as required by observations) can only be achieved with the combination of two highly questionable factors: an extended magnetosphere (ie. $q_{max} \rightarrow \infty$, no reconnection) and a very large diffusivity in the disc ($\beta \sim 1$).  

There is no indication whatsoever that anyone of these two requirements should be met.

\subsection{Numerical simulations}
%%%%%%%%%%%%%%%%%%%%

%%%%%%%
\begin{figure}[t]
\centering   \includegraphics[width=.8\textwidth]{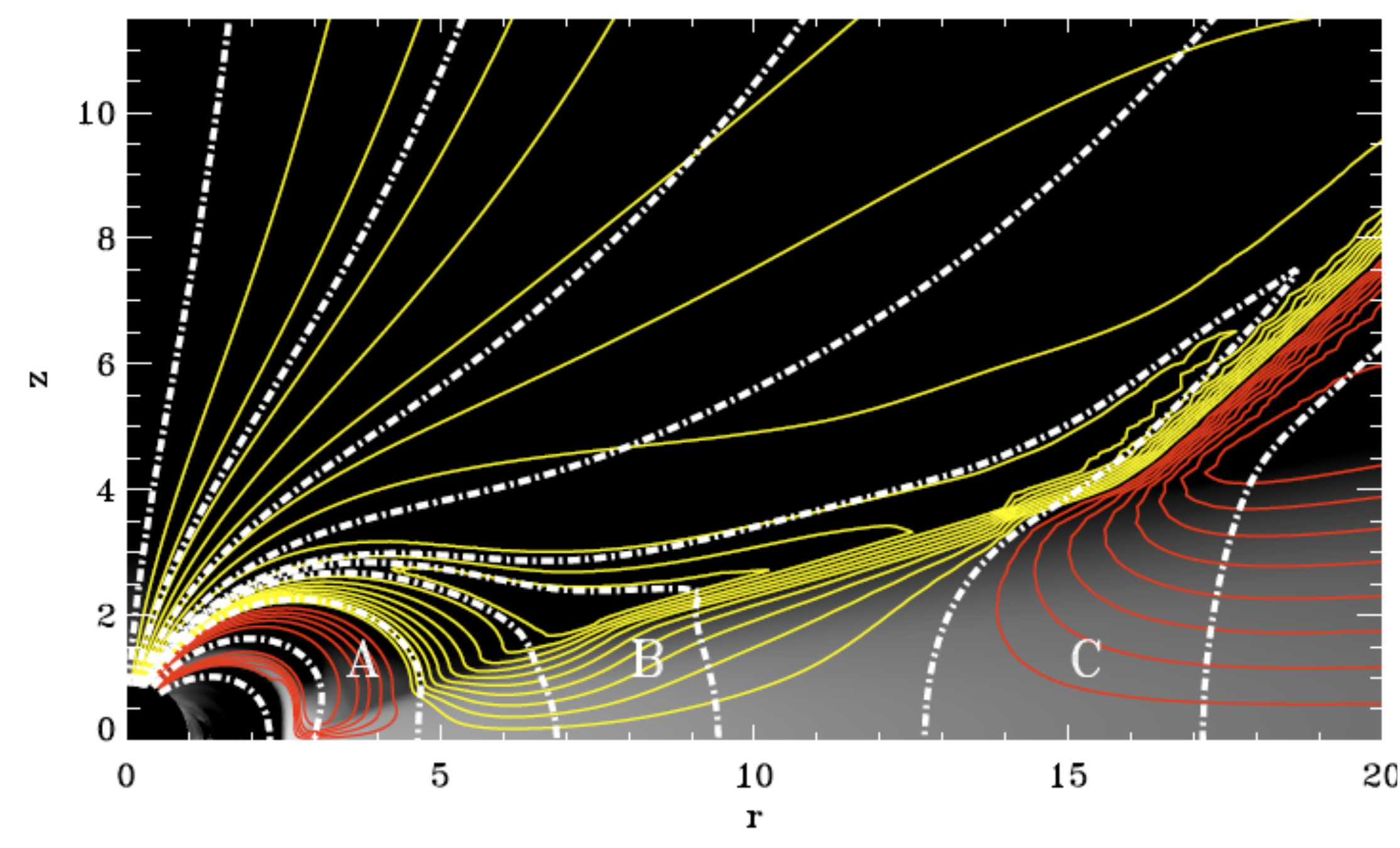}
  \caption{Isocontours of poloidal electric currents present at the star-disc interaction. Dark (red) circuits are circulating clockwise whereas light (yellow) circuits are counter-clockwise. Circuit A flows along magnetic flux tubes related to the accretion funnel and threading the disc up to the co-rotation radius: it spins up the star. Circuit B is related to the remaining stellar flux, threading the disc (closed, extended magnetosphere) but also to open field lines along which a stellar wind flows: it brakes down the star. Circuit C is not connected to the star but is instead the signature of a magnetic braking of the disc. Dot-dashed lines show a sample of poloidal magnetic field lines \citep{zann09}.}
   \label{fig_7}
\end{figure}
To make a long story short, 2D and 3D numerical simulations support most of the previous findings and it seems quiet safe now to discard the Ghosh \& Lamb picture as the main agent for spinning down a cTTS. 

%\subsubsection{2D}

2D simulations are particularly useful as they can be computed for quite a long time and the physical processes can be verified and understood a posteriori. The caveat is that they usually rely on an alpha prescription for both the viscosity $ \nu_v$ and  magnetic diffusivity $\nu_m$ in the disc and do not compute the full MHD turbulence.  However, they provide valuable insights into the physics of the star-disc interaction. 
%

%%%%%%%%%
\begin{figure}[t]
\[ \begin{array}{cc}
 \includegraphics[width=.4\textwidth]{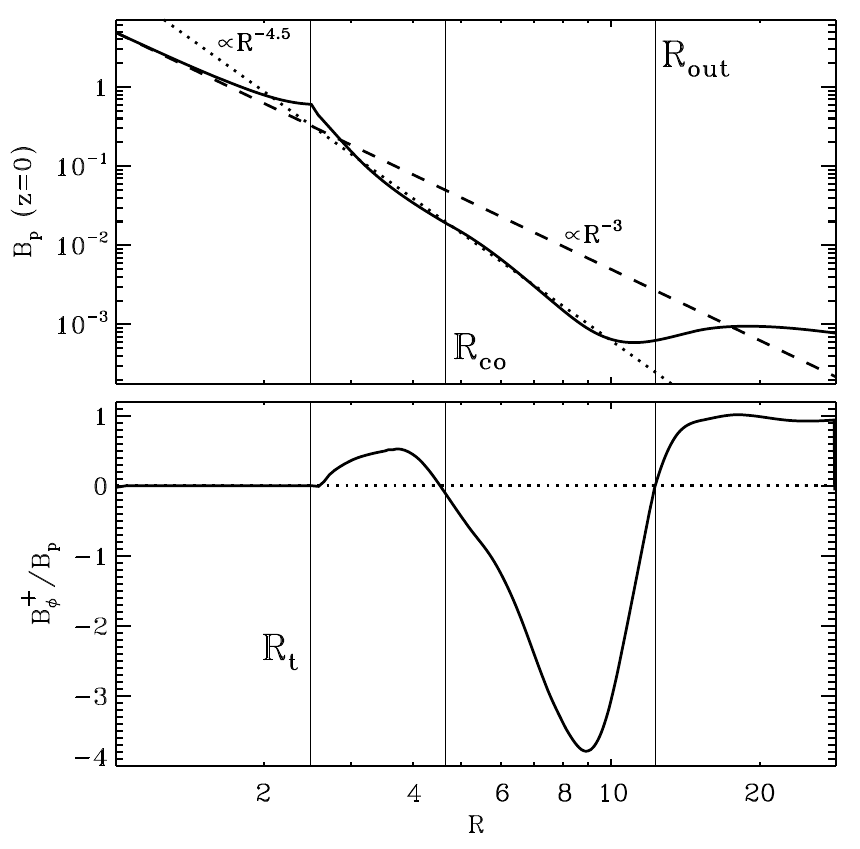}
  &    
 \includegraphics[width=.6\textwidth]{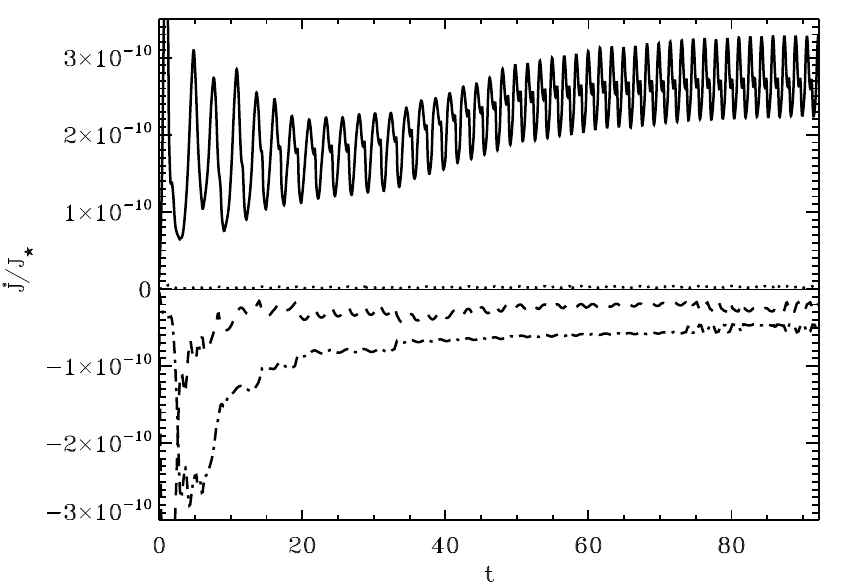}
\end{array} \]
  \caption{2D numerical simulations with a dipole field, with turbulence parameters $\alpha_m=\alpha_v=1 $ chosen to maximize the extent of the closed magnetosphere \citep{zann09}. {\bf Left:} Final magnetic field configuration. {\bf Right:} time evolution of torques acting on the surface of the star, normalized over the stellar angular momentum (defining the inverse of the braking time). A positive torque spins up the star.
The accretion torque is here the sum of the negligible kinetic torque (dotted, $>0$) and magnetic contribution (solid, $>0$) along the magnetosphere connected to the disk below the co-rotation radius. The magnetic contribution beyond $r_{co}$ is shown in dashed line ($<0$) and a stellar wind torque in dot-dashed line ($<0$). Time is given in units of the rotation period of the central star.}
\label{fig_8}
\end{figure}
%%%%%%%

Figures~(\ref{fig_7}) and (\ref{fig_8})  show the outcome of a numerical simulation whose initial state was a stellar dipole field threading a Keplerian accretion disc. This simulation has been chosen so as to maximize the extent of the closed magnetosphere connecting the star to the disc. After a transient phase where the magnetic field structure is dramatically readjusted, the system relaxes to a quasi steady state. It can be seen that the field gets compressed and decreases radially faster than a dipole field in vacuum, while the maximum shear $q_{max}$ seems indeed limited to a value around unity. These two limitations severely diminish the efficiency of the braking torque beyond $r_{co}$: the overall torque is actually {\em ten times smaller} than the \citet{matt05a} estimate.    
 
In simulations of that kind, one always gets opened field lines from the magnetic poles, along which flows a plasma that usually reaches super-Alfv\'enic speeds. This is a stellar wind, although both the mass loading and initial enthalpy of the wind are numerically controlled (or biased, depending on the point of view). We will come back to this effect later, but note that this is an unavoidable (though negligible here) feature of such a  magnetized configuration. 

The oscillations seen in the dominant (spin-up) torque in  Fig.~(\ref{fig_8})  are a direct signature of the oscillations in the mass accretion rate onto the star itself. They are due to a mismatch between the two torques responsible for accretion. Mass is indeed being provided through the Keplerian disc mostly thanks to a viscous torque (with $m_s \sim h/r )$, whereas in the funnel flow and in the magnetically dominated region just outside it, the magnetic torque provides $m_s$ around unity.  However, such regular oscillations (with a period of the order of a few stellar rotations) have never been observed. It is therefore likely that some feedback (to be found yet) or some damping process must be at play.

%\subsubsection{3D}

The facts that the disc gets truncated at a radius given by Eq.(\ref{eq:rt}) and that the magnetosphere cannot remain connected to the disc much beyond (because of a maximum shear of order unity) are now well firmly confirmed by 2D but also 3D numerical simulations. Clearly, reconnections seen in simulations are numerically biased, but there is no doubt that field lines will tend to inflate and eventually reconnect, even when partially loaded with some mass. The location of the truncation radius has been clearly confirmed in 2D simulations using alpha prescriptions. However, MRI-driven discs (in 2D and 3D), where no alpha prescription has been used neither for the viscosity nor the diffusivity, provide the same truncation radius \citep{roma11, roma12}. This is a very important result, that can be understood once we recall that arguments used to establish Eq.(\ref{eq:rt}) do not depend on the origin of the magnetic diffusivity and viscosity. 
%%%%%%%%%%%
\begin{figure}[t]
\centering   \includegraphics[width=.8\textwidth]{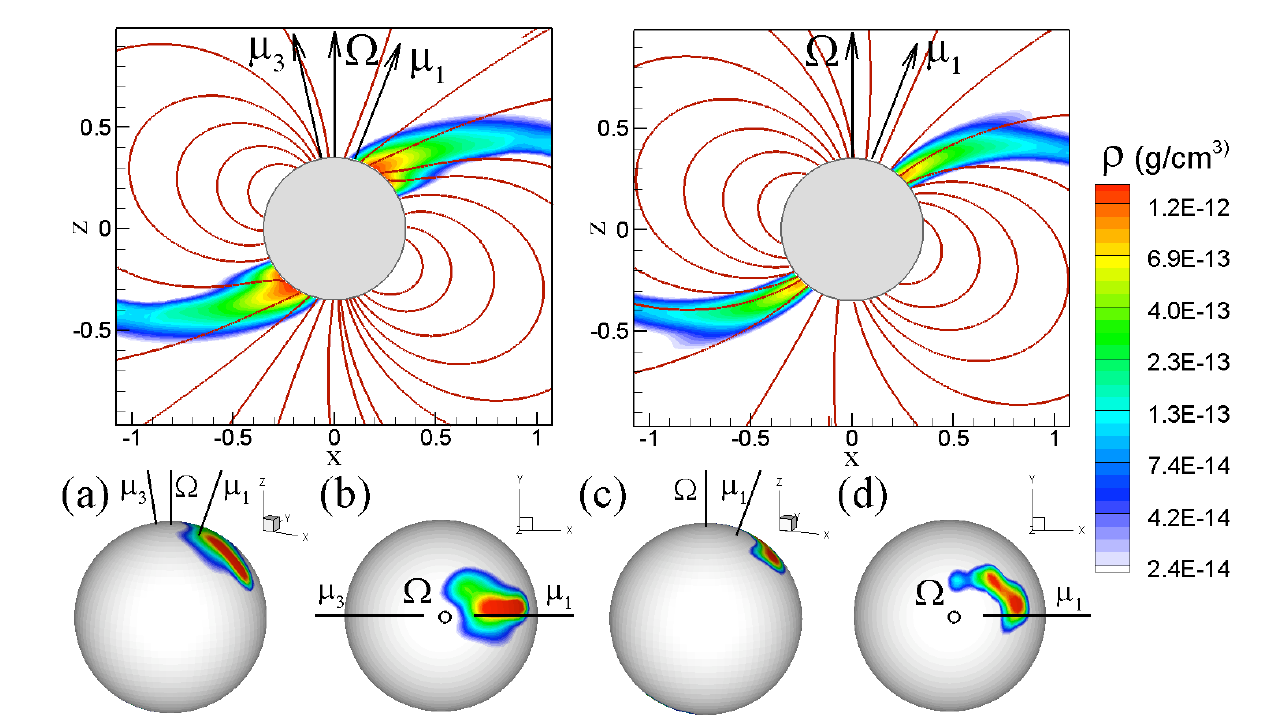}
  \caption{Comparison of matter flow and accretion spots in the dipole plus octuple model of BP Tau (left panels) and with a pure dipole case (right panels). The difference is mainly in the position and shape of the accretion shock (bottom panels).  This may have important consequences when deriving physical diagnostics from observations. From \citet{long11}.}
   \label{fig_9}
\end{figure}
%%%%%%%%%

But there are of course many other features that 3D simulations provide, most of them very useful to directly compare the outcome of simulations with real observations:
\begin{itemize} 
\item Since the truncation radius is defined by $\mu \sim 1$, it will be the locus of strong magnetic field instabilities, disturbing the laminar axisymmetric flow. The interchange instability is one of them and it is indeed observed in simulations, with mass pilling up at some places by expelling the magnetic flux and creating thereby massive fingers that fall into the star \citep{roma03b,roma08, kulk08}. 

\item The magnetic field topology can be much more complex than a simple aligned dipole (see J.F. Donati's contribution, this book).  Inclined dipoles modify of course the position of the accretion shock put introduce also a mass flux variability, due to their non axisymmetric  interaction with the disc \citep{roma04a}. Introducing for instance a quadrupolar or an octupolar component may totally modify the accretion flow. For instance, an aligned quadrupole field will not truncate the disc but allow instead the formation of an equatorial accretion belt \citep{long07, long08}. 

\item The variety of observed phenomena is quite rich and one gets the feeling that, in order to understand each individual star, one would need to know exactly its magnetic topology. This has been done for instance in the case of BP Tau, which is known to harbor an inclined dipole field of 1.2 kG and an octupole of 1.6 kG \citep{long11}. Although the octupole field is strongest at the surface of 
the star, it decreases more rapidly than the dipole field with the distance from the star. As a consequence, it determines the truncation radius $r_t$. However, the accretion funnel flow is deeply modified by the presence of the octupole (see Fig.\ref{fig_9}, seen also in analytical work of \citealt{moha08}). Consequences of complex fields are for instance the modification of the position, size and shape of accretion hot spots \citep{roma04a}, but also their possible disappearance when the accretion rate becomes too important \citep{long11}. In fact, multipolar fields are indeed very important in shaping the accretion shock.  

\item The dynamics of star-disc interaction, even with complex fields (inclined, several components), seems to be now quite at hand thanks to 3D simulations. Thorough comparisons with observations can even be done with a post-processing treatment \citep{kuro08, kuro11}. Indeed, using magnetic field topologies obtained from Doppler-Zeeman imaging technics, a full 3D star-disc magnetosphere can then be reconstructed and multidimensional non-LTE radiative transfer models of hydrogen and helium line profiles can then be computed in the resulting accretion flows. This is a fantastic tool and provides already great results. In particular, it allows to disentangle the various contributions to these lines, namely the accretion flow itself from lines coming from a hot stellar wind and the outer disc wind.  

\end{itemize}

To conclude this section, the physics of accretion funnel flows onto complex magnetospheres is well reproduced and (quite) well  understood. There is a need for more physical insight, in particular when considering how multipolar fields affect the accretion shock and if they do affect the angular momentum transfer between the star and the disc. But the global picture seems physically consistent: it is impossible for an accreting star to deposit its angular momentum back in the circumstellar disc. As conjectured earlier by \citet{shu88, shu94a}, the stellar angular momentum must be expelled away from the star-disc system. How can this be done?

Could the jets observed from young stellar objects be powered by the rotational energy of the central star? Could these confined supersonic flows carry away the excess angular momentum allowing the central protostar to maintain a almost constant angular velocity,  despite accretion and contraction?  We examine these questions in the following sections.

%%%%%%%%%%%%%%%%%%%%%%%%%%%%%%%%%%%%%%%%%%%%%%%%%%%%%%%%%%%%%
\section{Observational constraints and models of YSO jets}
%%%%%%%%%%%%%%%%%%%%%%%%%%%%%%%%%%%%%%%%%%%%%%%%%%%%%%%%%%%%%

\subsection{Jets from young stars: why a need for magnetic models ?}
%%%%%%%%%%%%%%%%%%%%%%%%%%%%%%%%%%%%

\begin{figure}[t]
\[ \begin{array}{cc}
\includegraphics[width=.5\textwidth]{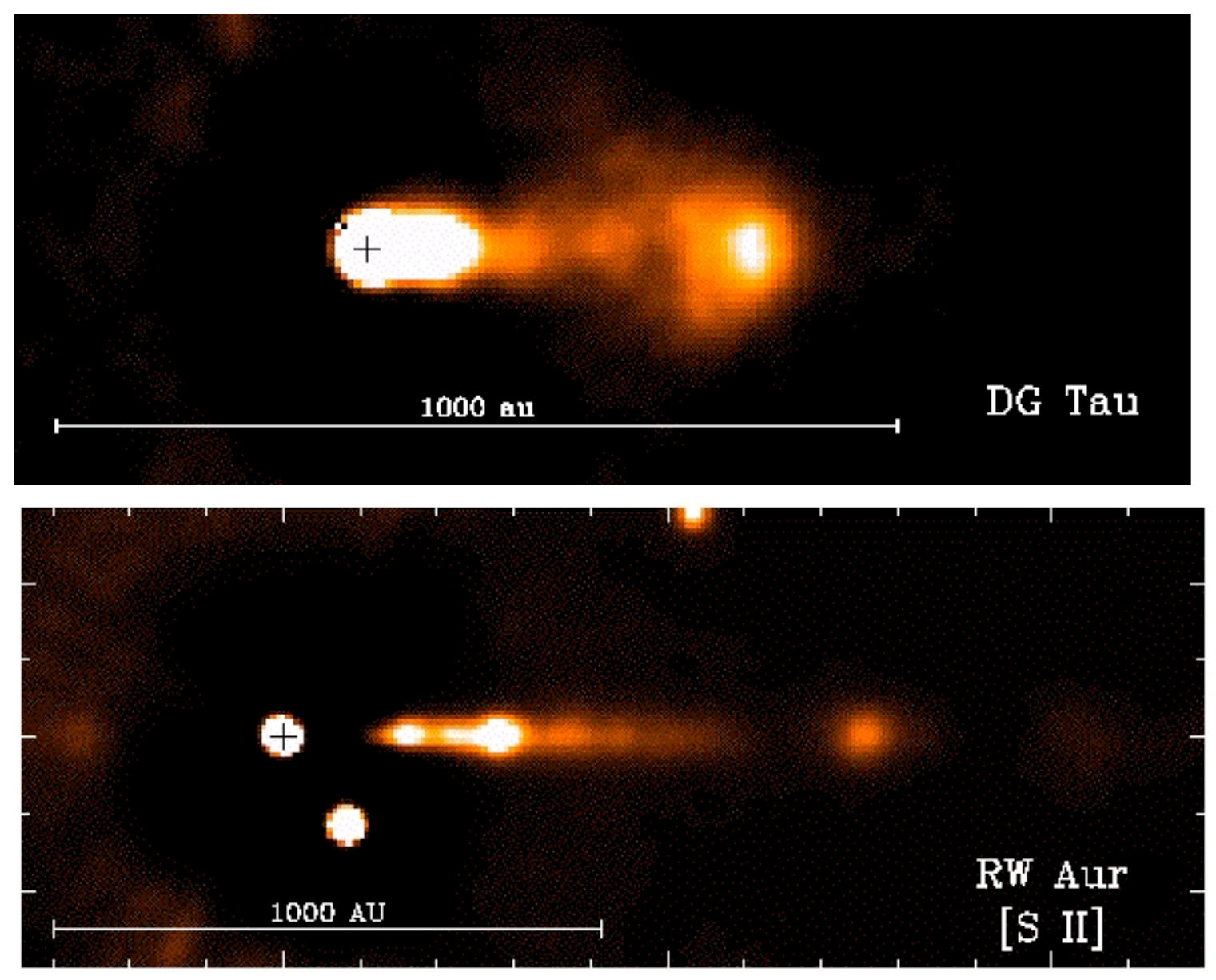}  
  &    
   \includegraphics[width=.45\textwidth]{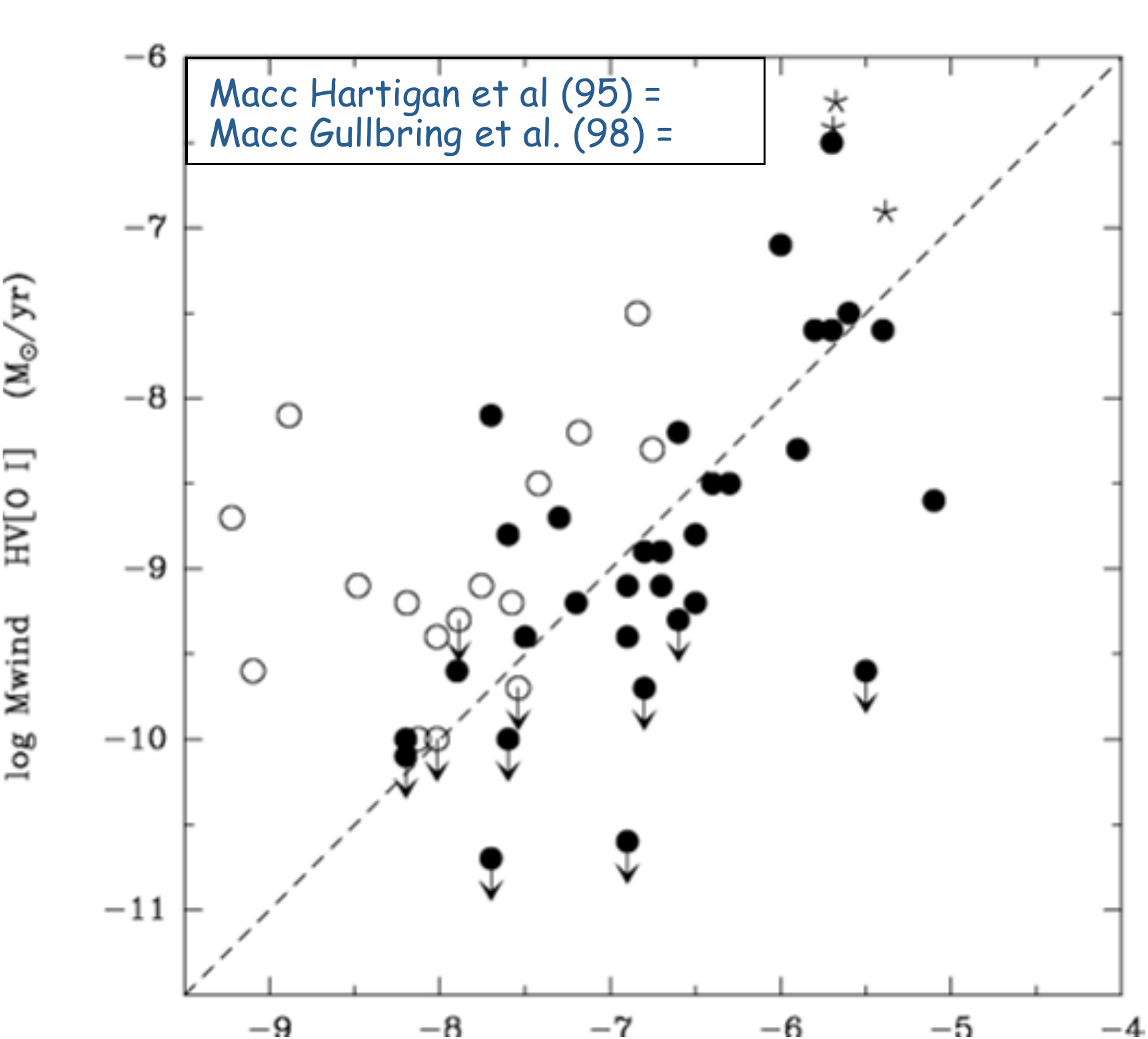}
\end{array} \]
  \caption{{\bf Left:} Two examples of microjets, detected by continuum subtraction in 30\% TTS, providing insights to internal regions, 30 to 100 AU from the source \citep{doug00}.  {\bf  Right:}  One-sided jet mass loss derived in TTS jets as function of the disc accretion rate (log, in $M_\odot/yr$). There is a correlation indicating a link between the two phenomena. The average one-sided value is roughly 10\% of the disc accretion rate, with a similar ratio in more embedded objects \citep{cabr07}.}
   \label{fig:obs}
\end{figure}

Young Stellar Objects (hereafter YSO) provide fantastic tools for investigating non-relativistic jets through images, Position-Velocity diagrams, line ratios etct... unveiling detailed kinematic maps, mass loss rates measurements and collimation/acceleration properties. Their most striking feature is certainly the fact that these supersonic jets are highly collimated (opening angle of a few degrees) very close to the source at the limit of our actual resolution (Fig.~\ref{fig:obs}). This hints to the necessity to link the collimation process to the acceleration one and only the presence of a large scale magnetic field anchored on the jet driving source has been proven to be effective enough. The interested reader will fruitfully benefit from the observational reviews of eg. \citet{cabr07, ray07}) and the JETSET book series. 

%%%%%%%%%%
%%%%%%%%%%

Observations clearly show that jets suffer strong disturbances due to shocks and possibly unsteady ejection events triggered at the source. However, the time scales inferred are long (at least years) with respect to the dynamical time scales at the innermost regions where jets are believed to be launched from ($\sim$ days). Moreover, within the first 100 AU, jets (termed "microjets", \citealt{doug00}) seem to be less messy and inferred shock speeds seem to be always much smaller than the main jet speed. Thus, while the jet phenomenon is intrinsically a time-dependent one, steady-state models certainly provide valuable tools to understand them.

\begin{figure}[t]
\centering   \includegraphics[width=\textwidth]{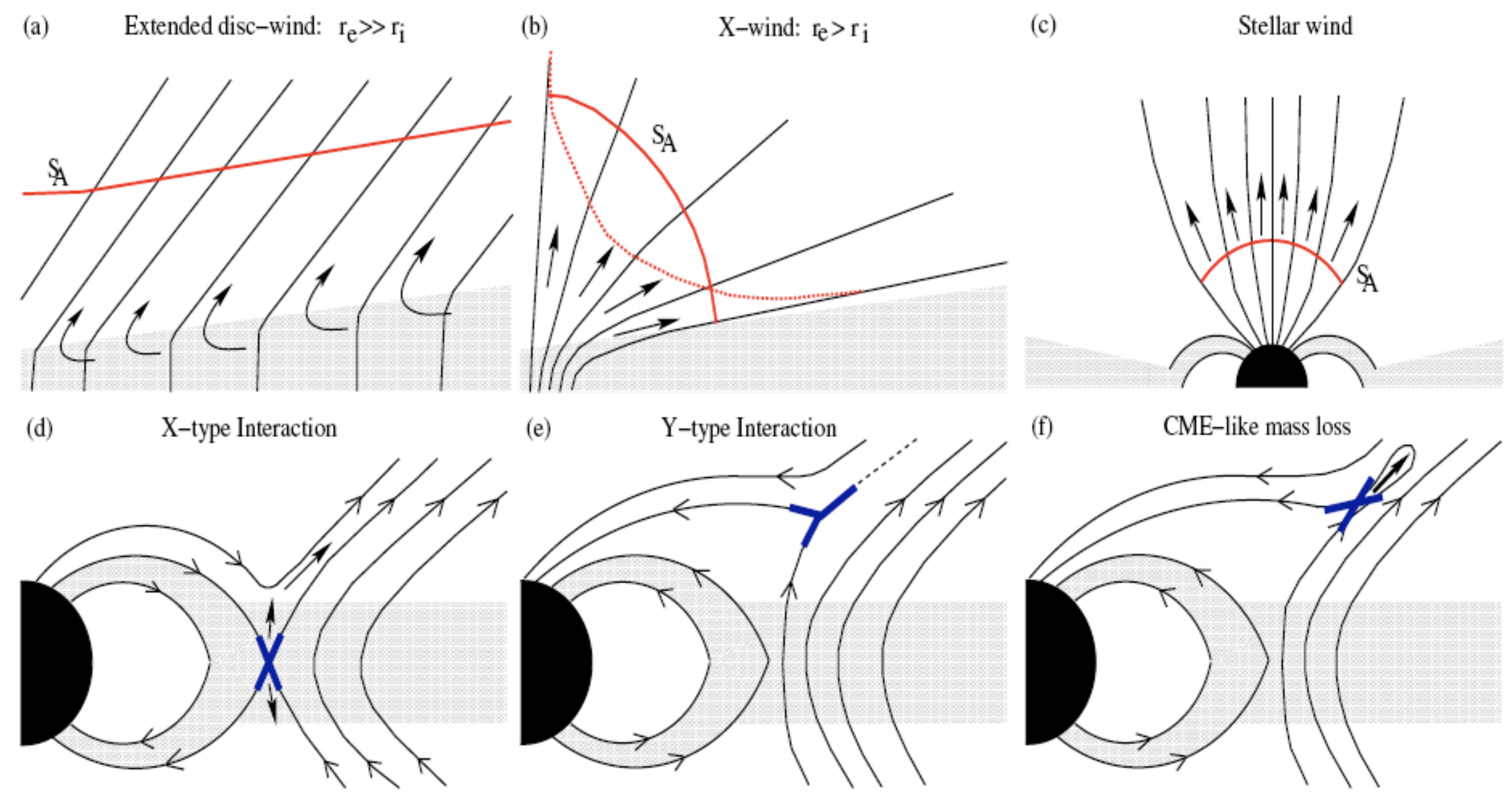}
  \caption{ {\bf Top:} Classes of published stationary MHD jets for YSOs. {\bf Bottom:} Sketch of possible axisymmetric magnetospheric configurations, steady and involving a X-type or a Y-type magnetic neutral point, or unsteady \citep{ferr06b}. }
   \label{fig:models}
\end{figure}

There are basically three main steady-state models of YSO jets in the literature (Fig~\ref{fig:models}): stellar winds \citep{saut04, matt05b}, X-winds \citep{shu94a,cai08} and magnetized disc winds \citep{blan82, ferr93a,ferr95,ferr04}. They are governed by the same set of MHD equations and share the same physics, the differences being the boundary conditions (see next section). They all rely on the existence of a large scale magnetic field, anchored on a rotating object (hence, spinning down this object). Stellar winds and X-winds assume somehow that there is no large scale magnetic field in the disc: all the relevant magnetic flux coming from the infalling envelope has been stored in the star. On the contrary, a magnetized disc wind requires another magnetic scenario, where a strong field remains present in the disc. How can these models be discriminated from observations?

%%%%%%%%%%
%%%%%%%%%

A review of both theoretical and observational constraints has been made by \citet{ferr06b}. There are actually five main pieces of evidences that, to our view, seem to favor magnetized disc winds:

{\em (1) Global mass and energy budgets:} disc winds tap the accretion power. This is the only physical situation where it is easy to extract energy and mass in amounts comparable to observations, naturally explaining the accretion-ejection correlation (Fig.~\ref{fig:obs}). Stellar winds suffer from a mass supply problem \citep{matt05b,zann11}, whereas some (unknown and theoretically challenging) process must be invoked to power X-winds with the stellar rotational energy \citep{shu94a}.

 {\em (2) Jet collimation scales:} the way a MHD jet opens is related to its transverse equilibrium (Grad-Shafranov equation) and linked to the poloidal motion along the magnetic surfaces (Bernoulli invariant). Contrary to some expectations, disc winds settled from the disc innermost radius to, say, 0.1 up to a few AU are consistent with all known images of optical jets \citep{garc01b,pese03,cabr07}.

{\em (3) Presence of dust in jets:} recent spectro-imaging studies of YSOs show a clear depletion of iron gas phase \citep{podi09,podi11, agra11}. This is an evidence of dust in jets, which cannot be explained if these are X-winds or stellar winds.

{\em (4) Molecular microjets:} small scale ($< 100$ AU) low-velocity molecular (CO in HH30 \citealt{pety06}, H$_2$ in DGTau, Agra-Amboage, in prep) flows are observed surrounding the inner optical jet. While these flows could in principle be due to the photo-evaporation of the outer disc, they could as well be the molecular counterpart of the magnetized disc wind, settled beyond 1 AU, since molecules would survive in the wind \citep{pano12}. 

\begin{figure}[t]
\centering   \includegraphics[width=.8\textwidth]{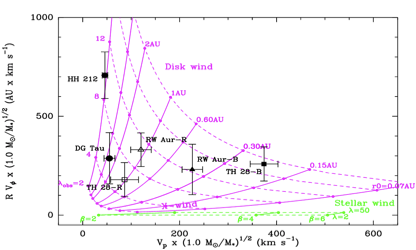}
  \caption{Comparison of predicted specific angular momentum vs.
   poloidal velocities with observations of T~Tauri microjets. Full
   and dashed curves show expected theoretical relations for MHD disc
   and stellar winds. Plotted in symbols are jet kinematics measured at distance
   $z \simeq$ 50~AU in the DG~Tau, RW~Aur, and Th~28 jets. The
   infrared HH~212 jet is also shown for comparison \citep{ferr06b}. }
   \label{fig:rot}
\end{figure}

{\em (5) Jet kinematics:} optical and IR emission line allow to derive jet kinematics, both along but also across the jet. It has been argued that rotation has been detected, allowing to determine the launching radius of the dominant (in terms of emission) streamline \cite{bacc02, ande03,pese04, coff12}. It is found that only extended disc winds can meet all known observational constraints, as long as the observed velocity shift is indeed a signature of jet rotation (Fig \ref{fig:rot}, see also \citealt{coff12}).

%%%%%%%%%%
%%%%%%%%%%

Magnetized disc winds can therefore constitute the main component, in terms of power and mass, of YSO jets. However, there are also several aspects that cannot be easily explained by such winds: 

{\bf (i) Variability:} a disc wind has no typical time scale (unless perhaps the orbital time scales at its inner and outer radii), whereas YSO jets harbour several time scales, from years (e.g. knot periodicity from 2.5 to tens of years \citep{hart07a,agra11}  to hundreds of years. 

{\bf (ii) Jet asymmetry:} YSO jets are bipolar but, while they seem to eject the same mass at each side, the terminal velocities differ by up to a factor 2 \citep{podi11,meln09}.

{\bf (iii) Presence of a hot inner flow}, detected in blue shifted HeI absorption lines \citep{edwa03, edwa06}. This is indicative of a component whose most probable origin is a stellar wind (which is unavoidable anyway).

{\bf (iv) X-ray emission} at the base of jets with a possible stationary component \citep{gued05,skin11}. This may be explained by a shock due to a very fast jet carrying not much mass flux \citep{boni07,boni11}. 
 
The picture is therefore more complex than just pure magnetized disc winds. Besides, disc winds are unable by nature to spin down the central star. Thus, either YSO jets are explained by another jet model or YSO jets are actually multi-component flows. In the following sections, we review the pros and cons of several steady-state wind models from a theoretical perspective. 
 
\subsection{Theory of steady-state MHD jets}
%%%%%%%%%%%%%%%%%%%%%%%

%%%%%%%%%%%
\begin{figure}[t]
\centering   \includegraphics[width=.3\textwidth]{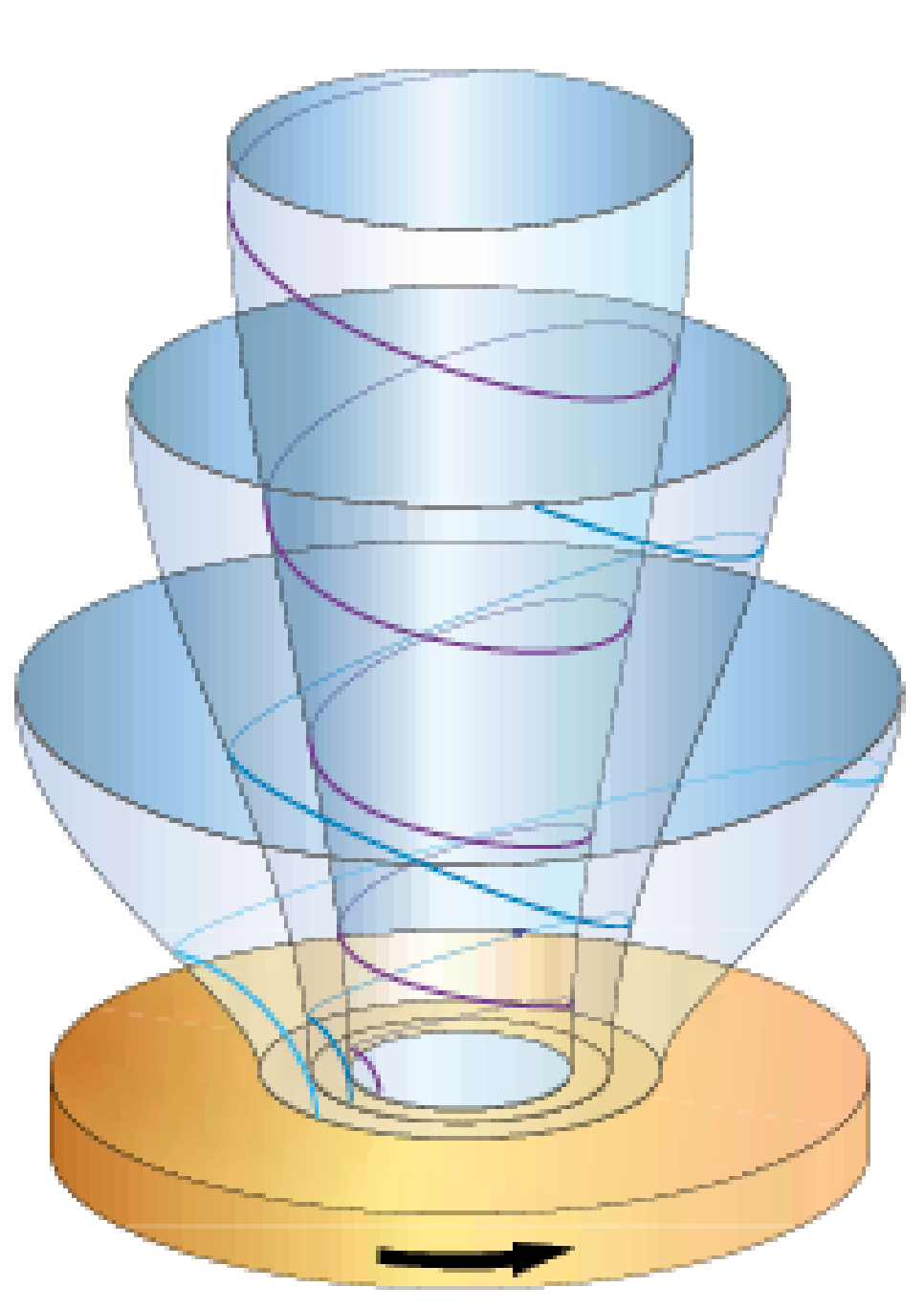}
  \caption{An axisymmetric MHD jet is made of a series of magnetic surfaces (in blue) of constant poloidal magnetic flux nested around each other and anchored on a rotating object (here a disc, in orange, but it could be a star). The rotation of the object imposes the field lines to rotate at the same angular velocity but, because they are carrying plasma and have thereby some inertia, they become trailing and the field lines take a helicoidal shape. The presence of such a helix is consistent with a toroidal magnetic field component $B_\phi$, hence a poloidal current $I= \int_0^r 2 \pi r J_z dr$ flowing along the plasma. Such a jet becomes then collimated by the hoop-stress $F_r= -J_z B_\phi$ (Z-pinch). } 
   \label{fig:smae}
\end{figure}
%%%%%%%%%%%%

The theory of non relativistic, steady-state MHD jets can be already found in several reviews or textbooks (e.g. \citealt{ferr02, tsin07}) so we just highlight here the main points. This theory takes advantage of the existence of MHD invariants in axisymmetry.

\subsubsection{MHD equations}
%%%%%%%%%%%%%%%

The magnetic field writes ${\bf B}= \nabla \times {\bf A}$ so that, choosing $a= rA_\phi$ in cylindrical coordinates, one can write the poloidal field components with one scalar only, namely   
\be
{\bf B}_p = \frac{1}{r} \nabla a \times {\bf e}_\phi
\ee
where $a(r,z)$ is related to the poloidal magnetic flux $\Psi= \int {\bf B}_p \cdot {\bf dS} = \int_0^r 2 \pi r B_z dr = 2\pi a(r,z)$ so that $a=Cst$ describes a magnetic surface (Fig.\ref{fig:smae}). In ideal MHD the flow is frozen in to the magnetic field so that a jet can be seen as a plasma enforced to flow along magnetic flux tubes whose cross-section vary along the z axis. 
In the same way, one has ${\bf J}= \nabla \times {\bf B}/\mu_o$ so that 
\be
{\bf J}_p = \frac{1}{r} \nabla b \times {\bf e}_\phi
\ee
where $b= rB_\phi$ and is related to the poloidal electric current that flows within a magnetic surface, namely $I= \int {\bf J}_p \cdot {\bf dS} = \int_0^r 2 \pi r J_z dr = 2\pi b(r,z)\mu_o$. Whether or not these flux tubes, that are nested around each other, remain in equilibrium depends on the jet transverse balance, which is then mainly controlled by the amount of poloidal electric current. Thus, for instance, when trying to understand asymptotic jet dynamics one has to address the radial distribution of this current.   
 
Since they will be useful for the next section on disc winds, let us  recall here the set of steady-state, axisymmetric, {\em resistive} MHD equations that will be used for dealing with accretion discs:

\noindent (1) Mass conservation equation
\be 
\nabla \cdot \rho {\bf u}  = 0
\label{eq1}
\ee

\noindent (2) Momentum conservation equation 
\be 
\rho {\bf u} \cdot \nabla {\bf u} = -\nabla P - \rho \nabla \Phi_G  + {\bf  J} \times {\bf B}  +  \nabla \cdot {\mathsf T}
\label{eq2}
\ee
where $\mu_o \vec J= \nabla \times {\bf B}$ is the electric current density and ${\mathsf T}$ the turbulent stress tensor which is related to a turbulent viscosity $\nu_v$ \citep{shak73}. Because this turbulence is assumed to vanish in the jets, this viscous term is set to zero there.

\noindent (3) Ohm's law and (4) the toroidal field induction equation
\begin{eqnarray}
  \eta_m {\bf J}_{\phi}  &=& {\bf u}_p \times {\bf B}_p
  \label{eq3}\\
  \nabla \cdot (\frac{\nu'_m}{r^2}\nabla rB_{\phi}) & = &
  \nabla \cdot \frac{1}{r}(B_{\phi} {\bf u}_p - {\bf B}_p\Omega r)
  \label{eq4} 
\end{eqnarray}
where $\eta_m=\mu_o \nu_m$ and $\eta'_m= \mu_o\nu'_m$ are anomalous (turbulent) resistivities \citep{ferr93a}. In the ideal MHD jet, these coefficients are set to zero. 

\noindent (5) Perfect gas law
\be
P= \rho \frac{k_B}{\bar \mu m_p} T = \rho C_s^2
\label{eq5}
\ee
where $C_s$ is the isothermal sound speed, $m_p$ is the proton mass and $\bar \mu$ is the mean molecular weight assumed to be 1/2 in a fully ionized ($n_i=n_e=n$)  and thermalized plasma ($T_i=T_e=T$).   

\noindent (6) An exact energy equation involves various physical mechanisms and an explicit form would be for instance
\be
\nabla \cdot (U {\bf u}_p\ +\ {\bf S}_{rad}\ +\ {\bf q}_{turb}) =  -P \nabla \cdot {\bf u}_p\ +\  q^+_{eff}
\label{eq:6}
\ee
where ${\bf S}_{rad}$ is the radiative energy flux, $q_{eff}=  \eta_m J^2_{\phi}\ +\ \eta'_mJ^2_p\ +\ \rho\nu_v
\left| r \nabla \Omega \right|^2$ an effective (turbulent) heating rate related to the turbulent transport coefficients, $U=P/(\gamma-1)$ is the internal energy, $\gamma$ the adiabatic index and ${\bf q}_{turb}$ some unknown turbulent energy transport term. This term arises from turbulent motions and introduces a redistribution of energy (its sign vary vary inside the volume). Indeed, using a kinetic description and allowing for
fluctuations in the plasma velocity and magnetic field, it is possible to show that all energetic effects associated with these fluctuations cannot be reduced to $q^+_{eff}$, namely only anomalous Joule and viscous heating terms. Therefore, a consistent treatment of turbulence would impose to take into account ${\bf q}_{turb}$ within the disc (see e.g. discussion in \citealt{shak78}). 
Because of all these effects, solving an exact energy equation for the disc is out of range and some approximation must be done. In the case of  jets, they are usually being computed using a polytropic  equation of state  
\be 
P= K \rho^{\Gamma} \ ,
\label{eq:6bis}
\ee 
\noindent where the polytropic index $\Gamma$ can be set to vary between 1 (isothermal case) and $\gamma=5/3$ (adiabatic case) for a monoatomic gas. Here $K$ is related to the entropy $S$ of the flow and  can be allowed to vary radially. But it remains constant along each field line in the adiabatic case (since $\rho T {\bf u}_p\cdot  \nabla S =0$).

For each magnetic surface, there are 7 variables ($\rho, P, u_r, \Omega, u_z, a, b$), 7 equations requiring 7 boundary conditions. One of them is the value of the angular velocity of the field lines $\Omega_*$ (see below): it is taken as the angular velocity of the rotating object (star or accretion disc). A steady-state wind requires that the plasma reaches velocities larger than the speed of waves. The flow must then cross the slow-magnetosonic (SM), Alfv\'en (A) and fast-magnetosonic (FM) critical points, which provides 3 more constraints. Thus, there are 3 free independent quantities that must be specified as boundary conditions along each magnetic surface. In the case of disc winds for instance, one usually choose radial distributions of $\rho(r), u_z(r), B_z(r)$ \citep{ouye99, pudr06, fend06, fend09}.

\subsubsection{MHD invariants}
%%%%%%%%%%%%%%%

In the ideal MHD regime relevant to jets, there are 5 MHD invariants along each magnetic surface. These are (i) the mass to magnetic flux ratio $\eta(a)$, (ii) the angular velocity of the field lines $\Omega_*(a)$, (iii) the total specific angular momentum $L(a)$, (iv) the total specific energy $E(a)$ and $(v)$ the entropy $K(a)$. Looking at these invariants provides good insights on jet physics.

% eta(a)
Ohm's law writes $\bf E + u \times B = 0$, namely  ${\bf E_\phi} = - {\bf u}_p  \times {\bf B}_p$. Now,  $E_\phi = - \frac{1}{r}\frac{\partial V}{\partial \phi} - \frac{\partial A_\phi}{\partial t} = 0$ for time independent, axisymmetric flows so that  $\bf u_p = \alpha B_p$. From mass conservation one derives $div \rho {\bf u_p} = div \rho \alpha {\bf B_p}= {\bf B_p}\cdot \nabla \rho \alpha + \rho \alpha div {\bf B_p} = 0$, namely $\eta \equiv \mu_o \rho \alpha$ is an invariant along any magnetic surface:
\be
\mu_o \rho {\bf u_p} = \eta (a) {\bf B_p} 
\label{eq:eta}
\ee
The mass flux at an altitude $z$ between $r, r+dr$ writes $d \dot M_w = 2 \pi r \rho u_z dr = 2 \pi r \frac{ \eta }{\mu_o}B_z dr = \frac{2 \pi}{\mu_o}\eta \frac{\partial a}{\partial r} dr$, allowing us to interpret $\eta = \frac{\mu_o}{2 \pi} \frac{\partial \dot M_w}{\partial a}$ as the mass flux to magnetic flux ratio, a measure of the mass loading into field lines.

% Omega(a)
The induction equation for the toroidal field writes
\begin{eqnarray}
\frac{\partial B_\phi}{\partial t} &=& -  \nabla \times {\bf E} |_\phi = \nabla \times ( {\bf u}\times {\bf B}) |_\phi \\
&=& \frac{\partial }{\partial z} \left ( \Omega r B_z - u_z B_\phi \right ) - \frac{\partial}{\partial r} \left ( u_r B_\phi - \Omega r B_r \right )\\
&=& - \frac{r}{r}\frac{\partial}{\partial r} r \left (u_r \frac{B_\phi}{r} - \Omega B_r \right) -  r \frac{\partial }{\partial z} \left (u_z \frac{B_\phi}{r}  -  \Omega B_z \right )
\end{eqnarray}
In steady state, this equation provides $div \left ( \frac{B_\phi}{r}{\bf u_p} - \Omega {\bf B_p} \right ) =div {\bf B_p} \left ( \frac{\eta}{\mu_o \rho r}B_\phi - \Omega  \right )= 0 $, so that we can identify 
\be
\Omega_* = \Omega - \eta \frac{B_\phi}{\mu_o \rho r}
\ee
as the angular velocity of the field line. This is a generalization of Ferraro's iso-rotation law when plasma inertia cannot be neglected. Note that the electric field writes ${\bf E}= -\Omega_* \nabla a$ which allows to interpret $\Omega_*$ as the (local) angular velocity of the reference frame where the electric field vanishes. 

% L(a)
The angular momentum equation writes
\begin{eqnarray}
\rho \left ( \frac{\partial u_\phi}{\partial t} + {\bf u} \cdot \nabla u_\phi \right) &= & F_\phi = {\bf J \times B}|_\phi = J_z B_r - J_rB_z \\
\frac{\rho}{r} \left ( \frac{\partial}{\partial t} \Omega r^2 + {\bf u_p}\cdot \nabla \Omega r^2 \right ) &=& \frac{1}{r} {\bf B_p} \cdot \nabla \frac{r B_\phi}{\mu_o}   
\end{eqnarray}
leading to 
\be
\frac{\partial}{\partial t} \rho \Omega r^2 +\, div \left ( \rho \Omega r^2 {\bf u_p} - \frac{r B_\phi {\bf B_p}}{\mu_o} \right ) =0   
\ee
Using Eq.(\ref{eq:eta}) and stationarity, this equation gives $L \nabla \cdot {\bf B_p} + {\bf B_p} \cdot \nabla L= 0$ showing that the total specific angular moment  
\be
L= \Omega r^2 - \frac{r B_\phi}{\eta}
\ee
is indeed an invariant along each surface. 

% E(a)
The last invariant is the specific energy carried along each field line. It is obtained following the usual procedure for the Bernoulli integral, namely
\be 
{\bf u_p} \cdot \nabla \left ( \frac{u^2}{2} + H + \Phi_G \right) = {\bf u_p} \cdot \frac{{\bf J \times B}}{\rho}
\ee
where $H= \frac{\gamma}{\gamma-1}\frac{P}{\rho}$ is the plasma enthalpy and the magnetic contribution
\begin{eqnarray}
{\bf u_p} \cdot \frac{{\bf J \times B}}{\rho} &= & \Omega_* r \frac{F_\phi}{\rho}  =  \Omega_* r  \frac{{\bf B_p}}{\mu_o \rho r} \cdot \nabla r B_\phi \\
&=& \frac{{\bf B_p}}{\mu_o \rho} \cdot \nabla \Omega_* rB_\phi =
\frac{{\bf u_p}}{\eta} \cdot \nabla \Omega_* rB_\phi = {\bf u_p} \cdot \nabla \frac{ \Omega_* rB_\phi}{\eta}
\end{eqnarray}
providing finally 
\be
E = \frac{u^2}{2} + H + \Phi_G  + \frac{ \Omega_* rB_\phi}{\eta}
\ee

\subsubsection{Some hints on jet physics}
%%%%%%%%%%%%%%%%%%%%

Combining the expressions of $\Omega_*$ and $L$ provides
\begin{eqnarray}
\Omega r^2 &=& \frac{m^2 L - \Omega_* r^2}{m^2 - 1} \\
- \frac{r B_\phi}{\eta} &=& \frac{\Omega_* r^2 - L}{m^2 - 1}
\end{eqnarray}
where $m= u_p/V_{A,p}$ is the poloidal Alfv\'en Mach number. The jet starts with $m<< 1$ so that $\Omega \simeq \Omega_*$ and  $- \frac{r B_\phi}{\eta} \simeq L$: the magnetic field enforces an iso-rotation and carries all the stored angular momentum.  Then, the plasma gets progressively accelerated until it becomes super-A (and super-FM). We can thus define the Alfv\'en radius $r_A$ as the cylindrical distance where $m=1$. The regularity condition at the Alfv\'en point provides $L= \Omega_* r_A^2$. One can interpret therefore $r_A$ as some magnetic lever arm which is acting on (braking down) the underlying rotating object. Much farther away, if the magnetic surface widens such that $r\gg r_A$ with $m \gg 1$, then almost all angular momentum has been transferred back to the plasma leading to a vanishing angular velocity $\Omega \simeq \Omega_* r^2_A/r^2 \ll \Omega_*$ and the asymptotic poloidal speed is just $u_{p,max} = \sqrt{2 E}$, where $E$ is determined at the base.

The magnetic forces can be written \citep{ferr97} 
\begin{eqnarray}
F_{\phi} & = & \frac{B_p}{2\pi r} \nabla_{\parallel} I \nonumber \\
F_{\parallel} & = & - \frac{B_{\phi}}{2\pi r}\nabla_{\parallel} I
\label{eq:force}\\
F_{\perp} & = & B_pJ_{\phi} - \frac{B_{\phi}}{2\pi r} \nabla_{\perp} I \ .
\nonumber
\end{eqnarray}
where $I = 2 \pi r B_{\phi}/\mu_o <0 $ is the total current flowing within this magnetic surface, $\nabla_{\parallel}\equiv (\vec{B}_p
\cdot\nabla)/B_p$ and $\nabla_{\perp}\equiv (\nabla a \cdot\nabla)/|\nabla a|$. These expressions show that the plasma is accelerated by the current leakage through this surface ($\nabla_{\parallel} I >0$). This effect gives rise to both a poloidal ($F_{\parallel}$) and a toroidal ($F_{\phi}$) magnetic force, accelerating azimuthally the matter and leading to a centrifugal force in the radial direction. When the current $I$ vanishes, or when it flows parallel to the magnetic surface, no magnetic acceleration arises anymore and the
plasma reaches an asymptotic state. 

Note also that the way this current is distributed across the magnetic surfaces is of great importance for the jet transverse equilibrium ($F_{\perp}$). This shows that one has to be careful when dealing with a particular current distribution, since it is of so great importance in both jet collimation and acceleration. It must also be noticed that {\em not all magnetic surfaces can be self-confined}. Indeed, because in MHD any electric current circuit must be closed ($div \, {\bf J}=0$), there must be an outer zone in the jet where the return current flows in. In this zone, the magnetic forces are de-collimating and global confinement relies therefore on the external pressure only, very much alike "magnetic towers" observed in Z-pinch discharges in laboratory experiments (see \citealt{ciar09, suzu10} and references therein).

The transverse equilibrium of each magnetic surface (hence of the whole jet) is obtained by projecting the momentum conservation equation across the surface \citep{ferr97}, namely  
\be
(1-m^2) \frac{B^2_p}{\mu_o {\cal R}}  -  \nabla_{\perp}\left(P + \frac{B^2}{2\mu_o}\right) - \rho \nabla_{\perp} \Phi_G 
+   (\rho\Omega^2r - \frac{B^2_{\phi}} {\mu_o r} )\nabla_{\perp}r = 0
\label{eq:fperp}
\ee
\noindent where $\nabla_{\perp}\equiv \nabla a \cdot \nabla /|\nabla a | $ provides the gradient of a quantity perpendicular to a magnetic surface ($\nabla_{\perp} Q <0$ for a quantity $Q$ decreasing with increasing magnetic flux) and ${\cal R}$, defined by
\be
\frac{1}{{\cal R}} \equiv \frac{\nabla a}{|\nabla a|} \cdot 
\frac{(\vec{B}_p\cdot\nabla)\vec{B}_p}{B^2_p} \ ,
\ee
\noindent is the local curvature radius of a particular magnetic surface. When ${\cal R} >0$, the surface is bent outwardly while for ${\cal R} <0$, it bends inwardly. The first term in Eq.(\ref{eq:fperp}) describes the reaction to the other forces of both magnetic tension due to the magnetic surface (with the sign of the curvature radius) and inertia of matter flowing along it (hence with opposite sign). The other forces are the total pressure gradient, gravity (which always acts to close the surfaces and deccelerate the flow, but whose effect is already negligible at the Alfv\'en surface), and the centrifugal outward effect competing with the inward hoop-stress due to the toroidal field. The flow undergoes therefore first an opening in the sub-A regime (due to the pressure gradient and centrifugal terms) despite the poloidal magnetic tension term and, once in super-A regime, gets confined mainly by the last term, the magnetic tension due to $B_\phi$ (the so called hoop-stress).  

Solving the full set of ideal MHD equations is a hard task, as the problem is 2D in nature. One nice way to grasp this is to use the Grad-Shafranov equation. Taking advantage of all MHD invariants, this equation (obtained from the previous one) writes for an adiabatic jet \citep{ferr02}
\begin{eqnarray}
  \nabla \cdot (m^2 -1) \frac{\nabla a}{\mu_o r^2}  &= & \rho \left \{
    \frac{d E}{d a} - \Omega \frac{d \Omega_*r_A^2}{d a} \right. \ +\  
  \left . (\Omega r^2 - \Omega_* r^2_A)\frac{d \Omega_*}{d a}    \right .   \nonumber\\ 
  & &  \ -\  \left .  \frac{C_s^2}{\gamma(\gamma-1)} \frac{d \ln K}{d a} \right \}
\ +\  \frac{B^2_{\phi} + m^2B^2_p}{\mu_o} \frac{d \ln \eta}{d a}  
  \label{eq:GS}
\end{eqnarray}
It is a PDE of mixed type providing $a(r,z)$ once the following quantities are specified: (1) all distributions of the invariants; (2) the localizations and shapes of the two limiting surfaces (SM and FM); (3) the boundary conditions there. Indeed, between the SM and FM surfaces the flow is elliptic whereas it becomes hyperbolic beyond the FM surface. This is an ill-posed mathematical problem, for which there is no method of resolution known.
  
This is the reason why so little is actually known on MHD jets. There are, on one side steady state jet solutions (from stars or discs) obtained with a self-similar ansatz. But because self-similarity introduces a strong symmetry, these solutions, albeit exact, are not general. On the other side, time-dependent numerical simulations allow to obtain full 2D or even 3D solutions that can, under some circumstances, converge to a final steady state \citep{ouye99, pudr06, fend06, fend09}. But they are computationally expensive and, because one needs to impose arbitrary distributions, do not allow to explore the full parameter range.

%%%%%%%%%%%%%%%%%%%%%%%%%%%%%
\section{Magnetized disc winds: The MAES model}
%%%%%%%%%%%%%%%%%%%%%%%%%%%%%
\label{sec:maes}

Since the seminal paper of \cite{blan82}, it is well known that magneto-centrifugally driven jets exert a torque on the underlying accretion disc. However, most of the studies devoted to the launching of such jets assumed either a platform\footnote{The disc structure is not considered and equations are  solved above the disc surface where ideal MHD applies.} or an underlying disc unaffected by the presence of the wind, namely where the dominant torque remains the "viscous" one (as in the Standard Accretion Disc, \cite{shak73}, hereafter SAD). Here, we describe instead Jet Emitting Discs (hereafter JED), which is a class of accretion solutions where all the local disc angular momentum is transported vertically via two jets. The physical origin of this torque is Lenz's law: once the ejected material reaches the Alfv\'enic radius, its inertia produces a feedback on the disc that acts to brake it down. The jets allow accretion and accretion feeds the jets: these interdependent dynamical structures have been termed Magnetized Accretion-Ejection Structures \citep{ferr93a, ferr95}.  

In a series of papers, the full set of dynamical equations describing both the resistive accretion disc and ideal MHD jets (Fig. \ref{fig:smae}) were simultaneously solved thanks to a self-similar Ansatz (see eg. \cite{ferr04} and references therein). It was shown that once a large scale magnetic field reaches a value smaller than but close to the equipartition value ($\mu \sim 1$), its torque becomes dominant wrt the usual viscous (turbulent) torque and accretion reaches almost sonic speeds. As a consequence, for a given accretion rate $\dot M$, a JED is much less dense than a Standard Accretion Disc or SAD. In these models, the disc accretion rate scales as $\dot M \propto r^\xi$ where $\xi$ is the local jet ejection efficiency. It is not a free parameter but a {\em result} of the trans-alfv\'enic jet condition. It is found to be typically around 0.01 for adiabatic or isothermal magnetic surfaces \citep{ferr97, cass00a}. When some heat deposition occurs at the disc upper layers, the disc mass loss may be drastically enhanced, reaching $\xi \sim 0.4$ \citep{cass00b}.

%%%%%%%%%
\begin{figure}[t]
\centering   \includegraphics[width=.6\textwidth]{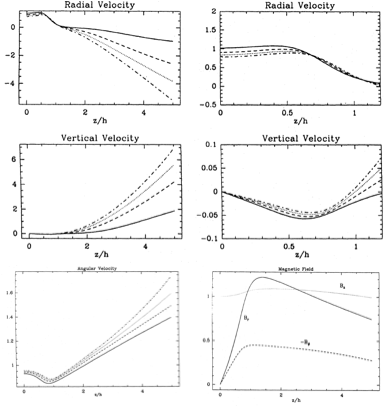}
  \caption{Vertical profiles of velocity for several solutions characterized by different ejection efficiencies $\xi$ and (bottom right) magnetic field components for a representative solution. Within a factor of 2, all components of the magnetic field are comparable at the disc surface. Here, the flow operates a transition around $x=z/h\simeq 1.2-1.6$ from a quasi-static resistive MHD disc to a super slow-magnetosonic ideal MHD jet. Figure adapted from \citet{ferr95}.}
   \label{fig:FP95}
\end{figure}
%%%%%%%%

Figure (\ref{fig:FP95}) shows a close up of the vertical profiles of the velocity and magnetic fields in a vertically isothermal disc. These profiles are quite typical of those found within a JED (see \citealt{ferr95} or \citealt{ferr02} for more details). The flow is accreting at a sonic speed (because of the quite strong magnetic field), with a slightly converging motion (ie $u_z <0$). Only the disc upper layers are being deviated into the jet. The slightly sub-keplerian angular velocity profile shows a typical arcade shape: the initial decrease is due to the increasing effect of the radial magnetic tension as the density decreases with height. It also exactly corresponds to a spin down torque exerted ($F_\phi \simeq - J_r B_z <0$) by the magnetic field. The following increase in $\Omega$ is the sign of an azimuthal magnetic acceleration ($F_\phi >0$). As a consequence, the centrifugal term appearing in the radial momentum conservation equation grows and becomes comparable to the radial magnetic tension (driving thereby a strong outward -ejection- motion).
The magnetic torque changes its sign roughly at the disc surface. When this occurs, the projection of the Lorentz force along a magnetic surface becomes also positive (Eq.~\ref{eq:force}). The jet is thus accelerated by magnetic means in both directions (azimuthally and in the poloidal plane). However, within the disc, the poloidal Lorentz force is acting against the flow. Note also that the disc is being vertically pinched by a strong magnetic compression. It is that element that limits the actual amplitude of the magnetic field (measured by the disc magnetization parameter $\mu = B_z^2/P$, where $P=\rho C_s^2$). While accretion is favoured for stronger $\mu$, the disc vertical balance forbids $\mu$ to be larger than about unity \citep{ferr95}. These two contradictory constraints are best satisfied when $\mu \sim 0.5$. 

All three magnetic field components are comparable at the disc surface. There, the jet magnetization, which is defined as the ratio of the MHD Poynting flux to the kinetic energy flux, namely
\be
\sigma = \left . \frac{- 2 \Omega_* r B_{\phi} B_p} {\mu_o \rho u^2 u_p} \right |_{z=h} \simeq \frac{1}{\xi}
\ee
is large for a low ejection index \citep{ferr97}. As a result, the matter which is frozen in the magnetic field, can be considered as "beads on a rotating wire" \citep{chan80}. A simple energetic requirement, based on the Bernoulli integral, as been derived by \citet{blan82}. Consider a field line anchored at a radius $r_o$ at the disc surface and making an angle $\theta$ with the vertical such that $\tan \theta = B_r^+/B_z$. Assuming isorotation and $\Omega=\Omega_* =\Omega_K$ and a negligible initial poloidal velocity, at which condition a cold flow (zero enthalpy) can be launched, namely develop a velocity going to $r= r_o + \delta r$ and $\delta z$ (with $\tan \theta = \delta r/\delta z$)? 
The Bernoulli invariant writes  
\begin{eqnarray*}
E - \Omega_* L&=& \frac{u^2}{2} + \Phi_G  - \Omega_*\Omega r^2 \\
&=& \frac{\Omega_{K,o}^2 r_o^2}{2} - \frac{GM}{r_o} - \Omega_{K,o}^2 r_o^2 = - \frac{3}{2}\frac{GM}{r_o}\\
&=&  \frac{u_p^2 + \Omega_{K,o}^2 r^2}{2} - \frac{GM}{\sqrt{r^2 + \delta z^2}} - \Omega_{K,o}^2 (r_o + \delta r)^2  
\end{eqnarray*}
which, after a second order Taylor expansion, gives
\be
u_p^2 = \Omega_{K,o}^2 \delta r^2 \left ( 3 - \tan^2\theta \right )
\ee
%%%%
or $\tan \theta > 1/\sqrt{3}$, namely an inclination angle with respect to the vertical larger than 30$^o$. All cold solutions obtained so far fulfill this constraint. When some heating is present at the disc surface this condition is relaxed \citep{cass00b}.

Note that although a thin disc approximation was used in \citet{ferr95}, subsequent works used the full expression for the gravitational potential of the central object. These works are still the only ones available in the literature that include {\it all dynamical terms}\footnote{Dynamo, disc self-gravity and radiation pressure have been however neglected.}, thereby allowing to study either geometrically thin or thick discs (the disc aspect ratio $\varepsilon=h/r$ is a free parameter). 
The JED parameter space (obtained with both slow and Alfv\'enic constraints) has been computed in \citet{ferr97} for isothermal and in \citet{cass00a} for adiabatic magnetic surfaces. It is shown that the ejection efficiency $\xi$ grows non-linearly with the disc magnetization $\mu$. The outcome of this parametric study is that only a tiny interval of JED physical conditions allows for steady-state ejection:
\begin{eqnarray}
&& 0.1 < \mu= \frac{B_z^2}{\mu_o P} < 1 \\  
&& 0.2 < \alpha_m = \frac{\nu_m}{V_A h} \sim 1 \\
&& \chi_m = \frac{\nu_m}{\nu_m'}  \simeq \frac{\alpha_m^2}{3}  \sim 1/3
\end{eqnarray}
where $\nu_m'$ is the turbulent magnetic diffusivity in the toroidal direction (suspected to be stronger than that in the poloidal direction, $\nu_m$). 

Although these solutions are solutions to {\em the full set of MHD equations}, they suffer of course from two caveats: (1) there are self-similar (power-laws of the cylindrical radius), so that no radial boundary condition can be imposed;  (2) a local $\alpha$-prescription for the turbulent torque and magnetic field transport has been made. 

The first caveat is actually not too serious: while astrophysical jets are definitely not self similar, the underlying physics revealed by these exact solutions remains valid, as far as the jet launching process is concerned. When it comes to jet propagation and asymptotic behavior, then the mathematical bias comes into play \citep{ferr97, ferr04}. In fact, most of the findings of self-similar calculations were confirmed by 2.5D time-dependent numerical simulations, using either the MHD VAC code \citep{cass02, cass04}, FLASH code \citep{zann07} or newer PLUTO code \citep{tzef09}. Among a quite large amount of available numerical simulations  of jet launching from {\it thin} accretion discs, only those cited above achieved a quasi-steady-state. The reason lies in the fact that they started their simulations with (i) an initial condition taking into account the radial and vertical magnetic forces, (ii) a field close to equipartition $\mu \sim 0.5$ and (iii) a turbulence mimicked with an $\alpha$-prescription with $\alpha_m \sim 1$. The parametric study done by \citet{tzef09} confirmed that a field larger or much smaller than equipartition would never allow steady-state ejection in a JED. To our view, these time-dependent experiments provide a firm ground to the analytical analysis. However, they also suffer from the $\alpha$-prescription used.

The second caveat, namely the use of an $\alpha$-prescription, is more serious and would require numerical simulations of global JEDs (these are yet to come). It is well known that whenever magnetic fields are present, the disc is unstable to the Magneto-Rotational Instability (hereafter MRI) and that a self-sustained turbulence sets in \citet{balb91}. This leads to a radial angular momentum transport very much alike an anomalous viscosity $\nu_v$ (see \citealt{balb03} and S. Fromang's contribution, this book). Within a JED, viscosity is useless but a magnetic diffusivity of turbulent origin ($\nu_m$ and $\nu_m'$) is needed. Thus, the problem remains. The initial conjecture of \citet{ferr93a} was that the relevant MHD instability in JEDs would also provide such a diffusivity, with an effective Prandtl number ${\cal P}_m= \nu_v/\nu_m \sim 1$ and some degree of anisotropy $\chi_m= \nu_m/\nu_m' \sim 1/3$.  

However, the MRI is quenched when the field approaches an equipartition level so that it might seem inconsistent with our assumption. This is actually not exactly the case: all trans-Alfv\'enic solutions found had a field {\em below} equipartition so that JEDs would be close to the limit of marginal stability (but still unstable). We therefore still expect MRI to be the main driver of the  turbulence in JEDs and recent works on MRI in shearing box seem to confirm this statement. 

Indeed, \citet{guan09} measured the decay time of an imposed field and derived the amplitude of the turbulent magnetic diffusivity $\nu_m$. It appears comparable to the "viscosity" $\nu_v$. Closer to our preoccupations, \citet{lesu09} imposed a current density inside $J_i$ the shearing box and verified that the turbulent term $< \delta \vec u \times \delta B>_i$ is actually proportional to  $J_i$. Thus, the turbulent magnetic  transport {\it does} seem to behave like a resistivity whose tensor is mostly diagonal. This is a major result that must be more generally established. Besides, they not only also report some degree of anisotropy (comparable to the one used in JED calculations) but the effective magnetic Prandtl number is close to unity (around 2-5). Finally, it is worth to stress that the scaling of the viscosity, derived from numerical experiments in shearing boxes \citep{pess07,lesu07}, is found to be
\begin{equation} 
\nu_v \equiv \alpha_v \Omega_k h^2 \sim V_A h
\end{equation}
namely $\alpha_m \sim 1$ as defined in \citet{ferr93a}. This is a remarkable result although verified only for $\mu \ll 1$. 

Recently, \citet{lesu13} showed with stratified shearing box simulations done with $\mu \sim 0.2$ that the outcome of MRI under these circumstances is not to lead to turbulence. Instead, trans-Alfv\'enic flows are launched as the result of a positive magnetic torque at eth disc surface. Hence, {\em jet launching appears as a natural consequence of MRI at large $\mu$}. Moreover, they showed that the accretion-ejection flow was unstable, with an instability resembling a Kelvin-Helmholtz instability involving radial scales. This is a very interesting result that deserves more studies, and in particular global simulations, to assess whether or not such instability could provide the source of the anomalous magnetic diffusivities required in analytical models and numerical simulations.   

We therefore believe that  the phenomenological scalings used for the turbulent transport coefficients are physically motivated. Besides, they have the great advantage of allowing us to grasp most of the averaged (ie within a mean field approach) behaviour of accretion-ejection structures. To summary, we have now a fairly well understanding of how self-confined jets can be launched from alpha discs and some indications from turbulent discs. But, of course, these MAES are not causally connected to the central object and cannot extract its angular momentum. Magnetized disc winds carry away the exact amount of disc angular momentum allowing disc material to accrete in a near Keplerian fashion. No hope, then, to spin down cTTS with such models.

%%%%%%%%%%%%%%%%%%%%%%%%%%%%%%%%
\section{The revival of stellar winds: The APSW model}
%%%%%%%%%%%%%%%%%%%%%%%%%%%%%%%%
\label{sec:apsw}

As \citet{mest87} wrote in their introduction, "{\em the idea of magnetic braking by a s stellar wind is now very familiar. In a seminal paper \citet{scha62} pointed out that if gas emitted from a star is kept corotating with the star by the magnetic torques out to large distances, it will carry off far more angular momentum per unit mass than gas retaining the angular momentum of the stellar surface}". This is the reason why using a stellar wind to spin down a protostar was the first idea that came into minds. The spin evolution of an accreting TTS is given by Eq.(\ref{eq:starevol}) which writes
\be
\frac{\partial I_* \Omega_*}{\partial t} = \Gamma_{acc} + \Gamma_{mag} - \Gamma_{w}
\label{eq:starevol2}
\ee
where the stellar wind torque contribution is given by 
\begin{eqnarray}
\Gamma_w &=& \oint_w \left ( \rho \Omega r^2 {\bf u_p} - \frac{r B_\phi {\bf B_p}}{\mu_o} \right ) \cdot {\bf dS}= 2 \int_w \rho {\bf u_p} L(a) \cdot {\bf dS}   \nonumber \\
&=& 2 \int_w \rho {\bf u_p} \Omega_* r_A^2 \cdot {\bf dS} = 2 \int_w \Omega_* r_A^2 d\dot M_w = 2 \dot M_w \Omega_* \bar r_A^2
\end{eqnarray}
In the above expression, $\bar r_A$ must be understood as a mass-weighted average Alfv\'en radius \citep{webe67,mest84}
. We already showed that within the steady-state GL paradigm the magnetic torque $ \Gamma_{mag}$ cannot balance the accretion torque $\Gamma_{acc}>0$. Neglecting the latter with respect to the former, a minimum requirement for spin equilibrium would then be $\Gamma_{acc}= \dot M_a \Omega_K r_t^2 = \Gamma_w = 2\dot M_w \Omega_* \bar r_A^2 $. This translates into the dimensionless form
\be
f J \delta =   \left(  \frac{r_t}{R_*}\right )^{1/2} 
\ee
where $f= \frac{ 2 \dot M_w }{ \dot M_a}$ is the mass flux ratio, $\delta = \frac{\Omega_*}{\Omega_K(R_*)}$ and $J= \frac{\bar r_A^2 }{R_*^2}$. Note that $r_t$ is given by Eq.(\ref{eq:rt}) so that $f J \delta \sim o(1)$ for typical cTTS parameters. Clearly, the more massive the wind and the better it is. 

%%%%%%%%%%
\begin{figure}[t]
\[ \begin{array}{cc}
\includegraphics[width=.4\textwidth]{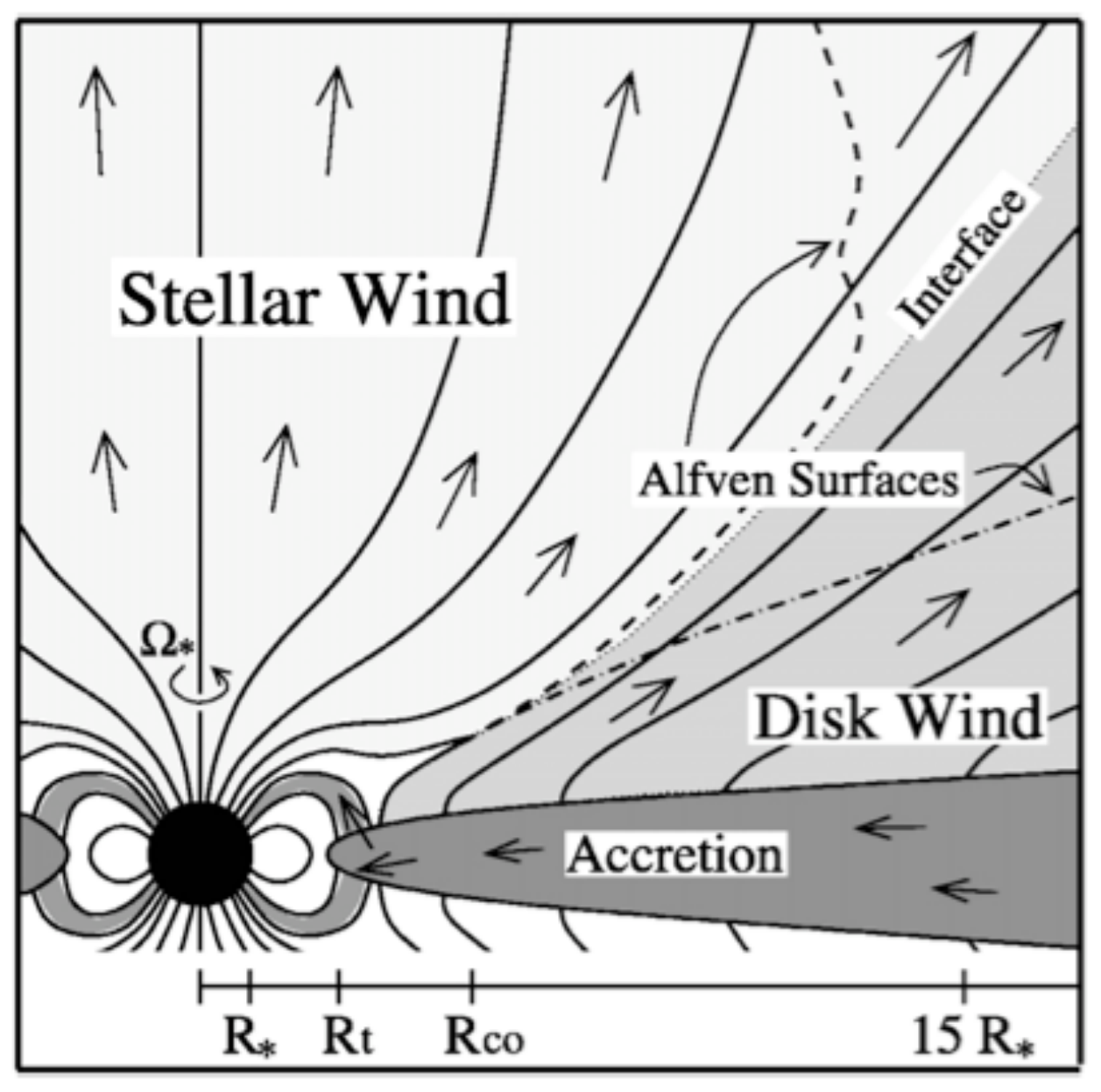}  
  &    
   \includegraphics[width=.5\textwidth]{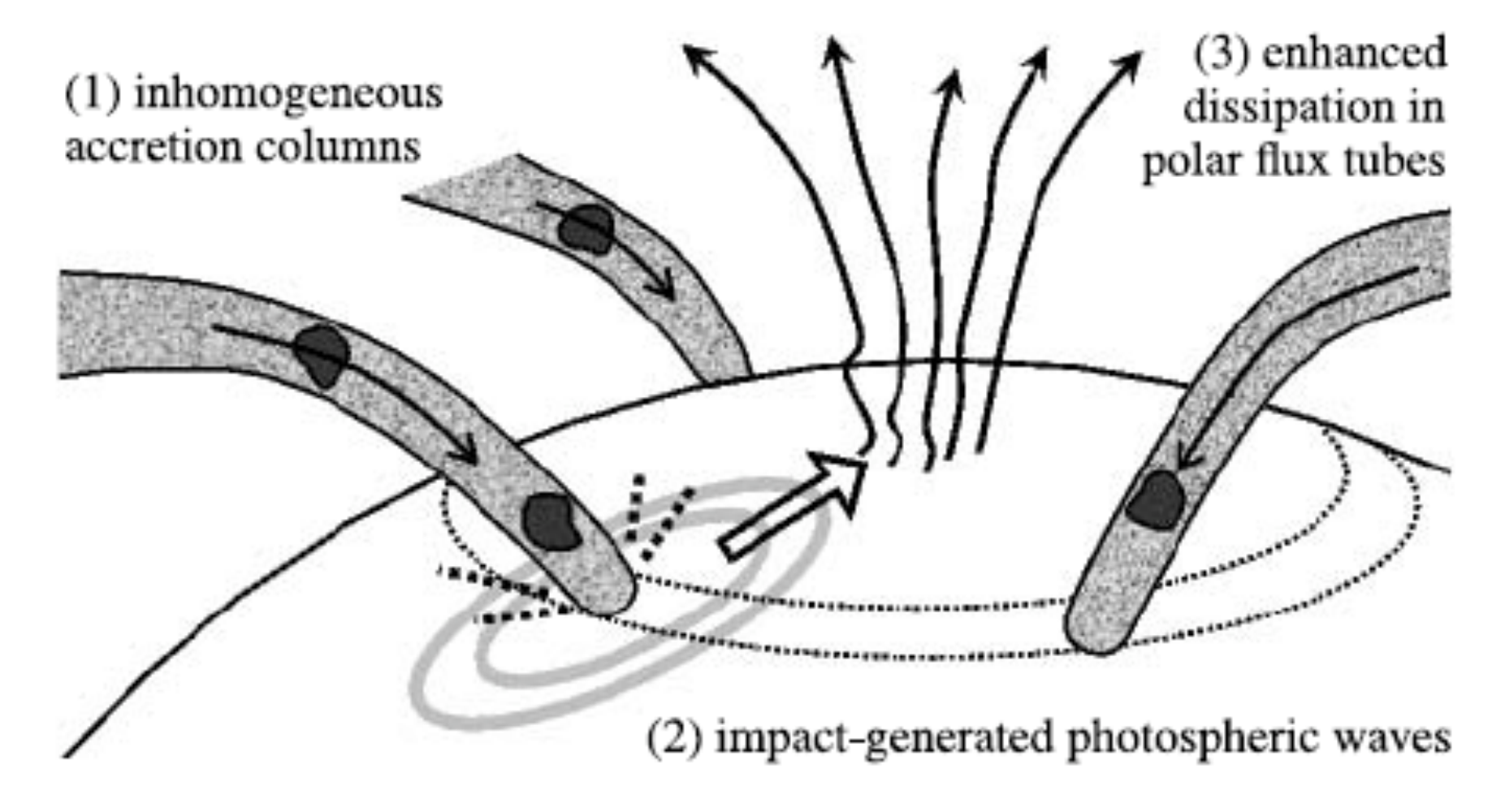}
\end{array} \]
  \caption{{\bf Left:} The APSW paradigm \citep{matt05b}. An outer disc wind would be present but the star would be spun down by the braking torque due to a massive stellar wind. Note that this magnetic interaction is of the Y-type \citep{ferr06b}. {\bf Right:}  The "magnetic cauldron" at the base of the APSW. In this region, multipolar magnetic fields would provide a complex situation allowing to convert thermal and kinetic energy released in the accretion shock into a turbulent magnetic energy \citep{cran08}.}
  \label{fig3_1}
\end{figure}
%%%%%%%%%%

However, in the case of cTTS, one has to face two severe observational issues:
\begin{itemize} 
 \item TTS are slow rotators with $\delta \sim 0.1$. Thus, contrary to magnetized disc winds, the centrifugal effect  can hardly help to expel off matter from the stellar surface. It has been therefore argued that enthalpy would play that role, with the caveat of the wind not being too massive. Indeed, escaping from the gravitational well requires $C_s^2 \sim \Phi_G$, which requires hot ($\sim 10^6$ K) material that would lead to excessive emission features that are not observed if it were too massive \citep{deca81, koni86}. Thus, any massive stellar wind model from a cTTS needs to rely on a non dissipative process, maintaining the outflowing material cold while providing nevertheless enough push to lift material out of the gravitational well. A pressure due to turbulent Alfv\'en waves is usually the best candidate invoked for this \citep{deca81, hart82a, hart82b}.
  
 \item The accretion-ejection correlation found in YSOs clearly states that, somehow, the wind energetics must be related to accretion. Moreover,  the energy powering stellar winds from TTS can hardly come from the stellar rotation alone. This can be readily seen by using the Bernoulli invariant. Indeed, the asymptotic speed reached along a magnetic field line anchored at a colatitude $\theta_o$ is
 \be
 v_p = \sqrt{\frac{G M_*}{R_o}} \sqrt{ \delta^2 \sin^2 \theta_o (2 \lambda - 1) -2 + \beta } 
 \label{eq:vit_stellar}
 \ee
where $R_o=R_*$ is the spherical radius ($r_o=R_o \sin \theta_o$), $\lambda= r_A^2/r_o^2$ and $\beta >0$ encompasses both thermal and turbulent pressure effects acting along the flow (see \citealt{ferr06b} for more details). Unless $\lambda \gg \delta^{-2}$ (namely an Alfv\'en radius $r_A \gg \delta^{-1} r_o$), the rotational energy is negligible and the flow must be fed by another source of energy (mediated through $\beta$). 
\end{itemize} 
%%%%%%%%%%
\begin{figure}[t]
\[ \begin{array}{cc}
\includegraphics[width=.45\textwidth]{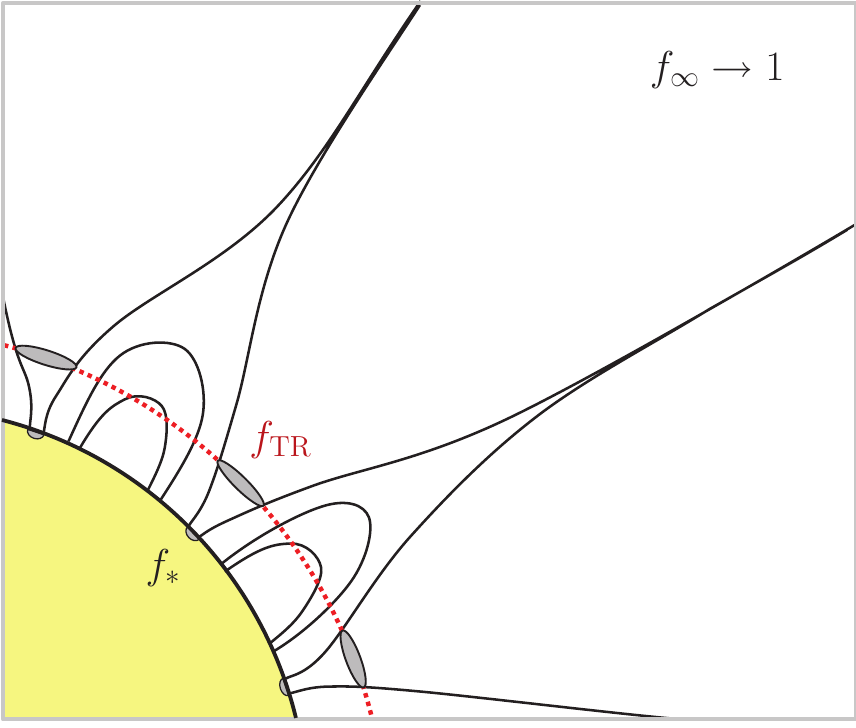}  
  &    
   \includegraphics[width=.45\textwidth]{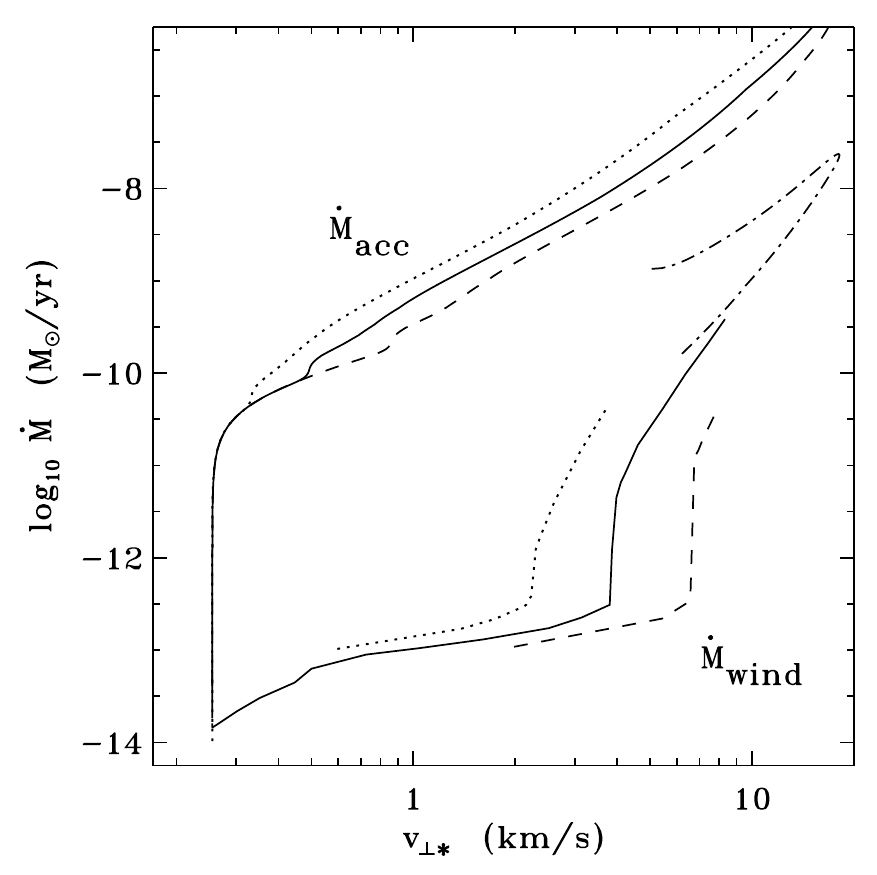}
\end{array} \]
  \caption{{\bf Left:} Sketch of flux tube expansion allowing the wind acceleration, as in a de Laval nozzle \citep{cran11}.  {\bf  Right:}   Mass accretion and mass loss rates for various models versus the photospheric Alfv\'en wave amplitude \citep{cran08}.}
   \label{fig3_2}
\end{figure}
%%%%%%%%%%

These two major issues can, in principle, be solved within the context of Accretion Powered Stellar Winds (hereafter APSW, \citealt{matt05b}). As depicted in Fig~(\ref{fig3_1}a), the idea is that the stellar dipole field truncates the disc at $r_t$ and enforces the matter to accrete along magnetospheric field lines. This infalling super-slow magnetosonic plasma forms a shock near the stellar surface and and releases its accretion power (leading to UV emission). The post-shocked plasma gets then redistributed along the stellar surface so as to become part of the star. Such a continuously maintained shock is expected to generate compressive waves that are able to propagate across the field lines and reach thereby the polar zone with open field lines (Fig~\ref{fig3_1}b). This situation is favorable to create and sustain a "magnetic cauldron"  that may lead to a pressure of turbulent Alfv\'en waves strong enough to drive a massive wind from the stellar surface. Hence, by this turbulent process one might be able to convert a fraction of the released thermal and kinetic energy of the shock to MHD turbulence and eventually to organized kinetic energy of a stellar wind. Such wind would then naturally account for both the accretion-ejection correlation and spin down the protostar. The question is: what are the allowed values of $f$ and $J$ for Accretion Powered Stellar Winds ?

Computing of much energy can be converted into the wind is a hard task, sharing many similarities with the coronal heating and wind acceleration issues in the solar wind. Using a code developed in the stationary one-fluid solar wind case, \citet{cran08} computed the ratio $f$ expected for typical cTTS. This has been done by computing the wind acceleration along a magnetic flux tube with a prescribed cross section (dipole background magnetic field and open flux streams at the poles) for a given spectrum and normalization of MHD fast waves that were injected at the lower boundary. The paper uses the analytic results of \citet{sche88} to set the energy flux in these waves. It is shown that this energy is sufficient to produce mass-loss rates with $f\simeq 0.01$ (Fig~\ref{fig3_2}, see also \citealt{cran09,cran11}). Note that the value of $f$ derived here highly depends on the turbulence injected but also on the flux tube opening, which controls how the flow gets ultimately accelerated. This last effect is crucial and critically depends on the whole transverse equilibrium of the magnetosphere. One should therefore not take this value of $f $ too seriously yet.     

%%%%%%%%%%%%%%
\begin{figure}[t]
\centering   \includegraphics[width=.9\textwidth]{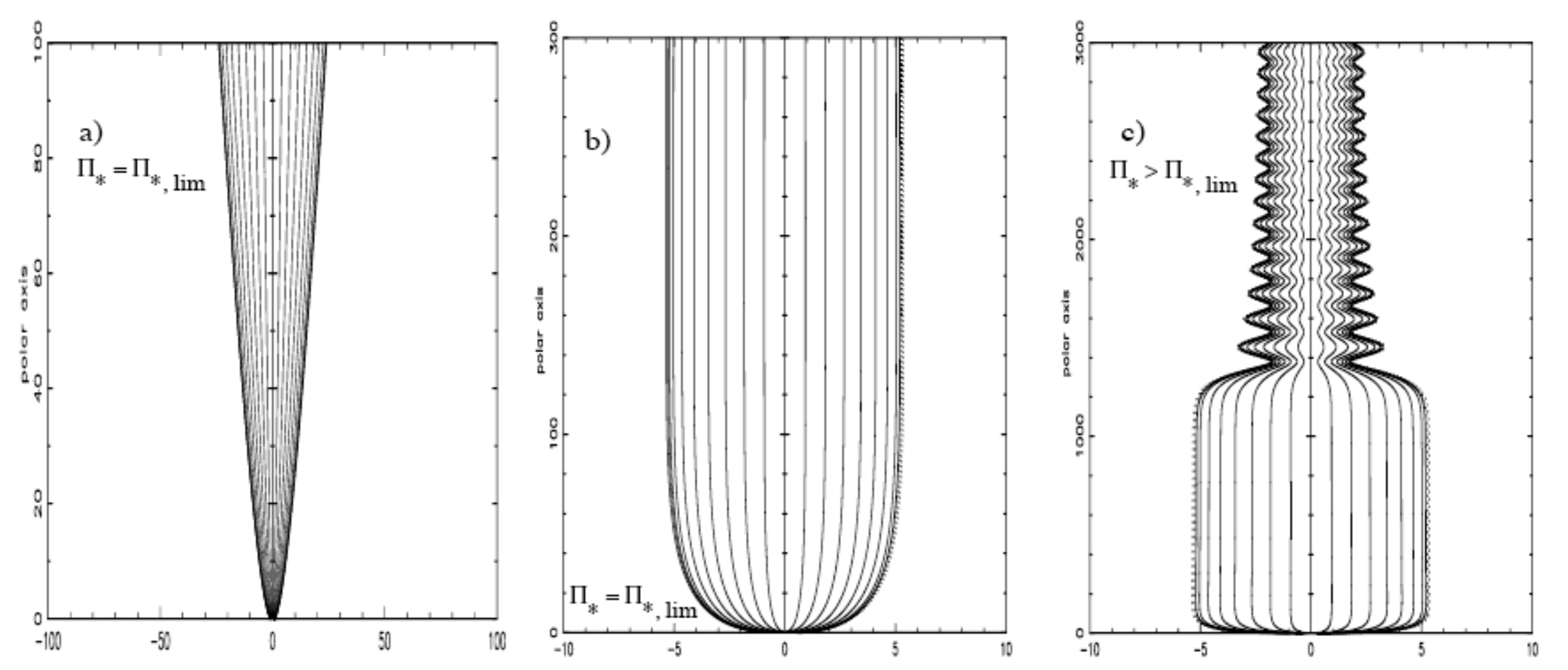}
  \caption{Meridional self-similar stellar wind models from \citet{saut02}. Different jet configurations can be established according to the transverse gradients of the specific pressure and energy.}
   \label{fig3_3}
\end{figure}
%%%%%%%%%%
Another way to tackle this difficult problem is to use $f$ as a free parameter and compute more accurately the magnetic wind structure as it will provide a link $J=J(f)$. Let us first use a simplified analytical approach. Assuming a spherical geometry (hence forgetting the 3D mass-weighted nature of $\bar r_A$), one can simply evaluate the mass flux $\dot M_w = 4 \pi r_A^2 \rho_A v_A$ at the Alfv\'en surface, with $v_A = \frac{B_A}{\sqrt{\mu_o \rho_A}}$ and $B_A = {B_*}\frac{R_*^2}{r_A^2}$. In the case of a centrifugally-driven outflow one can use $\omega_A = \Omega_* r_A/v_A$ as a constant for a given flow geometry \citep{mest84,pell92}, which provides
\be
\frac{r_A}{R_*} = \left ( \frac{4\pi}{\mu_o} \omega_A \frac{B_*^2 R_*^2}{\dot M_w \Omega_* R_*} \right )^{1/3}  
\ee
In the case of a thermally-driven wind, one would use instead $v_A \simeq v_{esc}$ and obtain a new scaling  \citep{kawa88}
\be
\frac{r_A}{R_*} \simeq \left ( \frac{4\pi}{\mu_o}  \frac{B_*^2 R_*^2}{\dot M_w v_{esc}} \right )^{1/2}  
\ee
while different exponents would be found if one were to use $B_A \propto r_A^{-n}$. Following \citet{matt08a} one can generalize this expression as 
\be
\frac{r_A}{R_*} = K \Psi^m  \, \, \, \mbox{   with } \, \, \,   \Psi = \frac{B_*^2 R_*^2}{\dot M_w v_{esc}}
\label{eq:rAwind1}
\ee
where $K$ and $m$ are factors depending on a specific wind model. Note that the disc truncation radius $r_t$ involves the same characteristic quantity $\Psi$ with $m=2/7$. Once $K$ and $m$ are specified, one can easily compute the spin evolution of a cTTS (see J. Bouvier's contribution). We can now turn our attention to stellar wind dynamical models and translate their findings into $K$ and $m$ values.   
%%%%%%%%%%
\begin{figure}[t]
\[ \begin{array}{cc}
\includegraphics[width=.35\textwidth]{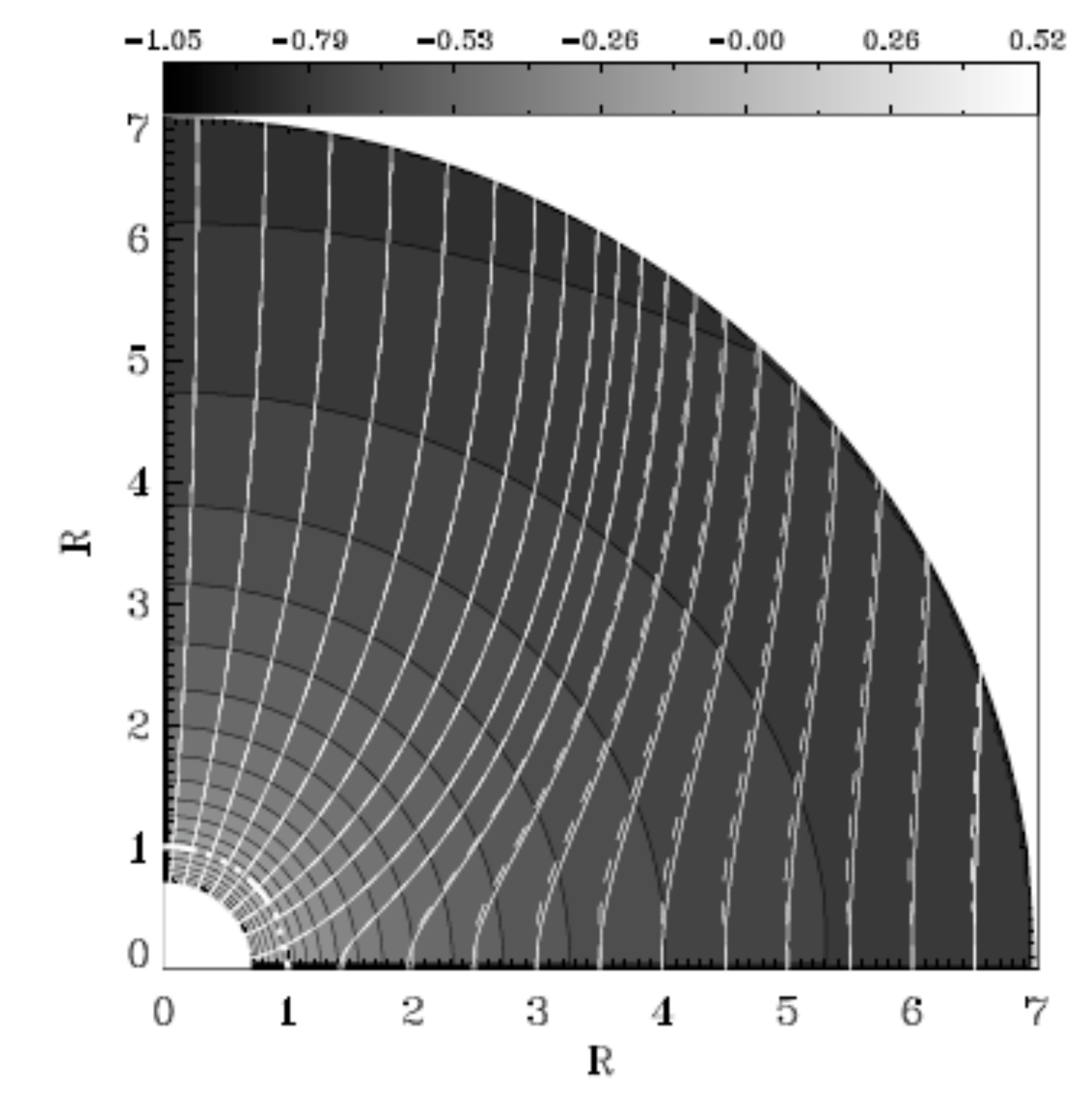}  
  &    
   \includegraphics[width=.65\textwidth]{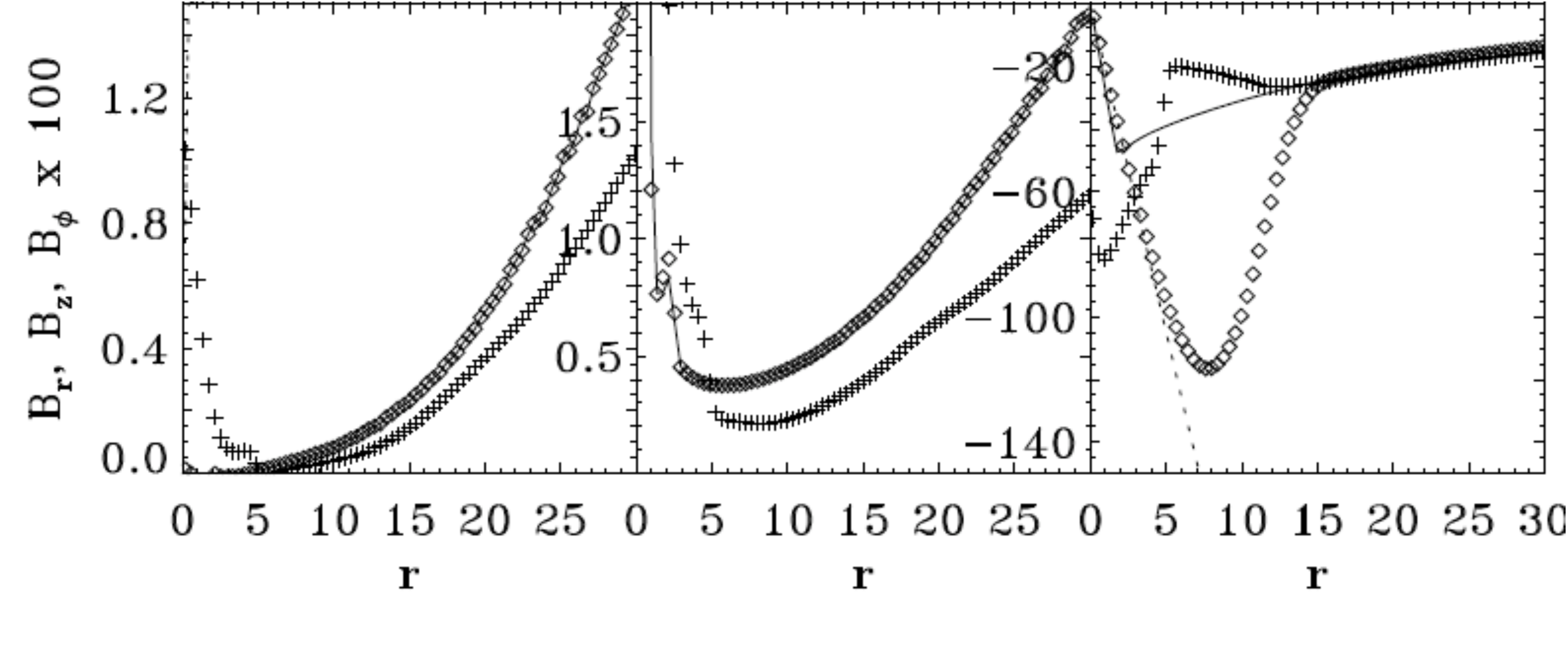}
\end{array} \]
  \caption{{\bf Left:} MHD numerical simulations using as initial condition (dashed lines) the "stellar" wind magnetic configuration of \citet{saut02} and relaxing towards a very similar steady-state (solid lines) \citep{mats08}. {\bf  Right:}  Radial distributions of the magnetic field components for a disc wind alone (solid lines), a stellar wind alone (dashed lines) and combined stellar+disc wind (diamonds= initial setup, crosses = final state) \citep{mats09}.}
  \label{fig3_4}
\end{figure}
%%%%%%%%%%

There have been several analytical fluid approaches to stellar winds (e.g. \citealt{park58, webe67, hart90}) but the more sophisticated ones used a meridional self-similar ansatz \citep{tsin92a,saut04}. 2D ideal MHD solutions can be found using this non linear variable separation, with the caveat that all critical surfaces must be spherical. The mass flux is assumed and the initial driving is due to a pressure gradient (assumed to be of turbulent origin). In the TTS case, the asymptotic speed is mostly provided by this initial thermal push (term $\beta$ in Eq.\ref{eq:vit_stellar}). Although these winds are "thermally" driven, the magnetic field plays an important role in the collimation and several interesting situations can occur, as illustrated in Fig~(\ref{fig3_3}).

These models require however some specific expansions around the axis which bring severe limitations. For instance, no published model ever obtained an Alfv\'en radius large enough to provide a significant torque on the star. This may be a consequence of imposing a spherical Alfv\'en surface (see below). Another important caveat is the collimation itself. Indeed, the expansions used imply and require that magnetic field lines are threading and outer differentially rotating object (interpreted as being the disc). It turns out that most of the collimating electrical current, responsible for the stellar wind collimation, is actually flowing towards the outer "disc" (see Fig.~\ref{fig3_4}). These wind solutions would not be maintained (the wind would become open and unsteady) without an outer disc wind to confine it \citep{mats09}. This is interesting but is an issue for the magnetic braking efficiency of the star.

\begin{figure}[t]
\[ \begin{array}{cc}
\includegraphics[width=.5\textwidth]{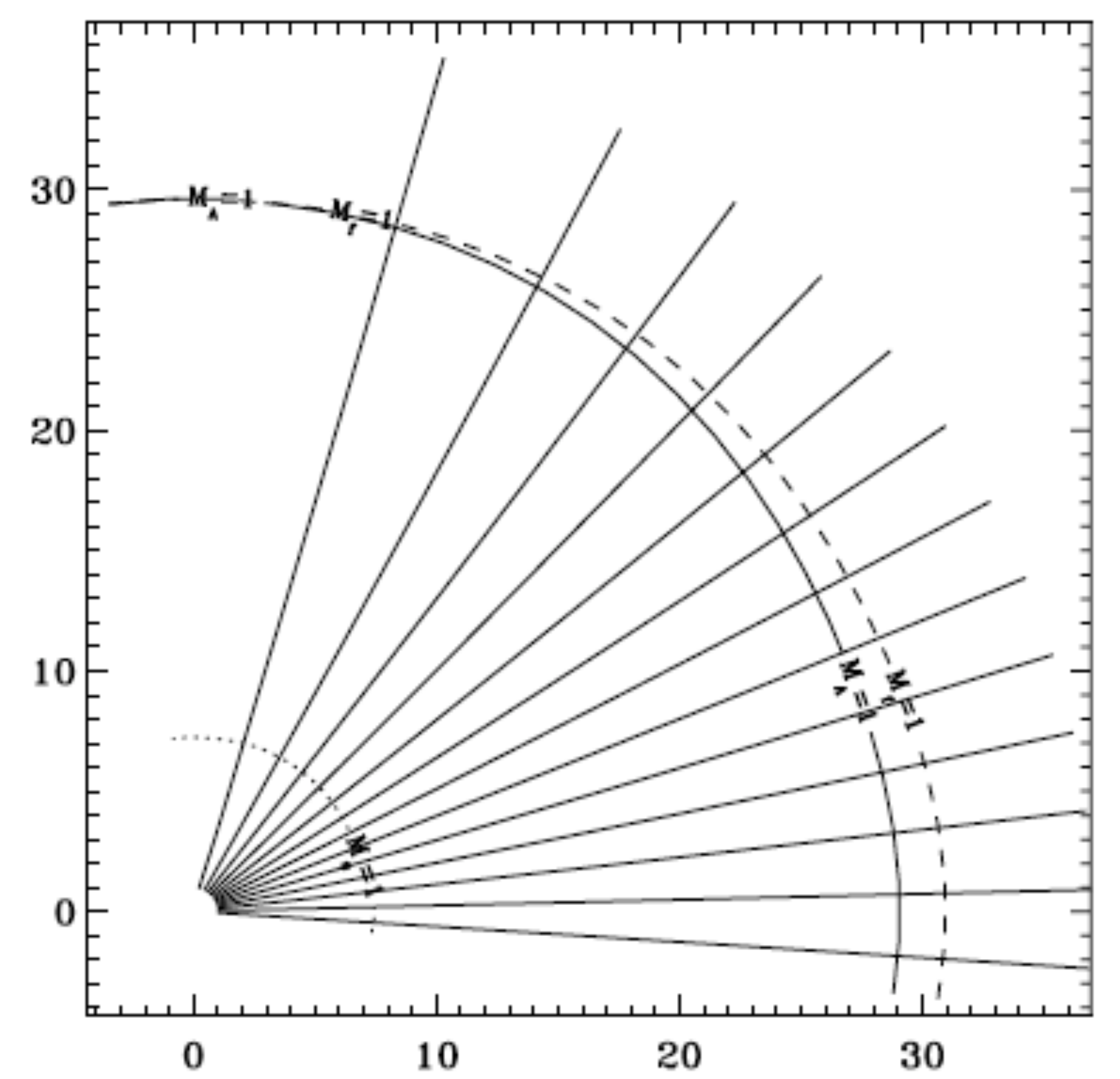}  
  &    
   \includegraphics[width=.4\textwidth]{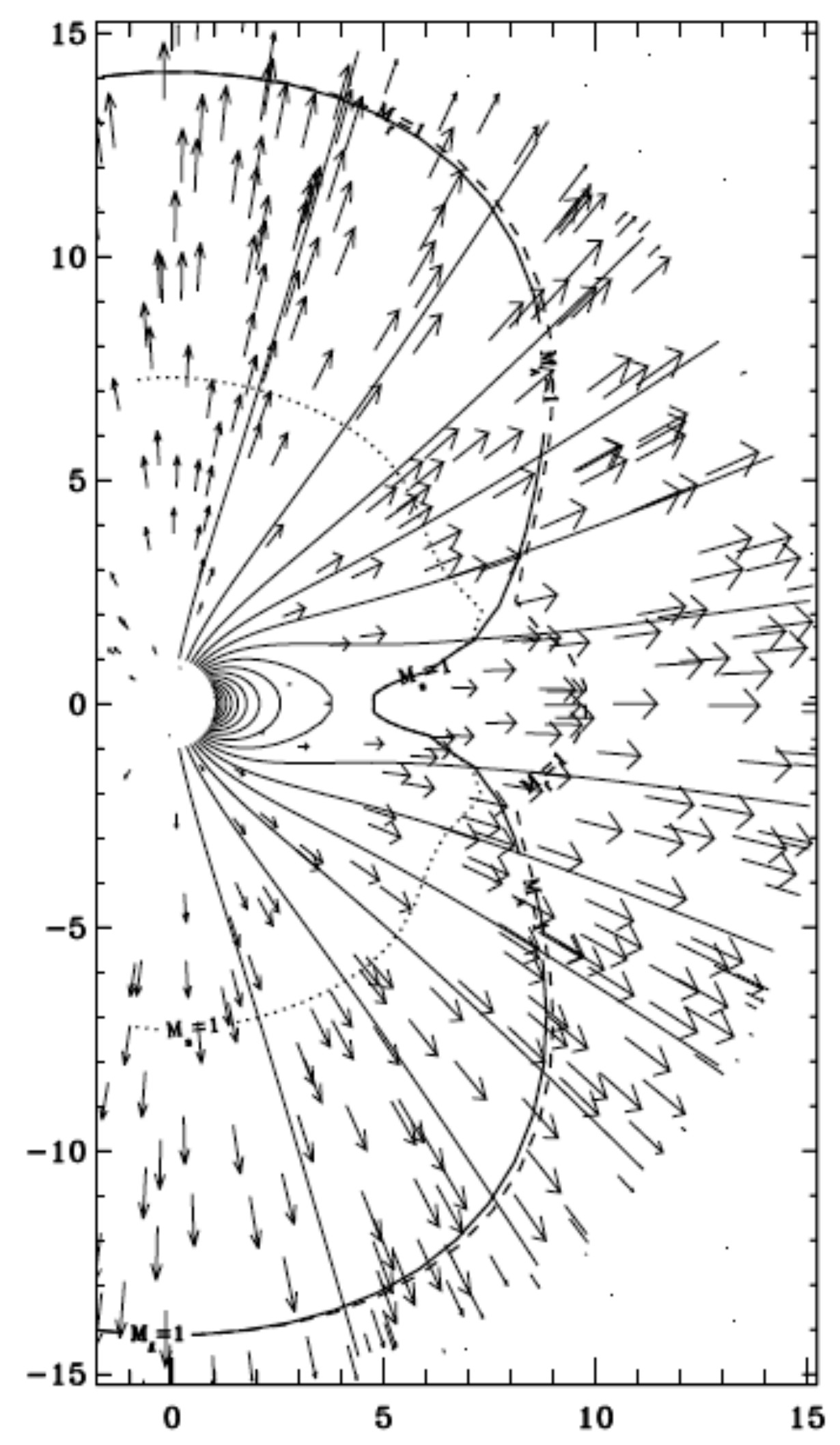}
\end{array} \]
  \caption{2D numerical simulations of a polytropic stellar wind in the low rotator case \citep{kepp99} {\bf  Left:} With a split-monopole. {\bf    Right:} With a dipole field.}
  \label{fig3_5}
\end{figure}

Numerical simulations of magnetized stellar winds are nowadays achievable and clearly show the effect of the magnetic collimation \citep{saku85, saku90}. Using a split-monopole as initial condition in a polytropic, thermally driven, axisymmetric wind \citet{kepp99} obtained a quasi spherical Alfv\'en surface. But when a dipole field is being used, the Alfv\'en surface is not spherical anymore, with critical surfaces displaced inwards with respect to the split-monopole model (Fig~\ref{fig3_5}). This is the sign of a better acceleration efficiency when dead zones are present (as it leads to a faster opening of the magnetic flux tubes). But lowering $r_A$ is probably not a very good thing for magnetic braking... 

%%%%%%%%%
%%%%%%%%%%

%%%%%%%%%%
\begin{figure}[t]
\[ \begin{array}{cc}
\includegraphics[width=.45\textwidth]{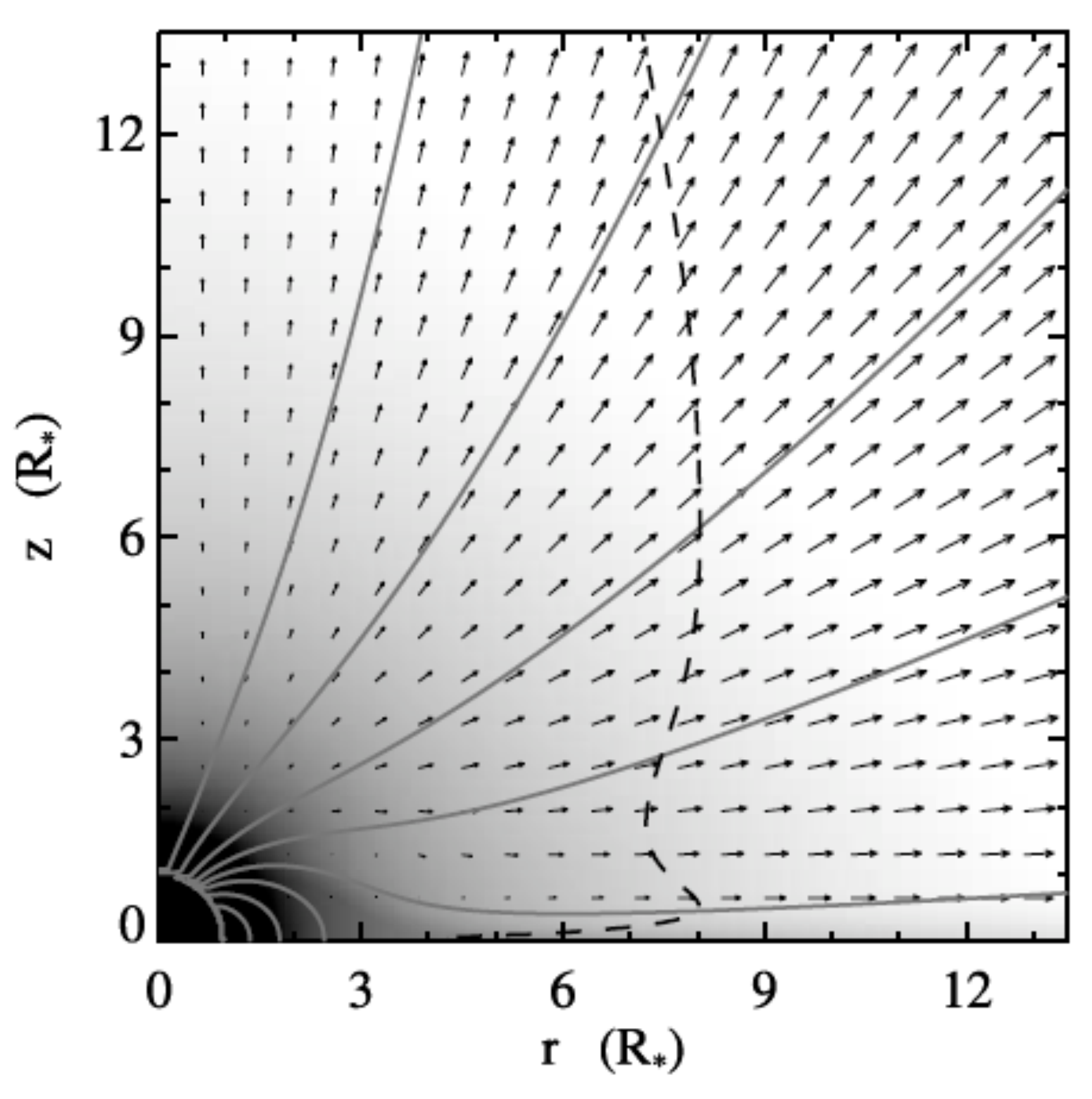}  
  &    
   \includegraphics[width=.45\textwidth]{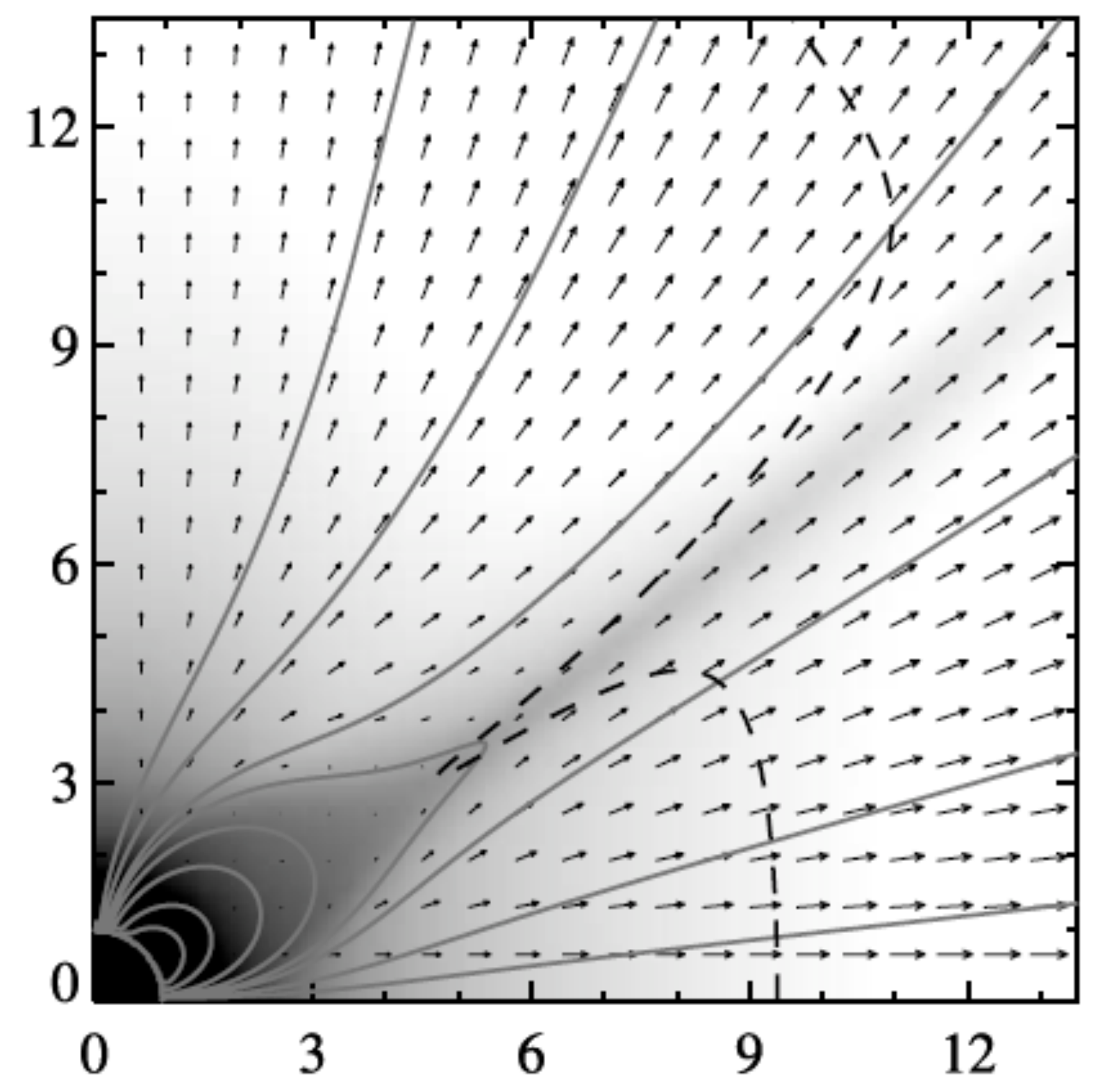}
\end{array} \]
  \caption{2D simulations of thermally driven stellar winds \citep{matt08a} showing the effect of the dead zone on the shape and position of the Alfv\'en surface (dashed line). {\bf  Left:} Dipole. {\bf  Right:}  Quadrupole.}
  \label{fig3_6}
\end{figure}
%%%%%%%%%%

\begin{figure}[t]
\[ \begin{array}{cc}
\includegraphics[width=.37\textwidth]{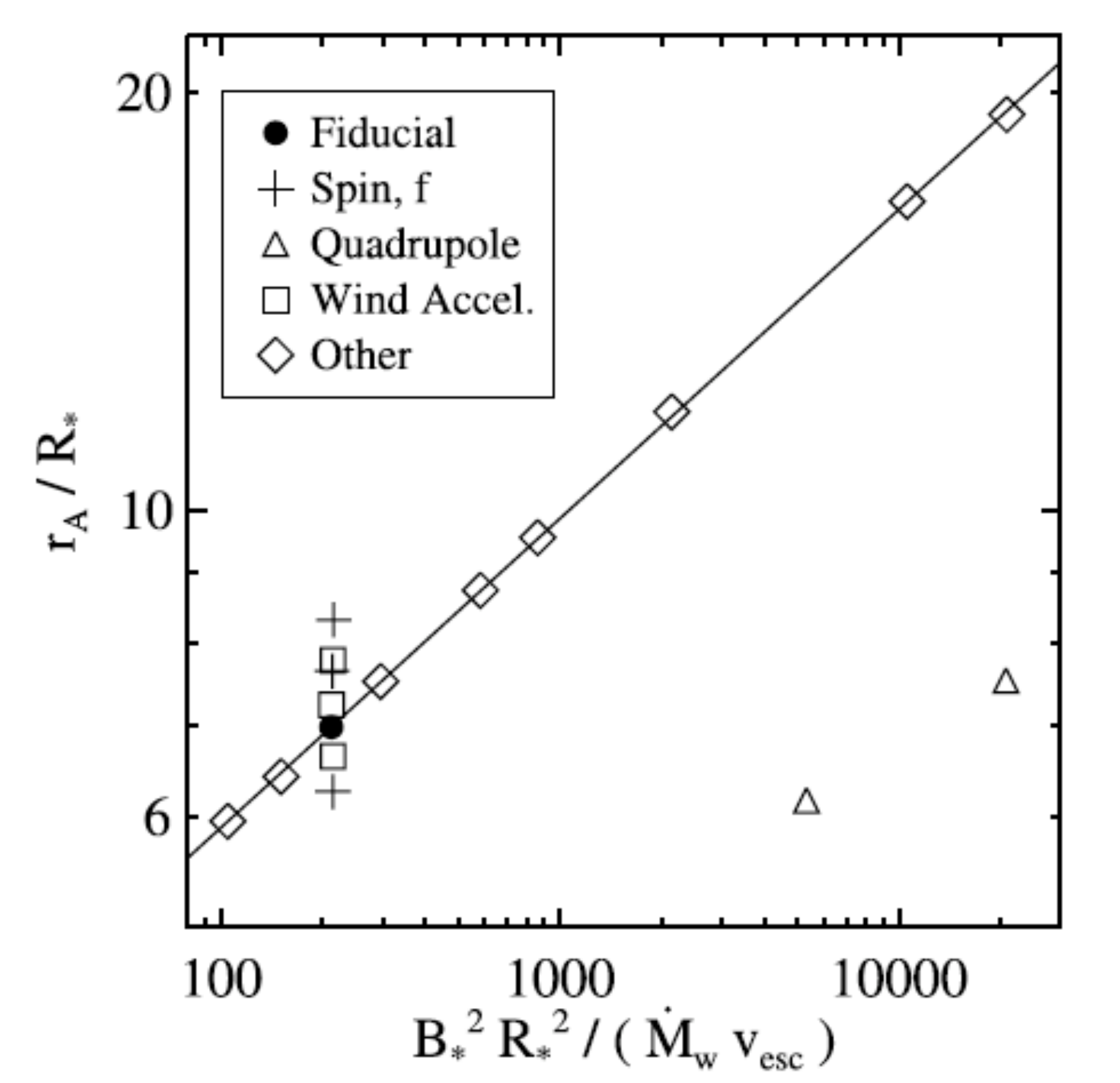}  
  &    
   \includegraphics[width=.5\textwidth]{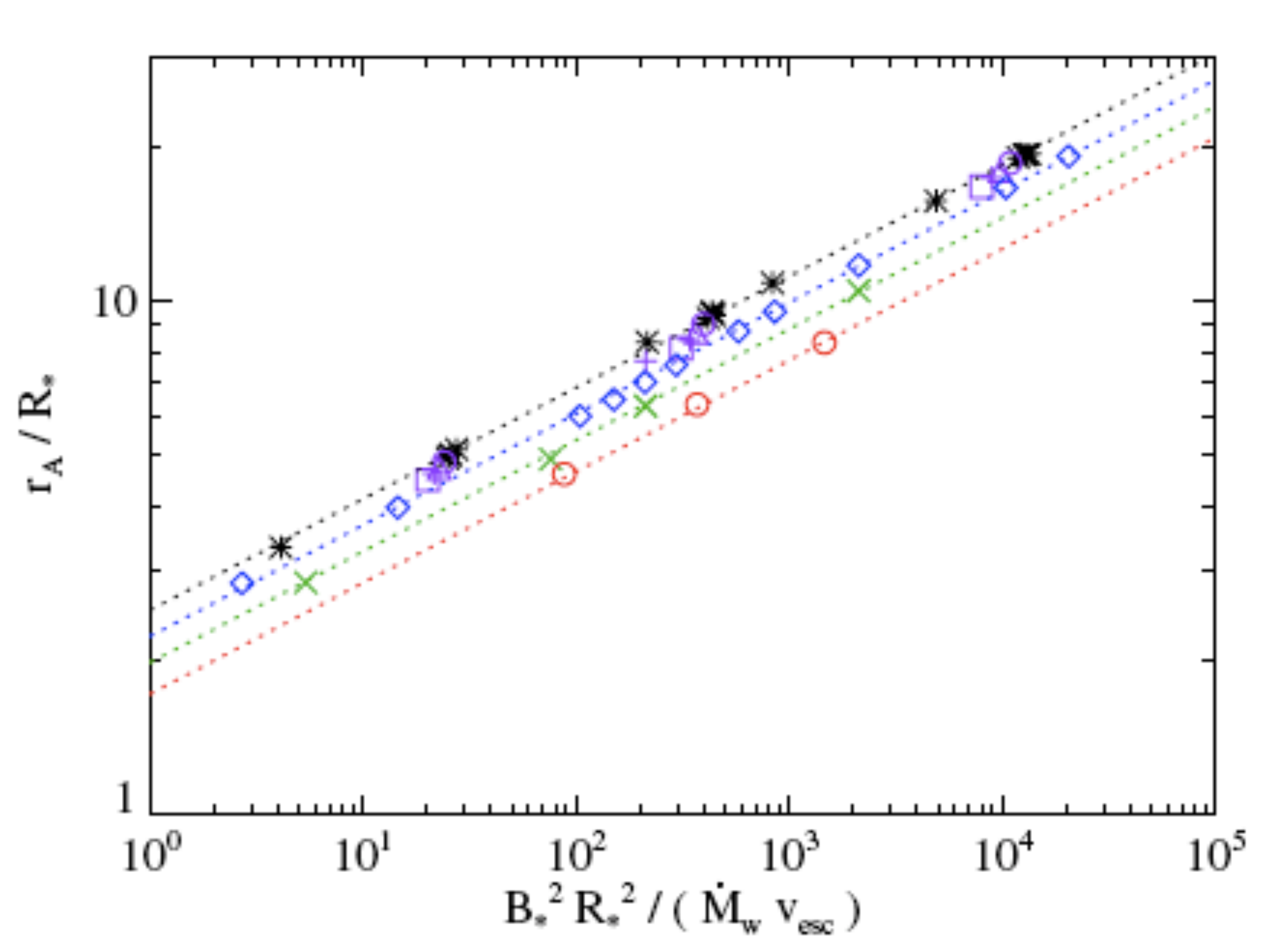}
\end{array} \]
  \caption{Set of numerical simulations of stellar winds and measure of the mass-weigthed Alfv\'en radius. {\bf Left:} Fitting formula given by Eq.(\ref{eq:rAwind1}) \citep{matt08a}. {\bf Right:}  Fitting formula Eq.(\ref{eq:rAwind2}) in the dipole case only, taking into account rotation with $\delta=0.001, 0.1, 0.2, 0.4$ (dashed lines, top to bottom) \citet{matt12b}.}
  \label{fig3_7}
\end{figure}

\citet{matt08a} did a battery of 17 runs in order to explore the effect of (mostly) mass loss, field strength and topology on the dynamics of thermally driven stellar winds (no disc). Using a 2D MHD code initiated with a Parker wind and either a dipole or a quadrupole field, they allowed the wind to relax to a steady-state. The wide-angle wind structure has been found to be self-confined, with a non-spherical Alfv\'en shape (Fig.~\ref{fig3_6}). With no surprise, the Alfv\'en surface goes from a sphere to almost a cylinder (apart close to the dead zone in the dipolar case) as the stellar rotation increases. Also, a quadrupole seems to be much less efficient in accelerating the wind. They then measured the global torque $\Gamma_{w}$ due to the wind and computed the average Alfv\'en radius using  
\be
\frac{\bar r_A}{R_*} \equiv \left (\frac{\Gamma_{w}}{2 \dot M_w \Omega_*} \right )^{1/2}
\ee
in order to compare it to the fitting formula given by Eq.(\ref{eq:rAwind1}). Surprisingly, the fit is quite good (see Fig. \ref{fig3_7}a) and allows to derive the values of $K$ and $m$. For a dipole, they found $K=2.11$ and $m=0.223$ (note that $m=1/3$ in the split-monopole centrifugal wind and $m=1/2$ in the thermal case). Using $B\propto r^{-n}$ and $m=0.22$ shows that the magnetic field exponent is closer to $n=3.2$, somewhere between the purely radial ($n=2$) and dipolar ($n=3$) cases. This is a strong indication of the 2D character of the solution (and limitations of analytical treatments). In the quadrupolar case, they found $K=1.7$ and $m=0.15$, showing that this configuration provides a much weaker torque than in the dipole case: a smaller fraction of the magnetic stellar flux is driving a wind. 

%%%%%%%%%%%
%%%%%%%%%%

Because an effect of the stellar rotation was observed on the Alfv\'en surface, \citet{matt12b} did 50 more simulations in the dipole field case, covering a wide range of relative magnetic field strengths and rotation rates (Fig.\ref{fig3_7}b). They provide another fitting formula
\be
\frac{r_A}{R_*} = K_1 \left [ \frac{\Psi}{\sqrt{K_2^2  + \frac{\delta^2}{2}  }} \right ]^m
\label{eq:rAwind2}
\ee
with $K_1= 1.3$, $K_2=0.05$ and $m=0.217$ and $\delta$ is the fraction of the breakup speed. Clearly, rotation becomes important only when $\delta$ becomes significantly larger than $\sim 0.07$ so that, for most cTTS, Eq.(\ref{eq:rAwind1}) is sufficient and $\Psi$ is the relevant parameter. Since there is a clear impact of the magnetic field geometry, one could argue that using a dipole field only is too simple (specially since the trend shown is a {\em decrease of the torque efficiency with field complexity}). And, indeed, we know that cTTS magnetic fields are composed of, at least, inclined dipole and octupole... 

%%%%%%%%%%%%%
\begin{figure}[t]
\[ \begin{array}{cc}
 \includegraphics[width=.4\textwidth]{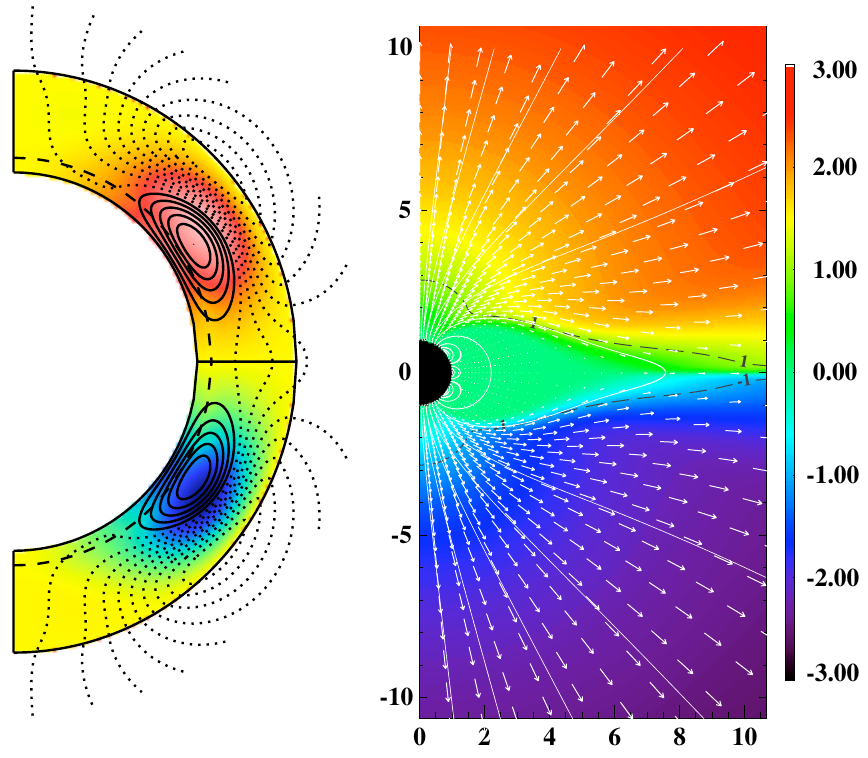}  
&
 \includegraphics[width=.4\textwidth]{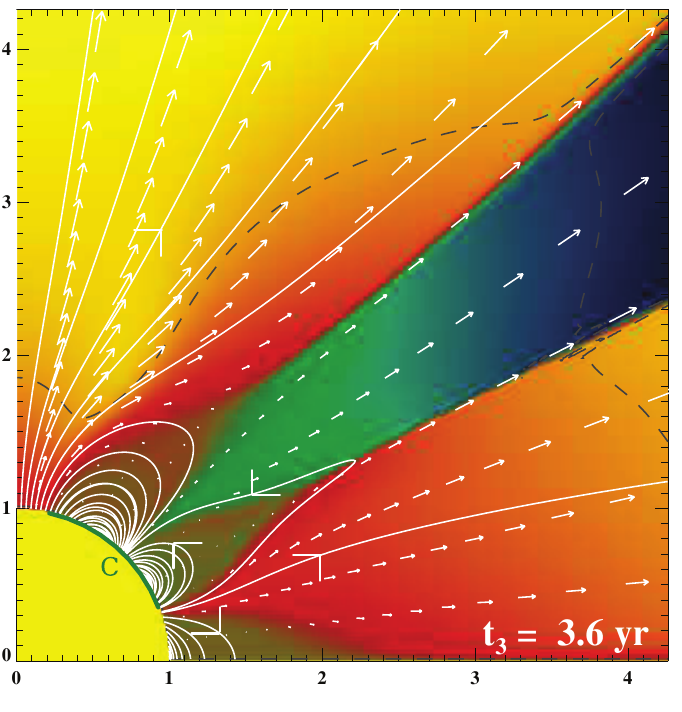} \\
\includegraphics[width=.4\textwidth]{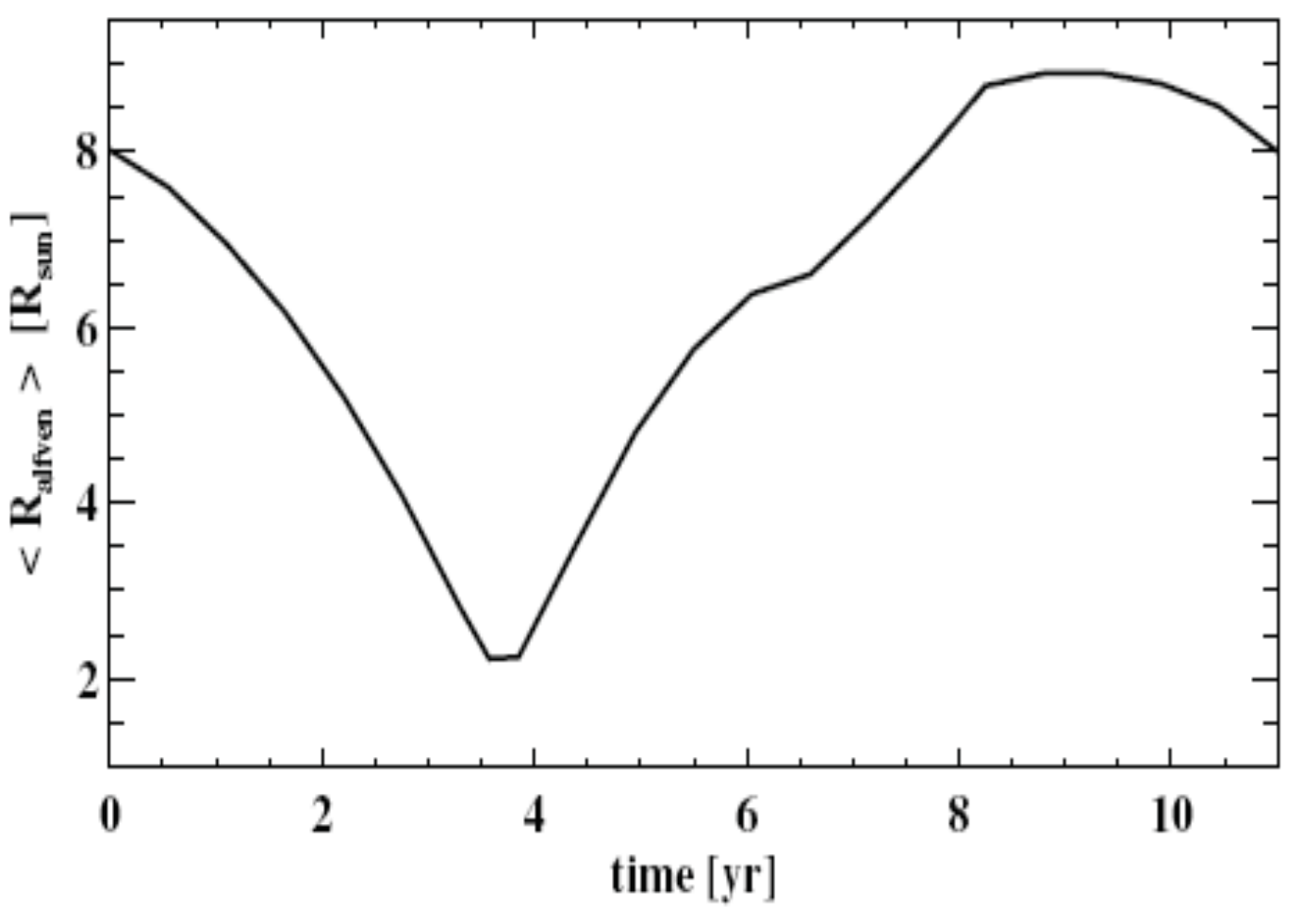}  
  &    
   \includegraphics[width=.5\textwidth]{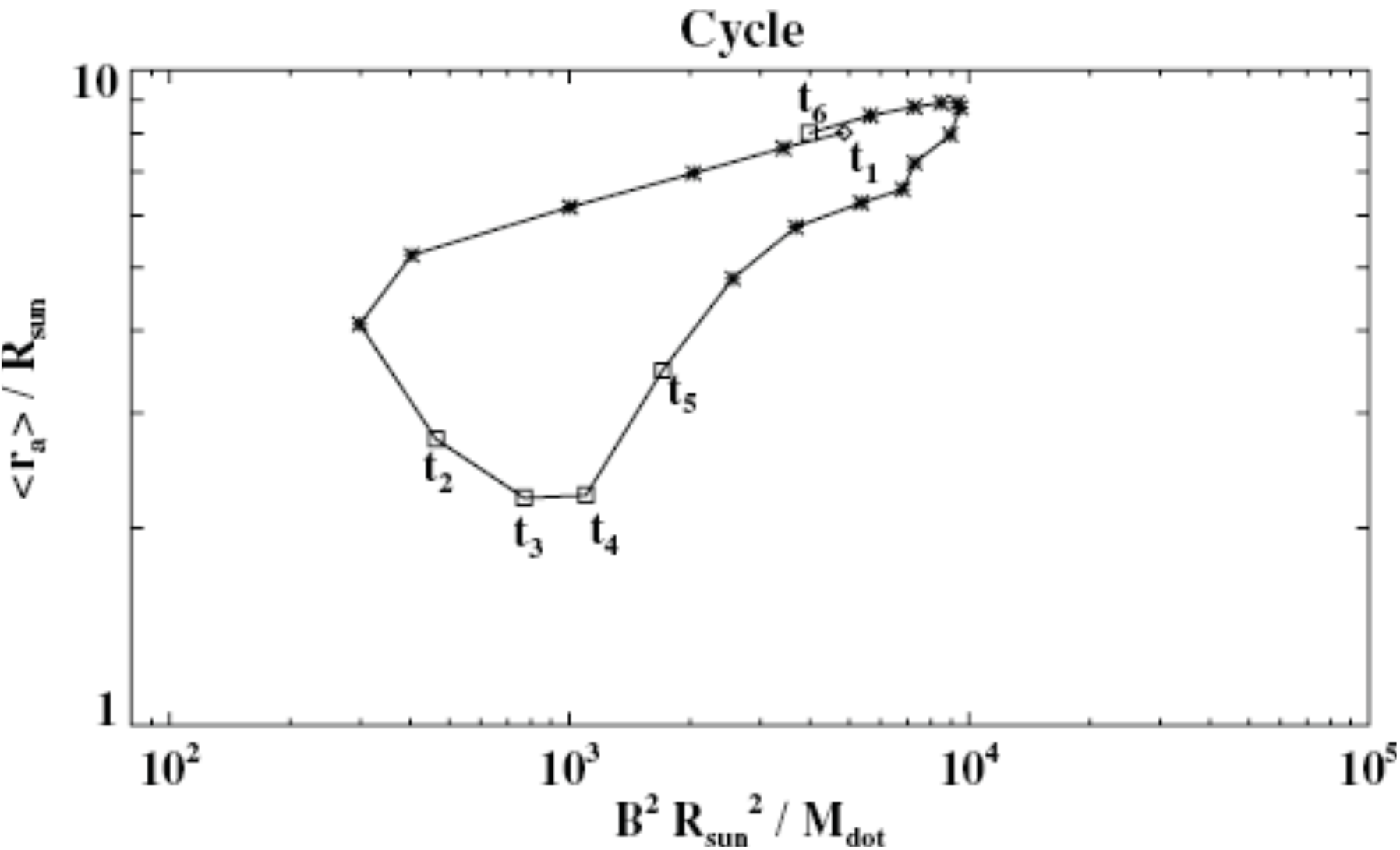}
\end{array} \]
  \caption{Coupling a solar dynamo with the wind \citep{pint11}. The kinematic dynamo-generated field is obtained by imposing a meridional circulation and differential rotation profiles in the convection zone, the corona being assumed to be a vacuum (top left). This field is then used as initial condition in a 2.5D code computing the thermally-driven wind structure (top right). Then the total torque on the star can be computed during a cycle, with the average Alfv\'en radius  (bottom left). When plotted against the $\Psi$ parameter, one observes an hysteresis (bottom right).}
  \label{fig3_8}
\end{figure}
%%%%%%%%%%
\citet{pint11} studied an example of an even more complex situation. They used an axisymmetric kinematic dynamo-generated field (obtained with one MHD code) as the input magnetic field (in a different MHD code) computing the wind acceleration (with no feedback of the wind on the dynamo). Of course, the dynamo-generated field is time-dependent and the "real" situation would be a mess to compute. So they studied a sequence of steady-state wind solutions along a dynamo cycle and computed the total torque as well as the Alfv\'en radius (Fig.~\ref{fig3_8}). Clearly the situation is complex, since Eq.(\ref{eq:rAwind1}) is not uniquely verified all the way through the cycle. On the other hand, the reason (and need) of such a scaling is to compute the spin evolution of a star on time scales much longer than the duration of dynamo cycles. 

That's exactly what \citet{matt12a} did. They solved Eq.(\ref{eq:starevol2}) assuming a Hayashi track for the protostar, an exponentially decreasing accretion rate $\dot M_a(t)$, the fitting formula for the APSW torque $\Gamma_w$ and the results from \citet{matt05a} on the magnetic torque $\Gamma_{mag}$ (in the case $\beta=0.01$ and $q_{max}=1$, see section\ref{sec:GL}). As in \citet{ferr00}, the stellar contraction was included. They used as free parameters the magnetic field strength $B_*$ (0, 500 G and 2 kG) , the initial fraction of breakup speed $\delta_o$ (0.3 and 0.06) and mass flux to accretion rate ratio $f$ (0.1 and 0.01). Their main result is that it is indeed possible to achieved low rotation rates of $\delta \sim 0.1$ at TTS ages (1 to 3 Myrs later) but {\em only for a very large mass flux $f=0.1$}. In fact, the APSW torque increases as $fJ \propto f^{1-2m}$ and thus increases when $f$ increases (as long as $m<1/2$). 

This important result raises several objections:
\begin{itemize}

\item from observations: a mass outflow rate to accretion rate ratio $f\sim 0.1$ means that the stellar wind would be the main component in YSO jets. This is contrary to current observational evidences \citep{cabr07}. 

\item from MHD theory: all these results are obtained assuming that a significant fraction of the turbulent energy released in the accretion shock is indeed being transferred to the outflowing wind, with no detectable losses. There is a long way to go from the estimate $f\simeq 0.01$ of \citet{cran08} to the required $f\simeq 0.1$. 

\end{itemize}

%%%%%%%%
\begin{figure}[t]
\centering   \includegraphics[width=.9\textwidth]{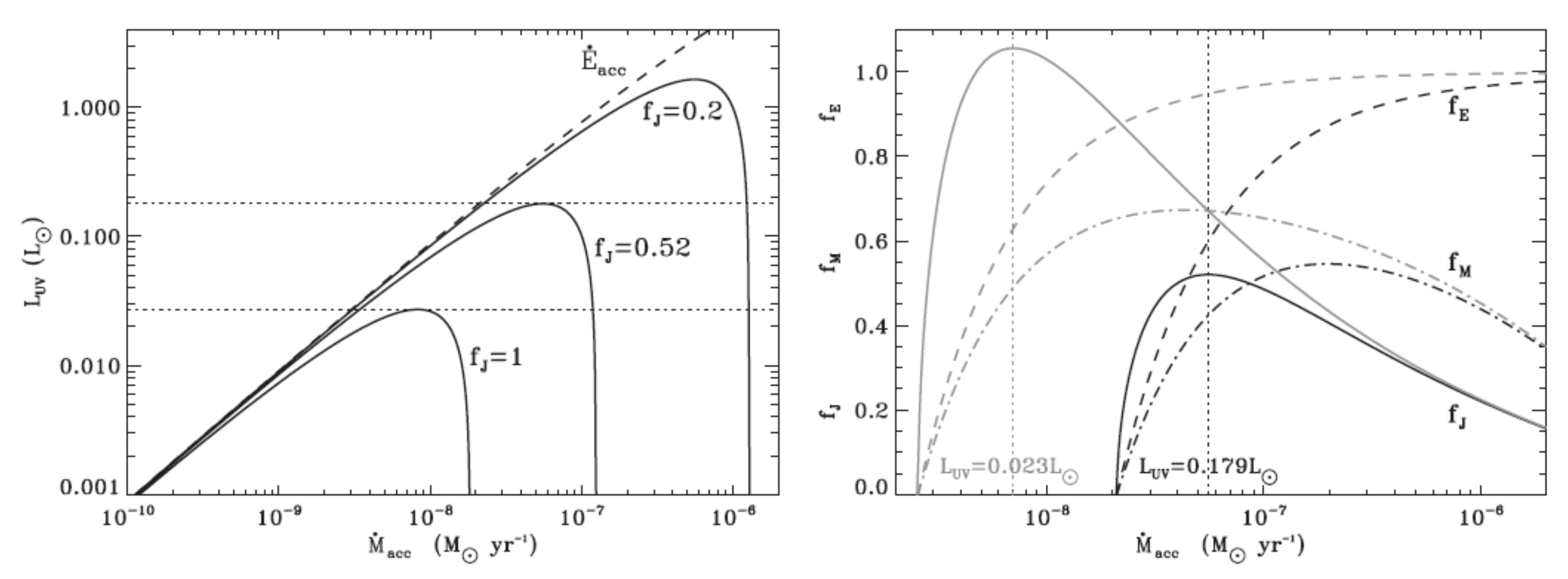}
  \caption{Left:  accretion shock luminosity $L_{UV}$ as a function of the real accretion rate for different values of the $f_J$ parameter. The dotted lines represent the maximum accretion luminosities compatible with a given $f_J$ value. Right: properties of the APSW as a function of the accretion rate for two possible accretion luminosities (gray and black lines); $f_J$ (solid lines), $f_M$ (dot-dashed lines), 
$f_E$ (dashed lines) \citep{zann11}.}
   \label{fig3_9}
\end{figure}
%%%%%%%%

Besides, one cannot "eat the cake and still having it". APSW are tapping the accretion energy: the more massive they are and the more energy they need. For each watt that an APSW takes, it is a watt that is not being radiated away in UV (assuming a 100\% conversion efficiency). Basically, one has
\be
P_{acc}= 2 P_{wind} + L_{UV}
\ee
where $P_{wind}= \dot M_w v_{esc}^2/2$ is the (one-sided) APSW power, $L_{UV}$ the accretion shock power that is radiated away and lost by the system (observed in the UV band) and $P_{acc} \propto \dot M_a$ the released accretion power, related to the real accretion rate $\dot M_a$ onto the star. Let us call $\dot M_{a,obs}$ the observationally derived accretion rate onto the star: it is 
\be
\dot M_{a, obs}= k \frac{L_{UV}}{GM_*/R*} = k \frac{P_{acc} - 2P_{wind}}{GM_*/R*}  < \dot M_a 
\ee
where $k$ is a factor of order unity (depending on $r_t, \delta$ etc...). Thus, the larger $f$, the larger $P_{wind}$, the smaller $\dot M_{a, obs}$. Let us define the following ratios
\begin{eqnarray}
f_M &=& \frac{2 \dot M_w}{\dot M_a}   \nonumber \\
f_E &=& \frac{2P_{wind}}{P_{acc}} \\ 
f_J &=& \frac{\Gamma_{w}}{\Gamma_{acc}} = \frac{2\dot M_w \Omega_* \bar r_A^2}{\dot M_a \sqrt{GM_* r_t}} = \delta f_M \left ( \frac{\bar r_A}{R_*}\right)^2 \left ( \frac{r_t}{R_*}\right )^{-1/2} \nonumber
\end{eqnarray}
($f_J=1$ describes a spin equilibrium solution when stellar contraction is ignored). For a given star, for which a UV luminosity is observed, we can plot the remaining UV power as function of the real accretion rate $\dot M_a$ for different values of the $f_J$ parameter (the importance of the APSW). This is illustrated in Fig~(\ref{fig3_9}) in the BP Tau case \citep{zann11}. For a given value of $f_J$, the larger $\dot M_a$ and the more efficient the APSW needs to be. Eventually, the APSW needs to tap most of the accretion power and nothing is left for radiation. Thus, for a given $L_{UV}$, there is a maximum value of $f_J$. 

In the case of BP Tau for instance, there have been two UV luminosities reported in the literature (see \citealt{zann11} and references therein). In one case, the maximum $f_J$ becomes larger than unity (net negative torque) but with $f_M=0.49$ and $f_E= 0.63$ ! In the other case, the maximum $f_J$ is 0.52 with $f_M=0.42$ and $f_E= 0.6$ (Fig~\ref{fig3_9}). In both cases, about 60\% of the released energy must be entirely converted into wind power (no losses have been taken into account here). \citet{zann11} applied this simple reasoning to a sample of cTTS and found that low luminosities stars ($L_{UV} \ll 0.1 L_\odot$) are compatible with APSW providing a net (braking) torque but at the expense of a fantastic energetic conversion and APSW being the main component of YSO jets. If one limits an energy conversion factor and mass loss $f_E\sim f_M\sim 0.01$, then $f_J\sim 0.1-0.2$ only. For more luminous objects, APSW always appear inefficient to reach spin equilibrium. 

We therefore conclude that APSWs are unlikely to be the main mechanism providing the required spin down torque for cTTS.

%%%%%%%%%%%%%%%%%%%%%%%%%%%%%%%%%%%%%%%%%%%%%%%%%%%%%%%%%%%%%
\section{Ejection from the star-disc interaction zone: The X-wind model}
%%%%%%%%%%%%%%%%%%%%%%%%%%%%%%%%%%%%%%%%%%%%%%%%%%%%%%%%%%%%%
\label{sec:xwind}

As pointed out by \citet{shu94a}, there is a numerical coincidence between the YSO jet energetics and the stellar spin equilibrium condition. The jet asymptotic speed along a field line is $v_w = v_{K,i} \left( 2 \lambda - 3 \right)^{1/2} \sim 100 \left( 2 \lambda - 3 \right)^{1/2}km/s$ if that line is anchored at a radius $r_i$ a few stellar radii. Thus, in order to explain jet velocities of 200-500 km/s, one needs $\lambda$ from 3 to 14. If that radius $r_i$ coincides with the disc truncation radius $r_t$, one has $\dot M_{a} \Omega_{K,i} r_i^2= 2 \dot M_w \Omega_* r_A^2$, (where $\Omega_*= \Omega_{K,i}$ is here the angular velocity of the magnetic field line), which translates into a dimensionless form $f \lambda = 1$. As a consequence, a magnetized wind expelled from the disc with the "correct" magnetic lever arm parameter $\lambda\sim 3-14$ and carrying a fraction  $f = 1/\lambda \sim 0.07-0.4$, will naturally account for both stellar spin equilibrium and YSO jet power. 

%%%%%%%%%%%%%%
\begin{figure}[t]
\[ \begin{array}{cc}
  \includegraphics[width=.5\textwidth]{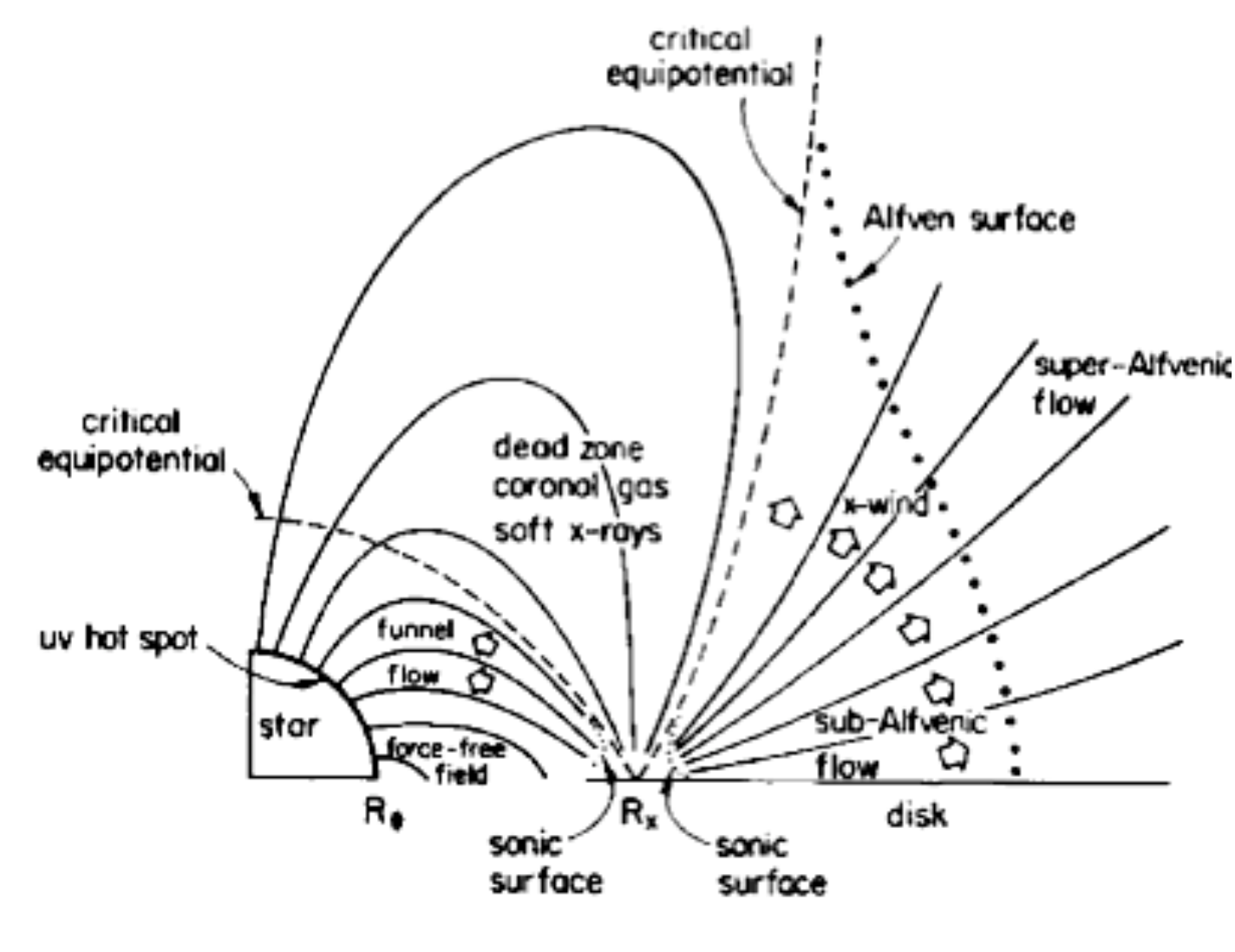}  
  &    
  \includegraphics[width=.5\textwidth]{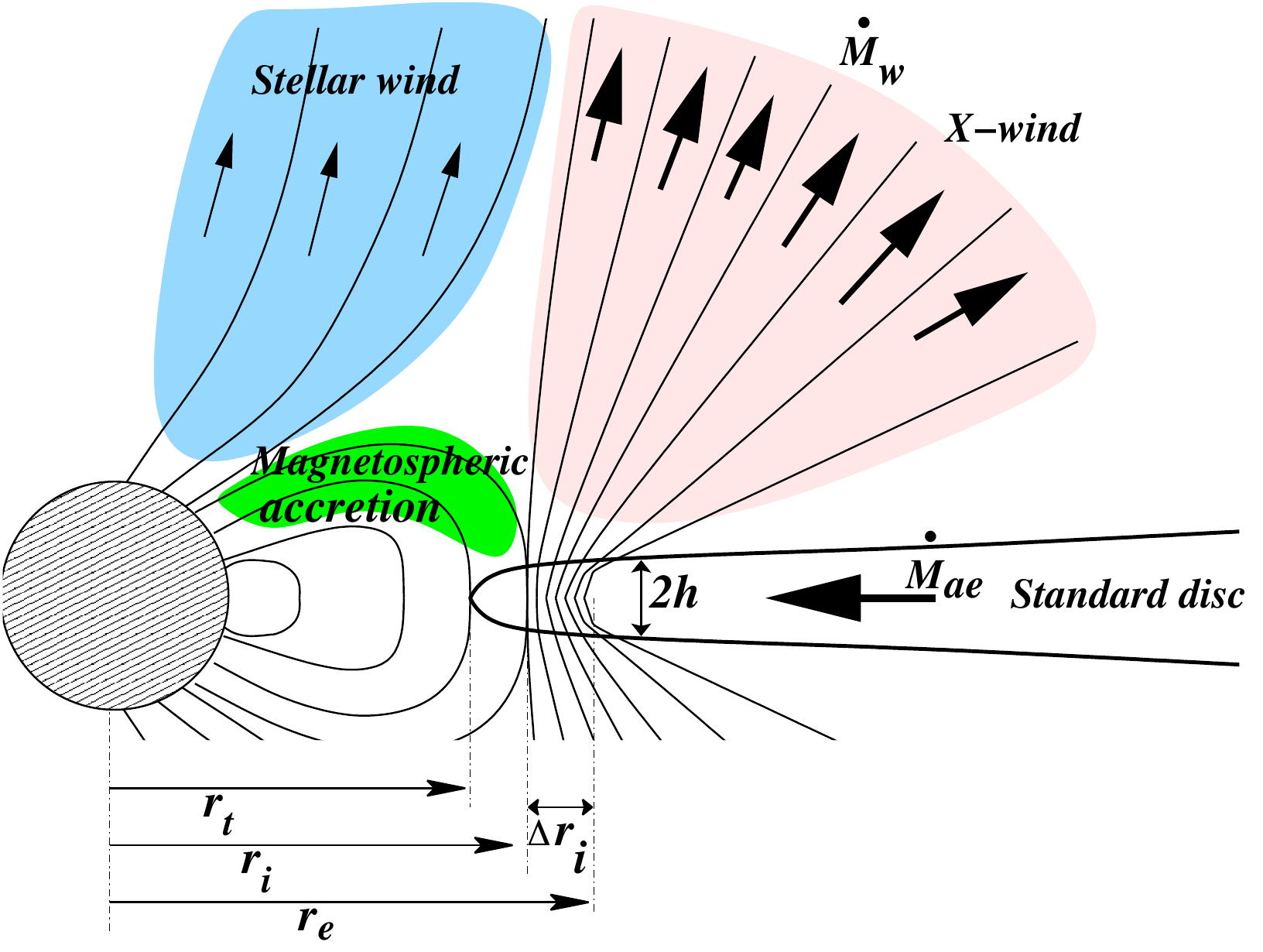}  
\end{array} \]
  \caption{{\bf Left:} Sketch of the X-wind paradigm \citep{shu94a, shu94b}. This is a Y-type magnetic interaction with no intrinsic disc magnetic field. The star-disc interaction has lead to the opening of the field lines and some stellar flux has been stored in the disc leading to this fan-shaped geometry. {\bf Right:} Highlight of some specific zones: the stellar wind (blue) is not taken into account as assumed negligible; the magnetospheric funnel flow (green) has been computed in ideal MHD assuming a potential field, dipolar \citep{ostr95} or multipolar \citep{moha08};  2D cold wind calculations have been undergone for the ideal MHD X-wind (red) \citep{naji94, shu95,cai08}, as well as some observational diagnostics \citep{shan98, shan02, shan04}. Picture courtesy F. Casse.}
  \label{fig4_1}
\end{figure}
%%%%%%%%%%%

The double goal sought by the X-wind model, as theorized in  \citet{shu94a}, is to explain YSO jets while simultaneously braking down the protostar. More precisely, X-winds should be the dominant component in YSO jets (in terms of appearance, mass flux and power) while carrying away enough angular momentum so that the star is not being spun up by the accreting material (zero net torque condition, contraction neglected). Since jets are believed to be magnetically launched, \citet{shu94a} argue convincingly\footnote{This can be simply understood as the divergence free condition for the magnetic field $\vec\nabla\cdot \vec B=0$ leads to $\frac{B_r^+} {\Delta r_i}  \sim \frac{B_z}{h}$, where $B_r^+$ is the radial component at the disc surface. Now, the \citet{blan82} criterion for cold jet production requires $B_r^+ \sim B_z$, which gives $\Delta r_i \sim h$.} that $\Delta r_i \sim h$ (see their \S2.4), where $h$ is the local disc scale height ($h/r \ll 1$) and the disc rotation law can be considered as nearly Keplerian in the X-wind launching zone. As formulated in the literature, the steady state {\em cold} X-wind paradigm requires the following conditions
\begin{enumerate}
\item a wind footpoint at $R_X \equiv r_i \simeq r_t \simeq r_{co}$
\item a launching zone of extent $\Delta r_i \sim h$
\item a Keplerian rotation law $\Omega(r)= \Omega_K(r)$
\item a large ejected mass flux $f=  2 \dot M_w/\dot M_a \sim 1/3$ (in fact between 0.1 and 0.4, see \citealt{naji94}) 
\item large asymptotic wind velocity $v_w > v_K = \sqrt{GM_*/r_i}$
\end{enumerate}
where the last two would in principle allow to interpret YSO jets as X-winds. This paradigm is based on a series of papers addressing distinctive points (Fig.~\ref{fig4_1}). The global picture is described in \citet{shu94a}. The axisymmetric, steady state ideal MHD wind structure has been explored in several papers: the fan shaped elliptic zone, going from the super slow-magnetosonic disc surface (a point source) to a prescribed Alfv\'en surface has been computed in \citet{shu94b, naji94}; this "inner" solution has then been matched to an "outer" asymptotic cylindrical jet solution \citep{shu95, shan98, shan02}. Note that in terms of  MHD wind theory, these wind calculations were not self-consistent. Indeed, it is not clear how exactly such a matching (interpolation of MHD invariants) has been done and the jet trans-field  (Grad-Shafranov) equation is most probably not satisfied. However, \citet{cai08}, by using a variational method that minimizes a functional computed with trial functions \citep{ross94}, did obtain a 2D cold (initially super-SM) fan-shaped super-Alfv\'enic flow consistent with MHD equations (and apparently not too different from the previous ones). 

In all these calculations the underlying disc is a mere boundary condition. However, no theoretical argument (let alone a calculation) ever showed the capability of a near-Keplerian annulus to display such huge ejection efficiencies (namely $f\sim 1/3$). Note also the following subtlety in the X-wind scenario:  if the X-wind carries away all the angular momentum lost by the accreting material (so that $\Gamma_{acc}=0$ on the star), how come the underlying disc still maintains a Keplerian rotation law? One would naively expect that the disc material would reach instead a zero angular velocity. The assumption that $\Omega = \Omega_K$ (essential for magneto-centrifugal acceleration) implies that the disc is being azimuthaly accelerated by the inner radii. Basically, the star would spin up the disc material just below  $R_X$ and that excess of angular momentum would be carried away radially by the viscosity beyond $R_X$ and eventually expelled away in the X-wind. Only by these means could the steady-state picture of X-winds be maintained (see discussion \S2.3 in \citealt{shu94a}).

Following exactly the same set of assumptions, \citet{ferr13} analyzed the impact of a large ejection fraction $f$ on the underlying disc dynamics. To do so, general properties of cold, fan-shaped winds have been first derived, linking in particular the jet power to the torque acting on the underlying portion of the disc (an annulus). Then the energy balance of this annulus has been computed within the framework of resistive (turbulent) MHD. Let us expose here these various steps, using the same notations ($B_X, \rho_X, V_X$) as in the X-wind theory.

%%%%%%%%%%%%%%
\begin{figure}[t]
\[ \begin{array}{cc}
  \includegraphics[width=.5\textwidth]{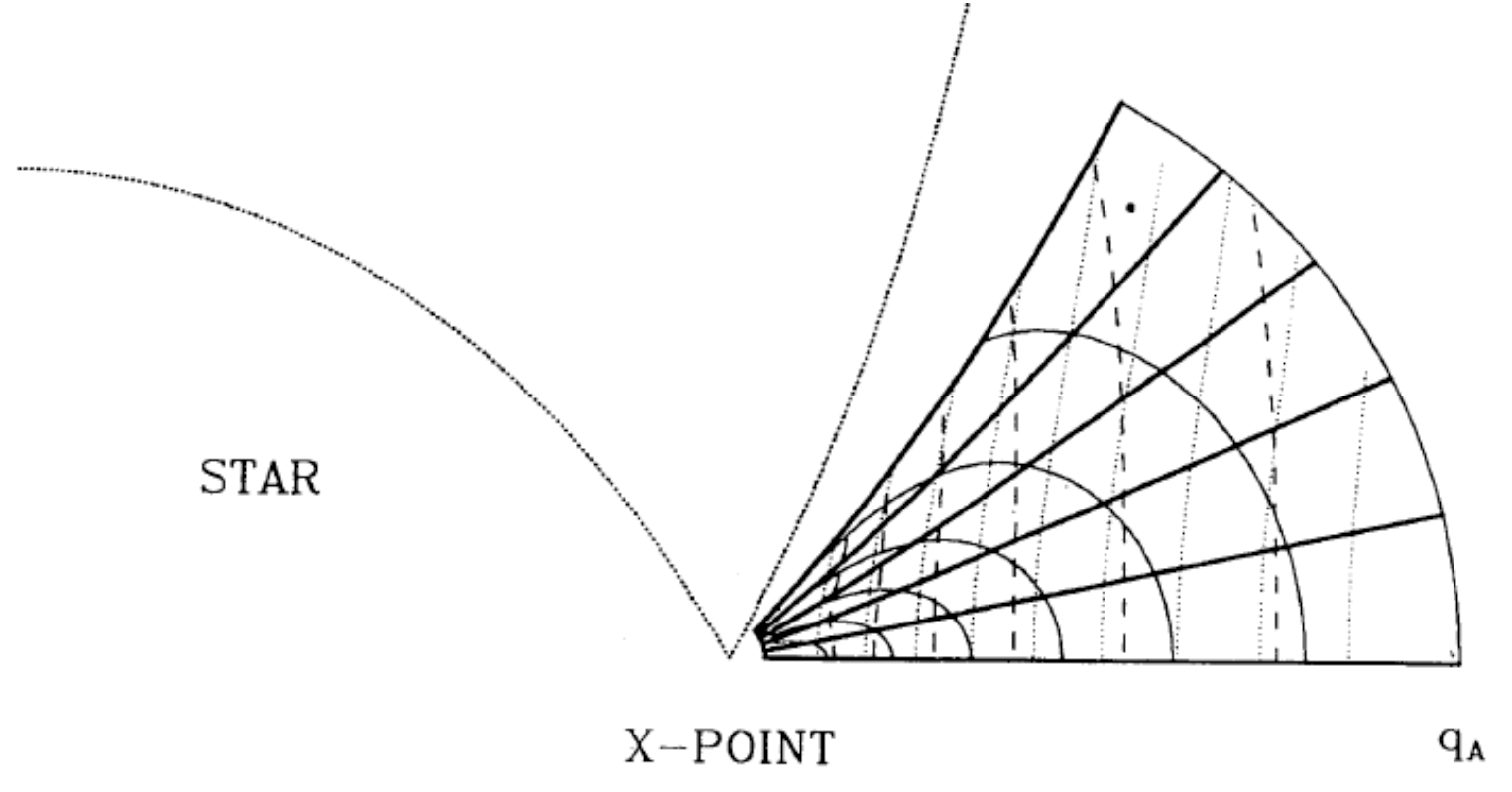}  
  &    
  \includegraphics[width=.3\textwidth]{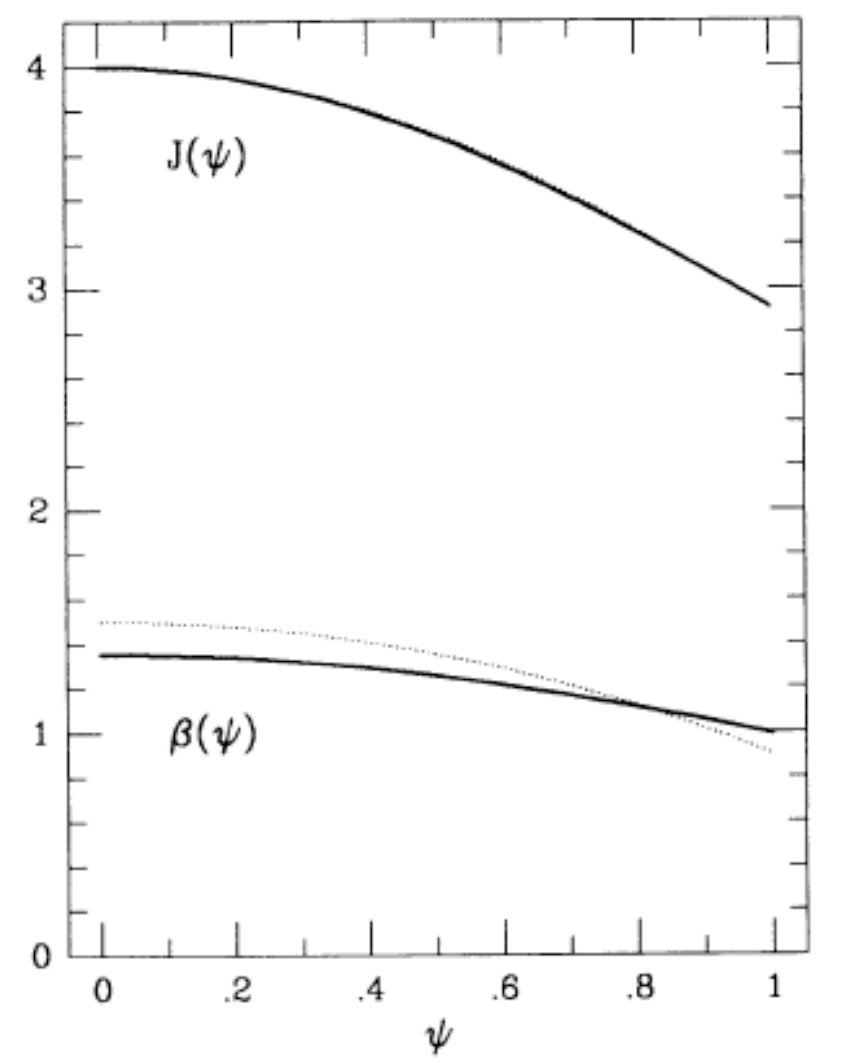}  
\end{array} \]
  \caption{{\bf Left:} Meridional plane contours of the variables ${\cal A}$ (light solid lines) and $\Phi$ (heavy solid lines) used to compute the solution in the sub-Alfv\'enic region in the case of straight upper boundary and spherical Alfv\'en surface. Light dotted lines and light dashed lines are contours of poloidal velocity and density respectively \citep{naji94}. {\bf Right:} Distributions of the invariants $\beta(\Phi)$ and $J(\Phi)$ used to compute this model.}
  \label{fig4_2}
\end{figure}
%%%%%%%%%%%%%%%%%%%%

The magnetic structure of an X-wind (or any other fan-shaped wind) experiences a spherical dilution from $r_i= R_X$. We label each magnetic surface with $\Phi$ and $\Phi_X\simeq 2\pi B_{z,i} r_i \Delta r_i$ is the total magnetic flux carried by the wind. Since  $\dot M_w =  \int 2 \pi r \rho u_z dr =  \int  \frac{\eta }{\mu_o}  d\Phi$ and using $\beta(\Phi) = \frac{\mu_o \rho_X V_X}{\eta B_X}$, we get $\Phi_X \simeq \bar \beta 2\pi B_X R_X^2$, where $\bar \beta \sim 1$ is an average value (see Fig.~\ref{fig4_2}). This allows to provide the following important relation   
\be
\mu_i \simeq \bar{\beta}^2 \left ( \frac{2 \dot M_w}{\dot M_{ai}}\right ) \left (  \frac{u_o}{\Omega_Kh} \right )_i \frac{r_i^2}{\Delta r_i^2} = \bar{\beta}^2 \frac{f}{1-f} m_{s,i} \frac{r_i^2}{\Delta r_i^2}
\label{eq:mui}
\ee
which relates the disc magnetization $\mu = B^2/\mu_oP$ and sonic Mach number $m_s$ at $r_i=R_X$ to the wind properties ($\bar \beta$ and $f$). Because of the wind, the disc accretion rate $\dot M_a(r) =  - \int_{-h}^{h} 2 \pi r \rho u_r  dz$ is a function of the radius. Defining $\dot M_{ai} = \dot M_a(r_i)$ and $\dot{M}_{ae}=  \dot M_a(r_e)$, mass conservation writes $ \dot{M}_{ai} =\dot{M}_{ae} - 2 \dot{M}_{w}= \dot{M}_{ae} (1-f)$. The specific angular momentum and energy along a field line write
\begin{eqnarray}
J(\Phi) &=& \frac{L(\Phi)}{\Omega_{K,i}r_i^2} \simeq \frac{r_A^2}{r_i^2} \simeq  1 + q \beta \bar \beta \frac{r_i}{\Delta r_i} \\
E( \Phi) &=& \frac{\Omega^2_{K,i} r_i^2}{2} \left ( 2 J - 3 \right )
\end{eqnarray}
where $q= - B_\phi^+/B_z$ measures the magnetic shear at the disc surface. The total power carried away by the cold wind through the disc surface  is
\be
2P_{wind}= 2 \int E(\Phi)\rho {\bf u}_p \cdot {\bf dS} =  2 P_{MHD} + 2 P_{kin}
\ee
The kinetic power carried by the outflowing cold plasma at the disc surface, namely
 \be
P_{kin} = \displaystyle\int \left(\frac{u^2}{2}+\Phi_G\right) \rho{\bf u}_p  \cdot  {\bf dS} 
\simeq -f\frac{GM_*\dot{M}_{ae}}{4r_i}
\ee
is initially negative: the wind power is mostly stored as magnetic power that will be eventually transferred to the plasma. The dominant contribution in $P_{wind}$ is the MHD Poynting flux ${\bf S}_{MHD} $ leaving the disc
\begin{eqnarray}
 P_{MHD}&=& \displaystyle\int - \Omega_* r \frac{B_\phi^+}{\mu_o} {\bf B_p} \cdot {\bf dS} \simeq 
  - \left . \frac{B_\phi^+ B_z}{\mu_o} \right |_i 2 \pi r_i^2 \Omega_K \Delta r_i 
 \nonumber \\
& \simeq & \frac{GM_* \dot{M}_{ai}}{4 r_i} \left . \frac{ 2 q \mu}{m_s} \right |_i \frac{\Delta r_i}{r_i}
\end{eqnarray}
It is of course related to the torque exerted on the underlying disc. Inserting Eq.~(\ref{eq:mui}) into this expression provides a wind power
\begin{eqnarray}
2P_{wind} &\simeq &\frac{GM_* \dot{M}_{ae}}{2 r_i} \left ((1-f) \left . \frac{ 2 q \mu}{m_s} \right |_i \frac{\Delta r_i}{r_i}- f \right ) \nonumber \\
&\simeq &  \frac{GM_* \dot{M}_{ae}}{2 r_i}  \left ( 2q \bar \beta^2 f \frac{r_i}{\Delta r_i} - f \right ) 
\label{eq:Pwind}
\end{eqnarray}
Defining the average maximum jet velocity $v_w$ as $P_{wind} = \frac{1}{2} \dot M_w v_w^2$, one obtains 
\be
v_w = \Omega_{Ki}r_i \left( 2q \bar \beta^2\frac{r_i}{\Delta r_i} - 1\right)^{1/2}
\ee
This expression is nothing more than $v_w = \Omega_{Ki}r_i (2 \bar J - 3)^{1/2}$, with an average magnetic lever arm $\bar J= 1 + q \bar \beta^2 r_i/\Delta r_i$. This result is general and verified by all X-wind calculations. Because these models use $\bar \beta \sim 1$, X-winds require a tiny magnetic shear, namely $q\sim \Delta r_i/r_i = h/r \ll 1$ \citep{shu94b, naji94}. But in magneto-centrifugally driven winds from accretion discs the amount of magnetic shear is related to the diffusion within the disc, namely $q \sim 1/\alpha'_m$, where the turbulent magnetic diffusivity in the toroidal direction has been parametrized by $\nu'_m= \alpha'_m \Omega_K h$. Thus, unless $\alpha'_m \sim r/h$, there is an inconsistency with the disc physics (see \citealt{ferr13} for more details).

%%%%%%%%%
The energy budget of the annulus writes
\be
P_{acc} + P_{vis} = 2P_{wind} + 2P_{rad}
\label{eq:Pacc}
\ee
where the total power carried away by the cold winds $2P_{wind}$ and the disc luminosity $2P_{rad}$ are fed by the power available in the disc between $r_i$ and $r_e= r_i + \Delta r_i$. This power has two possible origins: the released accretion power $P_{acc}$ and the power $P_{vis}$ brought in by viscosity. Note that this term is always neglected in accretion disc theory since the torque at $r_i$ is always assumed to vanish. Here that assumption cannot be done. Any other energy flux, like e.g. heat deposition by irradiation, is assumed negligible here. The accretion power is in this case
\be
P_{acc} = \frac{GM_*\dot{M}_{ae}}{2r_i}\left( \frac{\Delta r_i}{r_i} - f \right)
\ee
It can be readily seen that there is no mechanical power available if  $ f > \Delta r_i/r_i$ as assumed in X-winds: there is not enough radial contrast to feed large amounts of mass with mechanical power. The power feeding X-winds must arise from viscous stresses.

In a turbulent accretion disc, the inward flux of mass coexists with outwardly directed fluxes of energy and angular momentum. At each radius, energy and angular momentum are therefore being deposited from the inner radii, which translates here into a net power
\begin{eqnarray}
P_{vis} &=&  \left [ \int ({\bf u}\cdot{\mathbf T})\cdot{\bf dS} \right ]^{r_e}_{r_i} = \left [ \int_{-h}^{+h} 2 \pi r \Omega r T_{r\phi} dz \right ]^{r_e}_{r_i} = - \left [  \frac{3}{2} {\cal R}_e^{-1} \, \frac{GM_*\dot{M}_{a}}{r} \right ]^{r_e}_{r_i}  \nonumber \\
& \simeq & \frac{3}{2} {\cal R}_{e,e}^{-1} \, \frac{GM_*\dot{M}_{a,e}}{r_i} \left ( {\cal D} + \frac{\Delta r_i}{r_i} \right )
\end{eqnarray}
where 
\be
{\cal D} = (1-f)\frac{{\cal R}_{e,e}}{{\cal R}_{e,i}} -1 =  \frac{\Sigma_i}{\Sigma_e} \frac{\nu_{v,i}}{\nu_{v,e}} - 1
\label{eq:defD}
\ee
where ${\cal R}_e= r u_o/\nu_v$ is the Reynolds number, evaluated at $r_i$ or $r_e$. Since the only available source of energy is $P_{vis}$, the difference factor ${\cal D}$ needs to be positive and of the order unity.

Following the same approach, one can also evaluate the power that is dissipated within the annulus and lost as radiation $ 2P_{rad}$ so that $2P_{wind}$ can be computed using Eq.~(\ref{eq:Pacc}). This gives an expression that, combined with Eq.~(\ref{eq:Pwind}), provides the following relations
\begin{eqnarray}
q \bar \beta^2 f  &\simeq & {\cal D} \frac{\Delta r_i}{r_i}   \nonumber \\
 \bar J &= &1 + {\cal D}/f  \\
2P_{wind} &\simeq & \frac{GM_*\dot{M}_{a,e}}{2 r_i} \left ( 2 {\cal D} - f \right ) \nonumber
\end{eqnarray}
that are valid for any fan-shaped wind. In the case of X-winds, these constraints can be summarized that way
\be
 \frac{\bar J - 1}{\bar \beta^2}  = q  \frac{r_i}{\Delta r_i} = \frac{{\cal D}}{ \bar \beta^2 f} 
 \label{eq:bilan}
\ee
with the constraint  $f < 2 {\cal D}$. So far, $f$ (or ${\cal D}$) is unspecified and free as long as a full MHD calculation/simulation including the disc remains undone. X-winds, as they appear in the literature, require $ {\cal D}= 1-2f  \sim 1$ (zero torque condition) and $ q \sim \frac{\Delta r_i}{r_i}$ (total jet power). ${\cal D}$ measures the net energy deposition by viscosity in the zone of extent $\Delta r_i \sim h$. This deposit will eventually power the jets and that is why X-winds require ${\cal D}$ of order unity to be of any significance. The problem is the tiny width of the launching zone, which imposes that $ {\cal D}\sim 1$ requires a Reynolds number of order unity as well. Otherwise there is much more power lost at $r_e$ than the one gained at $r_i$ and ${\cal D}\simeq -1$, Eq.(\ref{eq:defD}).

To go further, one needs to consider the {\em disc} angular momentum equation at $r_i$ (an equation that has been discarded in \citealt{shu94a}).  The disc angular momentum equation ($\phi$-component of Eq.~\ref{eq2}) writes
\be
\frac{\rho {\bf u}_p}{r} \cdot \nabla \Omega r^2 = F_\phi +  \frac{1}{r^2} \frac{\partial}{\partial r} \rho \nu_v r^3 \frac{\partial \Omega}{\partial r} 
\ee
where $F_\phi=  J_z B_r- J_r B_z$. Integrating this equation over the disc thickness gives 
\be
\frac{\dot M_a \Omega_K}{4\pi r}  \simeq - 2 \frac{B_zB_\phi^+}{\mu_o} + \frac{3}{2 r^2} \frac{\partial}{\partial r}  \Sigma \nu_v \Omega_K r^2
\ee 
where both $\Omega \simeq \Omega_K$ and $h$ roughly constant over the extent $\Delta r_i$ have been assumed. Here, $\Sigma$ stands for the local disc surface density, defined as $\Sigma = \int_{-h}^{h}\rho dz$. Defining the  accretion sonic Mach number as $m_s= \dot M_a/2\pi \Sigma \Omega_K rh= u_o/C_s$, where $u_o$ is the average accretion velocity, one gets
\be
m_s =  m_s^{mag} + m_s^{visc} = 2 q \mu +  \zeta \alpha_v \varepsilon
\ee
where $m_s^{mag}$ and $m_s^{visc}$ are respectively the magnetic and viscous contributions and where
\be
\zeta =  3 \frac{d \ln (\Sigma \nu_v r^{1/2})}{d \ln r} = \frac{3}{2} - 3 \frac{d \ln {\cal R}_e}{d \ln r} + 3 \frac{d \ln \dot M_a}{d \ln r} 
\label{eq:zeta}
\ee
is a factor that depends on the radial distributions. In the X-wind paradigm, viscous torques must transport angular momentum from $r_t$ to (and beyond) $r_i$. This implies that at $r_i$ the viscous torque {\em accelerates} the plasma and, indeed, $\zeta_i \simeq - 3 r_i/\Delta r_i$. With the constraints that both $\mu_i$ and ${\cal D}$ being of order unity and $q \sim h/r$, one gets
\begin{eqnarray}  
&& \alpha_{v,i} \sim h/r \nonumber \\
&& \alpha'_m \sim  r/h \label{eq:turbXwind}\\
&& \alpha_m \sim (h/r)^2 \nonumber
\end{eqnarray}
where the last condition has been derived using ${\cal R}_m \sim \varepsilon^{-1}$ for cold ejection \citep{ferr13}. These transport coefficients are those required within the disc so that X-winds can indeed be launched with $ {\cal D}\sim 1$. The almost perfect matching between the two modes of angular momentum transport (turbulent/radial, wind/vertical) gives rise to a slight unbalance, allowing henceforth accretion.

These findings lead to several comments:
\begin{itemize}

\item from MHD theory:  the constraints imposed by all published super-Alfv\'enic X-wind flow models ($\bar \beta \sim 1$ and $f$ large) on the underlying disc structure lead to quite an extreme transport coefficients set-up. None of the various transport coefficients are actually in agreement with our current knowledge of MHD turbulence in near Keplerian discs (see S. Fromang's contribution, this book). 

\item from MHD simulations: X-winds have never been observed in numerical simulations of star-disc interaction. However, these simulations were done within alpha-prescriptions with $\alpha_v \sim \alpha_m$ of order unity. Moreover, no anisotropy in the magnetic diffusivity was introduced whereas $\nu_m/\nu'_m \sim (h/r)^3$ would have been necessary. 

\item from observations: YSO jet kinematics are inconsistent with current X-wind calculations. For instance, the observed range and steep radial decline of poloidal speeds are not explained with the chosen Alfv\'en surface shape  \citep{ferr06b, cabr07}. 

\end{itemize}

Although this analyses carries many caveats (e.g. MHD turbulence treated as a mere viscous stress with anomalous coefficients), it allows to unveil many unsolved and critical issues in the X-wind paradigm. It might therefore appear safer to look for alternative ways to spin down accreting protostars.

%%%%%%%%%%%%%%%%%%%%%%%%%%%%%%%%%%%%%%%%%%%%%%%%%%%%%%%%%%%%%
\section{Other ejecta from the star-disc interaction zone}
%%%%%%%%%%%%%%%%%%%%%%%%%%%%%%%%%%%%%%%%%%%%%%%%%%%%%%%%%%%%%

Until now, only steady-state models have been considered, with not much success. It is therefore timely to look at what numerical simulations can offer. In the following we analyze only two situations that have been fairly well understood.

\subsection{Conical winds}

%%%%%%%%%
\begin{figure}[t]
\centering   \includegraphics[width=0.6\textwidth]{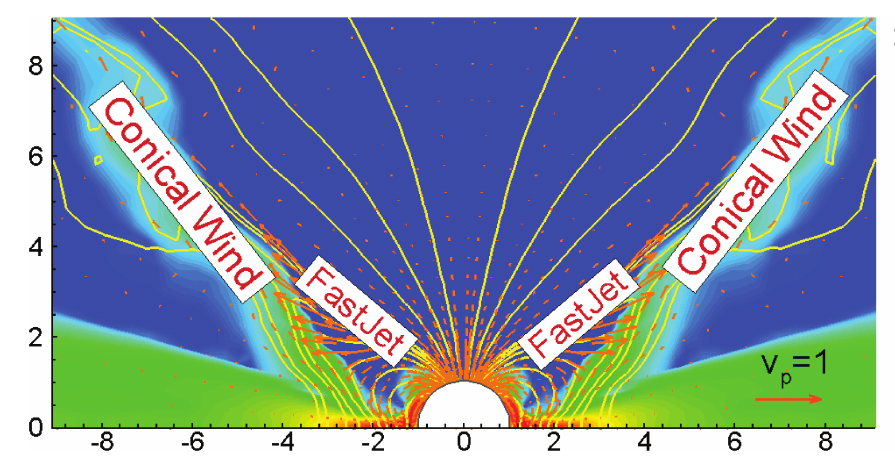}
  \caption{Two-component outflows observed for a slowly rotating magnetized star. The background shows the poloidal matter flux $\rho v_p$, the arrows are the poloidal velocity vectors and the lines are sample magnetic field lines \citep{roma09}. }
   \label{fig4_3}
\end{figure} 
%%%%%%%

\citet{roma09} reported the production of long lasting collimated outflows with the shape of a thin conical shell with 
a half-opening angle $\theta \sim 30^o - 40^o$ (Fig.~\ref{fig4_3}). These outflows have been termed "conical winds". These winds are massive, carrying about 10 to 30 percent of the disc accretion rate but are slow, with velocities typically around 30-50 km/s, i.e. 10 times too low to account for YSO jets. There is also a low-density, higher velocity component (termed the jet and being a stellar wind) in the region inside the conical wind due in part to the toroidal magnetic field pressure (as in Z-pinch experiments, see e.g. \citealt{ciar09, suzu10} and references therein).

\begin{figure}[t]
\centering   \includegraphics[width=\textwidth]{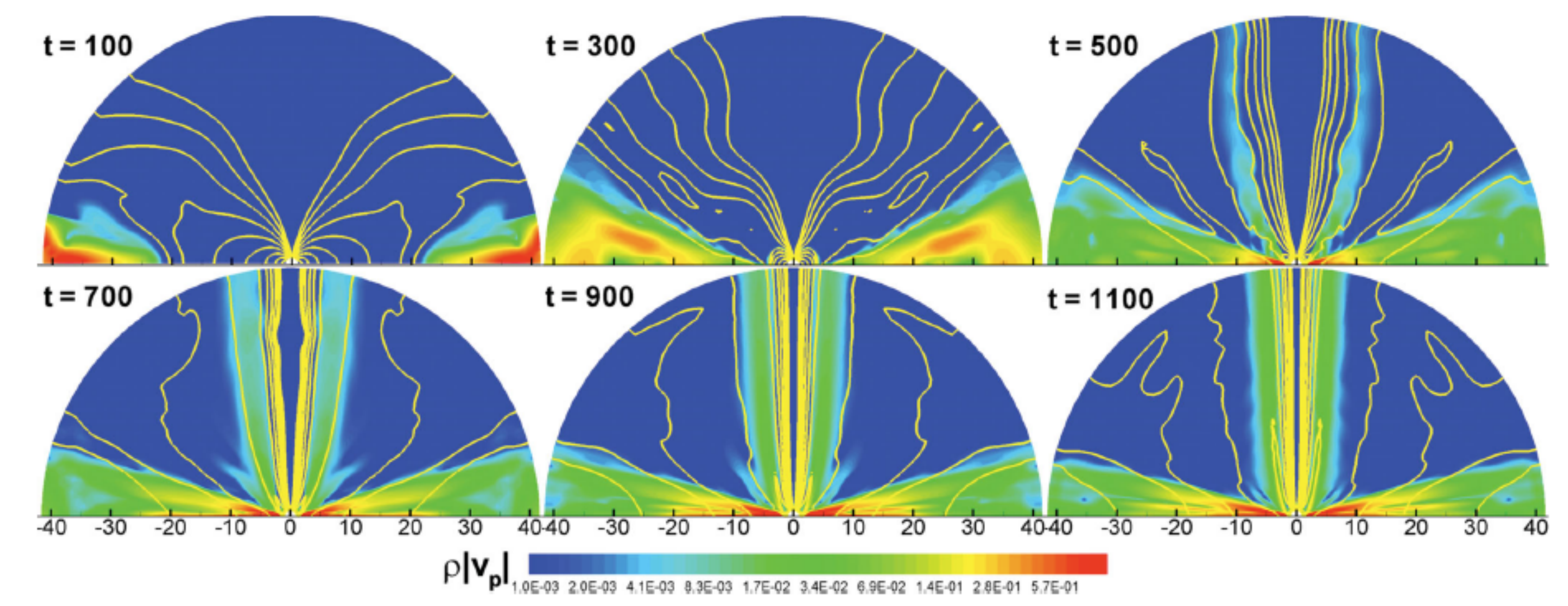}
  \caption{Time sequences, in units of the Keplerian period at $r=1$, of the star-disc interaction leading to conical winds
   \citep{lii12}. The disc material is injected from the outer boundary and builds up until it reaches the theoretical truncation radius $r_t$. The initially dipolar magnetosphere is strongly modified with opened field lines threading the disc. Significant magnetic flux is thereby trapped around $r_t$ from where are launched the winds. Note the much larger size of the computational box and the absence of stellar wind as compared to Fig(\ref{fig4_3}), allowing to see the collimation of the (finally not so much) conical wind.}
   \label{fig4_4}
\end{figure}

The conical winds appear in cases where the dipolar magnetic flux of the star is bunched up by the disc until the differential rotation leads to an opening of these field lines (Fig~\ref{fig4_4}). The disc truncation radius occurs precisely at the radius  $r_t$ given by Eq.(\ref{eq:rt}) and the star is being spun up. In the propeller regime (star rotating faster), conical winds still occur but the star is being spun down with no accretion. As said previously, since no cTTS seems to be at a non accreting stage, we discard this possibility.   
%%%%%%%%%%
%%%%%%%%%%%

Are conical winds a new class of (quasi steady-state) winds ? Using a computational box 3 times larger, \citet{lii12} investigated the dynamics and collimation properties of these winds. They used a viscous resistive disc with $\alpha_v=0.3$ and $\alpha_m=0.1$ and no stellar wind was allowed this time, leaving only the conical wind. Let us recall here the main properties: \\
(i) The wind is massive  with low velocities, consistent with $v_w= v_{K,i} \sqrt{2 \bar J - 3}$; \\
(ii) The wind is self-confined by the hoop stress whereas wind acceleration is done at the expense of the poloidal electrical current (electric current closure outside the computational box);\\
(iii) The launching zone has a disc magnetization $\mu$ around unity, which decreases quickly radially in the disc;\\
(iv) The wind exists even in the propeller case;\\ 
(v)  The wind remains axisymmetric even with an inclined dipole. 

All these arguments tend to show that conical winds are nothing else than classical ("warm" or dense) disc winds that are not causally connected to the star (as discussed in section~\ref{sec:maes}), albeit from a small radial extent. These winds are tapping the accretion energy and cannot spin down the central cTTS. And, unless they are launched from a much larger radial extent and with a smaller local ejection efficiency (larger magnetic lever arm), they cannot represent YSO jets because of their too low velocities (although they may be representative of FU Orioni outbursts, \citealt{koni11}). Thus, for our purposes here, conical winds are of not much help.

Steady-state conical winds are formed whenever ${\cal P}_m= \alpha_v/\alpha_m$ is about unity, although the viscosity coefficient needs to be high enough ($\alpha_v \geq 0.3$) in order to allow disc material to accrete towards the star. But when the magnetic diffusivity decreases, the stationary situation breaks down and oscillations were reported (see Appendix D2 in \citealt{roma09}). These were interpreted as unsteady versions of the conical winds but it turns out that they are actually a new class of ejection events from the star-disc interaction.  

\subsection{Unsteady Magnetospheric Ejections}
%%%%%%%%%%  
\begin{figure}[t]
\centering   \includegraphics[width=\textwidth]{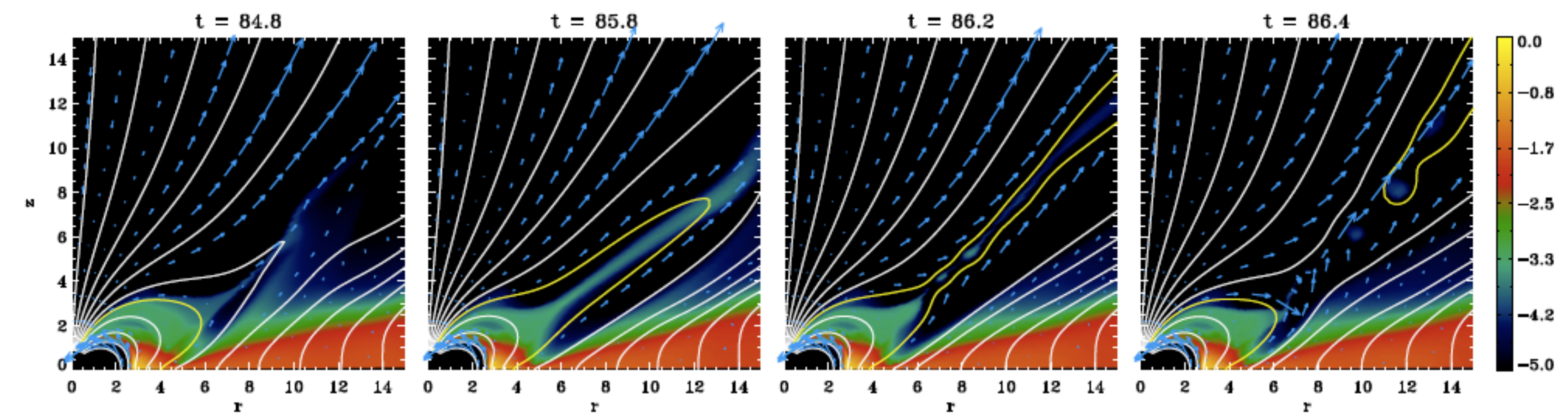}
  \caption{Time sequences in units of stellar rotation period of the periodic inflation/reconnection process which characterizes the magnetic slingshot dynamics of the magnetospheric ejections. The colored background is a logarithmic density map, white solid lines are sample field lines and blue arrows show poloidal speed vectors. The yellow solid lines follow the evolution of a single magnetic surface \citep{zann13}.}
   \label{fig4_5}
\end{figure} 
%%%%%%%%%% 

\begin{figure}[t]
\[ \begin{array}{cc}
  \includegraphics[width=.4\textwidth]{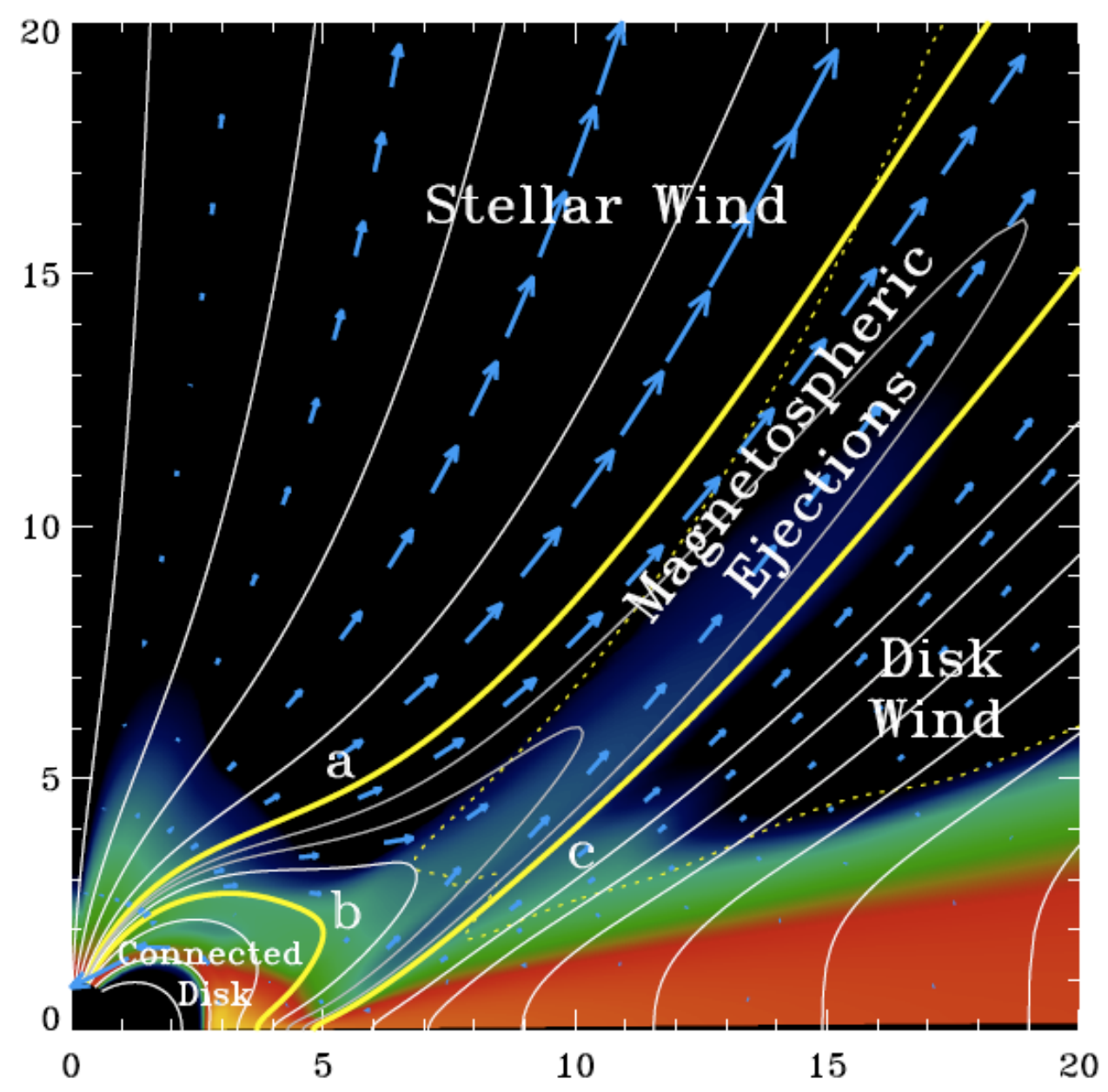}  
  &    
  \includegraphics[width=.5\textwidth]{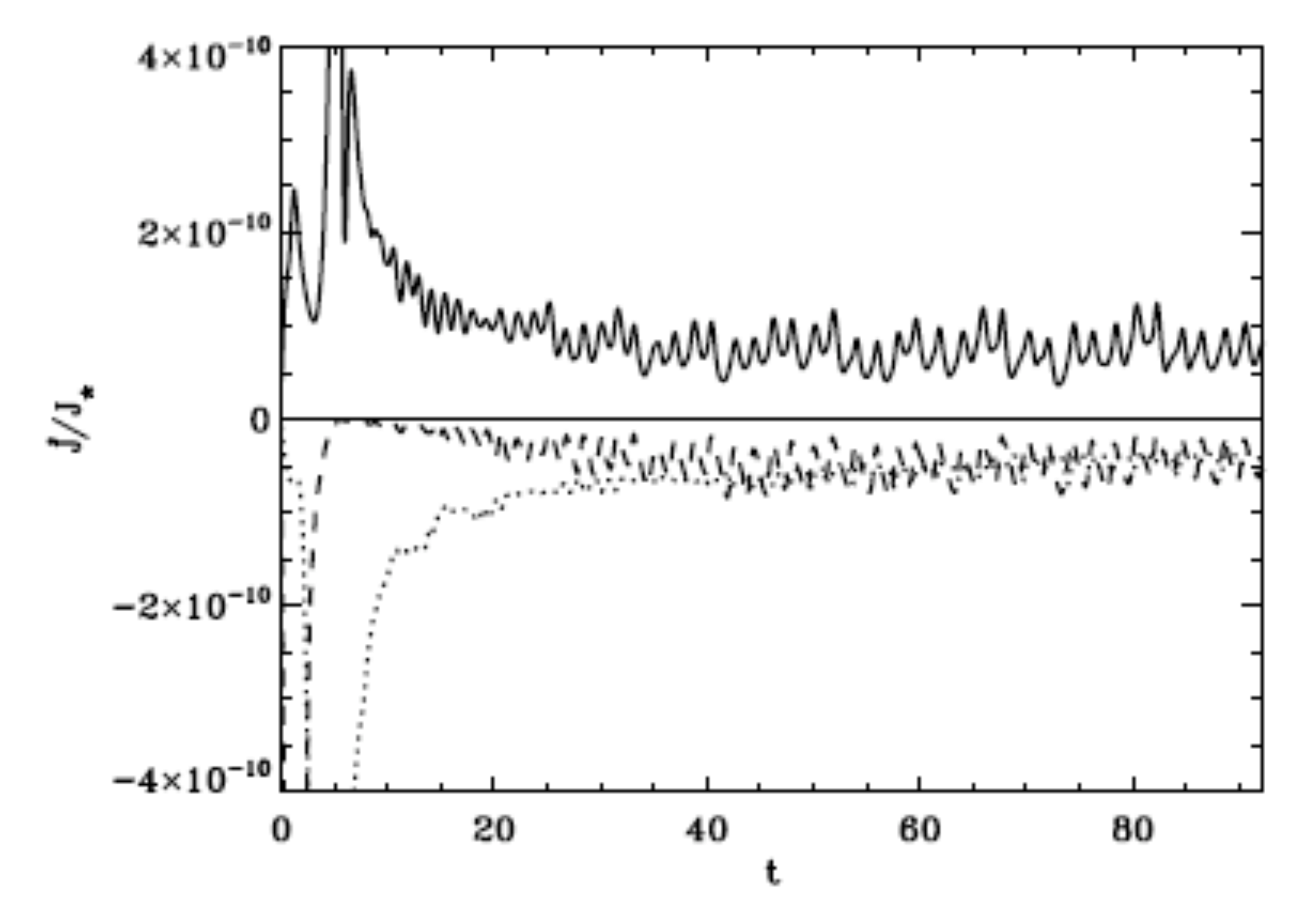}  
\end{array} \]
  \caption{Magnetospheric Ejections \citep{zann13}. {\bf Left:} Global view of the star-disc interacting system, obtained by averaging the simulation outcome over 54 stellar periods. Sample field lines are plotted with white solid lines. Thick yellow field lines, labeled as (a), (b) and (c), delimit the different dynamical constituents of the system indicated in the figure. The dotted line marks the Alfv\'en surface. 
{\bf Right:} Temporal evolution of the torques acting onto the star, normalized to the stellar angular momentum: accretion torque (solid line), stellar wind torque (dotted line) and the torque exerted by the MEs onto the stellar surface (dashed line). A positive (negative) torque spins the stellar rotation up (down).}
  \label{fig4_6}
\end{figure}

\citet{zann13} explored the dynamics of a dipolar magnetosphere interacting with a resistive viscous accretion disc with $ \alpha_v= 0.3, \alpha_m=0.1$, using a simulation box with twice the resolution of \citet{roma09}. The disc gets truncated at the usual radius $r_t$, accretion columns are formed but the magnetic diffusivity in the disc is too low to allow for a steady-state situation. As depicted in Fig.~(\ref{fig4_5}), stellar field lines located at the base of the funnel flow keep on inflating and reconnecting in a quasi periodic fashion. These reconnections give rise to plasmoids very much alike coronal mass ejections in the Sun. They propagate almost ballistically in a narrow channel delimited between two quasi-steady MHD flows, namely the inner stellar wind and an outer disc wind (Fig.~\ref{fig4_6}). The simulation box is too small to follow the collimation properties of these ejecta. However, they carry no electrical current able to confine them and rely therefore on the interplay between the two MHD flows (stellar wind and disc wind). 
   
But what is remarkable is the strong influence of these MEs on the stellar spin evolution. Not only they provide a strong braking torque, but they also decrease significantly the accretion torque. In fact,  {\em the combination of the MEs and the stellar wind does allow a net spin down torque on the star} (Fig.~\ref{fig4_6}). This is the first time were such a result has been reported. 
%%%%%%%%%%%%%%
%%%%%%%%%%%%%%%%%%%%

The reason for this braking efficiency lies on the dynamics of ME production. The innermost field lines connecting the star to the disc are loaded with too much mass and form classical accretion funnel flows. But field lines that are slightly farther away have less inertia and are thereby more controlled by the dynamics of the magnetic field. Because of the low magnetic diffusivity $\alpha_m$ in the disc, the toroidal magnetic field grows (due to the differential rotation between the star and the disc) and these field lines inflate. They are nevertheless loaded with disc material and exert thereby a strong braking torque onto the underlying disc. Simultaneously, stellar material is also being (thermally) pushed away from the star along the same field lines, until these two flows meet at the apex (cusp) of the lines. One reaches therefore a brand new situation, with material rotating more slowly than both the star and the underlying disc but nevertheless connected by the magnetic field. This is nothing else than a "{\em magnetic slingshot}" mechanism, leading to the radial expulsion of this material. The braking torque exerted on both the star and the disc is maintained as long as the field line topology is maintained. In numerical simulations, numerical reconnection comes into play much sooner than what microphysics would provide. We may thus expect even larger torques in real systems.  

In fact, MEs are doing exactly what X-winds were designed for: they extract almost all the disc angular momentum at $r_t$ so that $\Gamma_{acc}$ becomes much smaller. However, this is done at the expense of unsteady events that cannot be described in the framework of a steady-state MHD wind. Moreover, these ejecta are not self-confined and cannot, alone, represent and explain the YSO jet phenomenology. 

Time dependent star-disc interactions with quasi-periodic oscillations of the magnetosphere have been already reported in the literature  \citep{good97, good99b, matt02}. Are these early Episodic Magnetospheric Inflation the same as MEs ? It is hard to tell because the reported simulations did not allow to make a proper physical analysis of what exactly drives the in/out-flowing plasma. The numerical resolution at the star-disc interface was usually not high enough to accurately measure the forces and torques. We note for instance that the phenomenological interpretation involves a poloidal inflation (and reconnection) of the magnetic field lines but that the magnetic slingshot effect was not described. This might be because the focus of these papers was the explanation of the origin of YSO jets. Indeed, these simulations reported the formation of a dense, axially collimated jet, fueled during the inflation phases by mass coming from the accretion funnels and focused on the rotation axis by the expansion of the closed magnetosphere. Such an axial jet was not observed in \citet{zann13}, the axis being filled in with a stellar wind. Clearly, more numerical experiments of this kind should be undertaken.

%%%%%%%%%%%%%%%%%%%%%%%%%%%%%%%%%%%%%%%%%%%%%%%%%%%%%%%%%%%%%
\section{Concluding remarks}
%%%%%%%%%%%%%%%%%%%%%%%%%%%%%%%%%%%%%%%%%%%%%%%%%%%%%%%%%%%%%

There are a number of physical processes at play in a star-disc interaction that have been now understood, albeit only in simplified configurations. They nevertheless allowed to design a new paradigm for the magnetic braking of accreting  young stars.\\

{\em (1) The disc-locking paradigm \`a la Ghosh \& Lamb is ruled out}. The physics of 2D accretion funnel flows are now quite well understood and the disc truncation radius can be safely computed given the stellar dipole field strength and the disc accretion rate. Contrary to initial expectations, an accreting young star cannot deposit its angular momentum back onto the outer disc and is, instead, being spun up by the accreting material. Recent 3D simulations support the underlying physical understanding but more work should be done with fully turbulent discs. Moreover, the effects of higher multipolar field components on shaping the accretion funnels as well as on the efficiency of angular momentum transfer remain to be clarified. But given these caveats, when seeking for an explanation for the stellar spin-down, it seems quite safe now to rely on a star-disc interaction mode that would allow mass, energy and angular momentum to be expelled away from the star+disc system.\\ 

{\em (2) Steady-state models, such as X-winds, conical winds or Accretion Powered Stellar Winds, cannot spin down a protostar. Unsteady, Magnetospheric Ejection events triggered at the star-disc interface are the best candidates}. The drawback of such a solution is that it requires thorough analyses of accurately controlled time-dependent numerical simulations. And, because reconnection is so important in limiting the braking torque, one should implement a sub-grid model for reconnection in MHD simulations. It has been shown that the combined action of MEs and stellar winds is probably needed. However, stellar winds in numerical simulations are badly controlled as the mass flux comes from a boundary condition. A more physical setup should be therefore sought.\\    

{\em (3) There is no single model that can represent alone the YSO jet phenomenology: YSO jets are probably made of several components interacting with each other}. The early tempting suggestion that the stellar rotational energy is powering jets is facing overwhelming theoretical and observational difficulties.\\ 

Putting all these constraints together leads to the following, more complex, picture:
\begin{enumerate}
\item The bulk of the mass in YSO jets is carried by an extended magnetized disc wind, providing also a magnetic sheath allowing to confine other outflows coming from the central regions.
\item  A magnetic star-disc interaction, mainly through the (inclined) stellar dipole, allows to efficiently extract angular momentum that is expelled away in (uncollimated) plasmoids by the magnetic slingshot mechanism. These intermittent, massive bullets are then forced to follow a magnetic channel, as long as the outer magnetized disc wind is present. MEs would thus be the answer to the long lasting "disc-locking" issue in cTTS. They would also naturally provide a source of jet variability, compression and finally heating, necessary ingredients to make YSO jets shine. 
\item  An Accretion Powered Stellar Wind is certainly always present in cTTS, filling in the central axial part of YSO jets with a hot, probably tenuous, spine. Such a spine would carry nevertheless a fraction of the accretion energy released in the shock onto the star and provide an extra (and necessary) braking torque. How much accretion energy can be transferred, through waves, from the shocked region to the open field lines region, remains an issue that deserves further theoretical investigations.
\item This accretion shock and all related consequences critically depend on the complexity of the magnetic field topology near the stellar surface. This is a fascinating topic, where observations (spectro-polarimetry technics, lines and photometric monitoring) and 3D MHD simulations can be fruitfully combined. 

\end{enumerate}    

One might, at first sight, consider this paradigm as too complex. But note that no simple explanation addresses the ensemble of observational data. In the above paradigm there would be no jets (i.e. self-confined supersonic flows) if the outer magnetized disc wind is not present. Now, winds of that kind require a large scale magnetic field whose origin is probably related to the early in-falling stage (see P. Hennebelle's contribution, this book). The magnetic flux threading the disc would therefore be reminiscent of the initial conditions of stellar formation and may well vary from one TTS to another. Besides, for a given star, the disc magnetic flux is expected to evolve in time. But this is another story.

%%-----------------------------
%%      your bibliography
%%-----------------------------

%\begin{thebibliography}{99}

%\bibliographystyle{/sw/share/texmf-dist/tex/latex/aa/bibtex/aa}
%\bibliography{/Users/ferreira/mesDocuments/recherche/tex/references}

\end{document}